\documentstyle[prd,aps,psfig,twocolumn,amstex,floats]{revtex}
\input epsf

\renewcommand{\d}{\mbox{d}}
\newcommand{\ie}{\mbox{\rm i.e.}}
\newcommand{\eg}{\mbox{\rm e.g.}}
\newcommand{\ifm}[1]{\relax\ifmmode #1\else $#1$\fi}
\newcommand{\etal}{{\it et al.}}
\newcommand{\Dzero}{D\O}
\newcommand{\uaone}{UA1 Collaboration}
\newcommand{\uatwo}{UA2 Collaboration}
\newcommand{\cdf}{CDF Collaboration}
\newcommand{\dzero}{\Dzero\ Collaboration}
\newcommand{\alephc}{ALEPH Collaboration}
\newcommand{\opalc}{OPAL Collaboration}
\newcommand{\PRL}{Phys. Rev. Lett.}
\newcommand{\PL}{Phys. Lett.}
\newcommand{\PR}{Phys. Rev.}
\newcommand{\NP}{Nucl. Phys.}
\newcommand{\NIM}{Nucl. Instrum. Methods in Phys. Res.}
\newcommand{\ZP}{Z.~Phys.}
\newcommand{\GEAN}{{\sc geant}}

\newcommand{\HERW}{{\sc herwig}}
\newcommand{\PDFL}{{\sc pdflib}}

\newcommand{\abseta}{\ifm{|\eta|}}

\newcommand{\qt}{\ifm{q_T}}
\newcommand{\px}{\ifm{p_x}}
\newcommand{\py}{\ifm{p_y}}
\newcommand{\pz}{\ifm{p_z}}
\newcommand{\mt}{\ifm{m_T}}

\newcommand{\pt}{\ifm{p_T}}
\renewcommand{\pt}{\ifm{p_T}}

\newcommand{\mpt}{\mbox{$\rlap{\kern0.1em/}\pt$}}
\newcommand{\mptv}{\mbox{$\rlap{\kern0.1em/}\vec\pt$}}
\newcommand{\PM}{\ifm{\pm}}
\newcommand{\lt}{\ifm{<}}
\newcommand{\gt}{\ifm{>}}

\newcommand{\Eg}{\ifm{E(\gamma)}}
\newcommand{\phig}{\ifm{\phi(\gamma)}}
\newcommand{\etag}{\ifm{\eta(\gamma)}}

\newcommand{\ut}{\ifm{u_T}}
\newcommand{\utv}{\ifm{\vec\ut}}
\newcommand{\upar}{\ifm{u_{\parallel}}}
\newcommand{\uper}{\ifm{u_{\perp}}}

\newcommand{\phir}{\ifm{\phi(R)}}
\newcommand{\wb}{\ifm{W}}

\newcommand{\wev}{\ifm{W\to e\nu}}
\newcommand{\wegam}{\ifm{\wev\gamma}}
\newcommand{\wtv}{\ifm{W\to \tau\nu}}
\newcommand{\wte}{\ifm{W\to \tau\nu\to e\nu\overline\nu\nu}}
\newcommand{\wth}{\ifm{W\to \tau\nu\to \hbox{hadrons}+X}}

\newcommand{\mw}{\ifm{M_{W}}}
\newcommand{\wwidth}{\ifm{\Gamma_{W}}}
\newcommand{\ptw}{\ifm{\pt(W)}}

\newcommand{\zb}{\ifm{Z}}
\newcommand{\zee}{\ifm{Z\to ee}}

\newcommand{\mz}{\ifm{M_{Z}}}

\newcommand{\dphiee}{\ifm{\Delta\phi(ee)}}
\newcommand{\ptee}{\ifm{\pt(ee)}}

\newcommand{\pee}{\ifm{p(ee)}}
\newcommand{\peev}{\ifm{\vec\pee}}
\newcommand{\mee}{\ifm{m(ee)}}

\newcommand{\tev}{\ifm{\tau\to e\nu\overline\nu}}
\newcommand{\qqbar}{\ifm{q\overline{q}}}
\newcommand{\bbbar}{\ifm{b\overline{b}}}
\newcommand{\ee}{\ifm{e^+e^-}}
\newcommand{\ppbar}{\ifm{p\overline{p}}}

\newcommand{\GeV}{\mbox{GeV}}

\newcommand{\GeVm}{\mbox{GeV}}

\newcommand{\MeVm}{\mbox{MeV}}

\newcommand{\Ee}{\ifm{E(e)}}

\newcommand{\pe}{\ifm{p(e)}}
\newcommand{\pev}{\ifm{\vec\pe}}
\newcommand{\pte}{\ifm{\pt(e)}}
\newcommand{\ptev}{\ifm{\vec\pte}}
\newcommand{\peonev}{\ifm{\vec p(e_1)}}
\newcommand{\petwov}{\ifm{\vec p(e_2)}}
\newcommand{\phie}{\ifm{\phi(e)}}
\newcommand{\te}{\ifm{\theta(e)}}
\newcommand{\etae}{\ifm{\eta(e)}}

\newcommand{\xcal}{\ifm{x_{\rm cal}}}
\newcommand{\ycal}{\ifm{y_{\rm cal}}}
\newcommand{\zcal}{\ifm{z_{\rm cal}}}
\newcommand{\xtrk}{\ifm{x_{\rm trk}}}
\newcommand{\ytrk}{\ifm{y_{\rm trk}}}
\newcommand{\ztrk}{\ifm{z_{\rm trk}}}
\newcommand{\xvtx}{\ifm{x_{\rm vtx}}}
\newcommand{\yvtx}{\ifm{y_{\rm vtx}}}
\newcommand{\zvtx}{\ifm{z_{\rm vtx}}}
\newcommand{\fiso}{\ifm{f_{\rm iso}}}

\newcommand{\sigm}{\ifm{\sigma_{\rm trk}}}

\newcommand{\pnuz}{\ifm{\pz(\nu)}}

\newcommand{\ptnu}{\ifm{\pt(\nu)}}
\newcommand{\ptnuv}{\ifm{\vec\ptnu}}
\newcommand{\phinu}{\ifm{\phi(\nu)}}

\newcommand{\rrec}{\ifm{{\rm R_{rec}}}}
\newcommand{\sigrec}{\ifm{\sigma_{\rm rec}}}
\newcommand{\dupar}{\ifm{\Delta \upar}}
\newcommand{\mdupar}{\ifm{\langle \dupar \rangle}}
\newcommand{\alphaem}{\ifm{\alpha_{\rm EM}}}
\newcommand{\deltaem}{\ifm{\delta_{\rm EM}}}
\newcommand{\alphamb}{\ifm{\alpha_{\rm mb}}}
\newcommand{\srec}{\ifm{s_{\rm rec}}}
\newcommand{\alpharec}{\ifm{\alpha_{\rm rec}}}
\newcommand{\betarec}{\ifm{\beta_{\rm rec}}}

\newcommand{\regam}{\ifm{\Delta R(e\gamma)}}
\newcommand{\rzero}{\ifm{R_0}}

\newcommand{\fss}{\ifm{f_{\rm ss}}}
\newcommand{\alphacdc}{\ifm{\alpha_{\rm CDC}}}

\newcommand{\alphacc}{\ifm{\alpha_{\rm CC}}}
\newcommand{\gtwo}{\ifm{g_2}}

\newcommand{\cem}{\ifm{c_{\rm EM}}}
\newcommand{\sem}{\ifm{s_{\rm EM}}}
\newcommand{\nem}{\ifm{n_{\rm EM}}}
\newcommand{\lqcd}{\ifm{\Lambda_{\rm QCD}}}

\begin{document}
\pagestyle{myheadings}
\onecolumn

\title{ A Measurement of the \wb\ Boson Mass}
\author{                                                                      
B.~Abbott,$^{30}$                                                             
M.~Abolins,$^{27}$                                                            
B.S.~Acharya,$^{45}$                                                          
I.~Adam,$^{12}$                                                               
D.L.~Adams,$^{39}$                                                            
M.~Adams,$^{17}$                                                              
S.~Ahn,$^{14}$                                                                
H.~Aihara,$^{23}$                                                             
G.A.~Alves,$^{10}$                                                            
N.~Amos,$^{26}$
% ...
\\                                                                            
\vskip 0.50cm                                                                 
\centerline{(D\O\ Collaboration)}                                             
\vskip 0.50cm                                                                 
}                                                                             
\address{                                                                     
\centerline{$^{1}$Universidad de los Andes, Bogot\'{a}, Colombia}             
\centerline{$^{2}$University of Arizona, Tucson, Arizona 85721}               
\centerline{$^{3}$Boston University, Boston, Massachusetts 02215}             
\centerline{$^{4}$Brookhaven National Laboratory, Upton, New York 11973}      
\centerline{$^{5}$Brown University, Providence, Rhode Island 02912}           
\centerline{$^{6}$Universidad de Buenos Aires, Buenos Aires, Argentina}       
\centerline{$^{7}$University of California, Davis, California 95616}          
\centerline{$^{8}$University of California, Irvine, California 92697}         
\centerline{$^{9}$University of California, Riverside, California 92521}      
\centerline{$^{10}$LAFEX, Centro Brasileiro de Pesquisas F{\'\i}sicas,        
                  Rio de Janeiro, Brazil}                                     
\centerline{$^{11}$CINVESTAV, Mexico City, Mexico}                            
\centerline{$^{12}$Columbia University, New York, New York 10027}             
\centerline{$^{13}$Delhi University, Delhi, India 110007}                     
\centerline{$^{14}$Fermi National Accelerator Laboratory, Batavia,            
                   Illinois 60510}                                            
\centerline{$^{15}$Florida State University, Tallahassee, Florida 32306}      
\centerline{$^{16}$University of Hawaii, Honolulu, Hawaii 96822}              
\centerline{$^{17}$University of Illinois at Chicago, Chicago,                
                   Illinois 60607}                                            
\centerline{$^{18}$Indiana University, Bloomington, Indiana 47405}            
\centerline{$^{19}$Iowa State University, Ames, Iowa 50011}                   
\centerline{$^{20}$Korea University, Seoul, Korea}                            
\centerline{$^{21}$Kyungsung University, Pusan, Korea}                        
\centerline{$^{22}$Institute of Nuclear Physics, Krak\'ow, Poland}            
\centerline{$^{23}$Lawrence Berkeley National Laboratory and University of    
                   California, Berkeley, California 94720}                    
\centerline{$^{24}$Louisiana Tech University, Ruston, Louisiana 71272}        
\centerline{$^{25}$University of Maryland, College Park, Maryland 20742}      
\centerline{$^{26}$University of Michigan, Ann Arbor, Michigan 48109}         
\centerline{$^{27}$Michigan State University, East Lansing, Michigan 48824}   
\centerline{$^{28}$Moscow State University, Moscow, Russia}                   
\centerline{$^{29}$University of Nebraska, Lincoln, Nebraska 68588}           
\centerline{$^{30}$New York University, New York, New York 10003}             
\centerline{$^{31}$Northeastern University, Boston, Massachusetts 02115}      
\centerline{$^{32}$Northern Illinois University, DeKalb, Illinois 60115}      
\centerline{$^{33}$Northwestern University, Evanston, Illinois 60208}         
\centerline{$^{34}$University of Notre Dame, Notre Dame, Indiana 46556}       
\centerline{$^{35}$University of Oklahoma, Norman, Oklahoma 73019}            
\centerline{$^{36}$University of Panjab, Chandigarh 16-00-14, India}          
\centerline{$^{37}$Institute for High Energy Physics, 142-284 Protvino,       
                   Russia}                                                    
\centerline{$^{38}$Purdue University, West Lafayette, Indiana 47907}          
\centerline{$^{39}$Rice University, Houston, Texas 77005}                     
\centerline{$^{40}$Universidade do Estado do Rio de Janeiro, Brazil}          
\centerline{$^{41}$University of Rochester, Rochester, New York 14627}        
\centerline{$^{42}$CEA, DAPNIA/Service de Physique des Particules,            
                   CE-SACLAY, Gif-sur-Yvette, France}                         
\centerline{$^{43}$Seoul National University, Seoul, Korea}                   
\centerline{$^{44}$State University of New York, Stony Brook,                 
                   New York 11794}                                            
\centerline{$^{45}$Tata Institute of Fundamental Research,                    
                   Colaba, Mumbai 400005, India}                              
\centerline{$^{46}$University of Texas, Arlington, Texas 76019}               
\centerline{$^{47}$Texas A\&M University, College Station, Texas 77843}       
}                                                                             

\maketitle

\vspace{0.5in}
\begin{abstract}
We present a measurement of the \wb\ boson mass using data collected by the
\Dzero\ experiment at the Fermilab Tevatron during 1994--1995. We identify \wb\
bosons by their decays to $e\nu$ final states. We extract the \wb\ mass, \mw, by
fitting the transverse mass and transverse electron momentum spectra
from a sample of 28{,}323 \wev\ decay candidates. We use a sample of
3{,}563 dielectron events, mostly due to \zee\ decays, to constrain our model of
the detector response.  From the transverse mass fit we measure $\mw = 80.
44\pm0.10(stat)\pm0.07(syst)$ GeV. Combining this with our previously published
result from data taken in 1992--1993, we obtain $\mw = 80.43\pm0.11$ GeV.
\end{abstract}


\twocolumn

\section{ Introduction }
\label{sec-intro}
In this article we describe the most precise measurement to date of the mass of
the \wb\ boson, using data collected in 1994--1995 with the \Dzero\ detector at
the Fermilab Tevatron \ppbar\ collider 
\cite{IAdamthesis,Flattum_thesis,1bprl}. 

The study of the properties of the \wb\ boson began in  1983 with its discovery
by the UA1 \cite{UA1_W_discovery} and UA2 \cite{UA2_W_discovery} collaborations
at the CERN \ppbar\ collider. Together with the discovery of the \zb\ boson in
the same year \cite{UA1_Z_discovery,UA2_Z_discovery}, it provided a direct
confirmation of the unified model of the weak and electromagnetic interactions
\cite{SM}, which -- together with QCD -- is now called the Standard Model.

Since the \wb\ and \zb\ bosons are carriers of the weak force, their properties
are intimately coupled to the structure of the model. The properties of the
\zb\ boson have been studied in great detail in \ee\ collisions \cite{mz}. 
The study of the \wb\ boson has proven to be significantly more difficult, since
it is charged and therefore can not be resonantly produced in \ee\ collisions.
Until recently its direct study has therefore been the realm of experiments at 
\ppbar\ colliders which have performed the most precise direct measurements of
the \wb\ boson mass \cite{UA2,CDF,D0}. Direct measurements of the \wb\
boson mass have also been carried out at LEP2 \cite{OPAL,DELPHI,L3,ALEPH} using
nonresonant \wb\ pair production. A summary of these
measurements can be found in Table~\ref{tab:mw} at the end of this article.

\begin{figure}[htb]
\vspace{-0.5cm}
\centerline{\psfig{figure=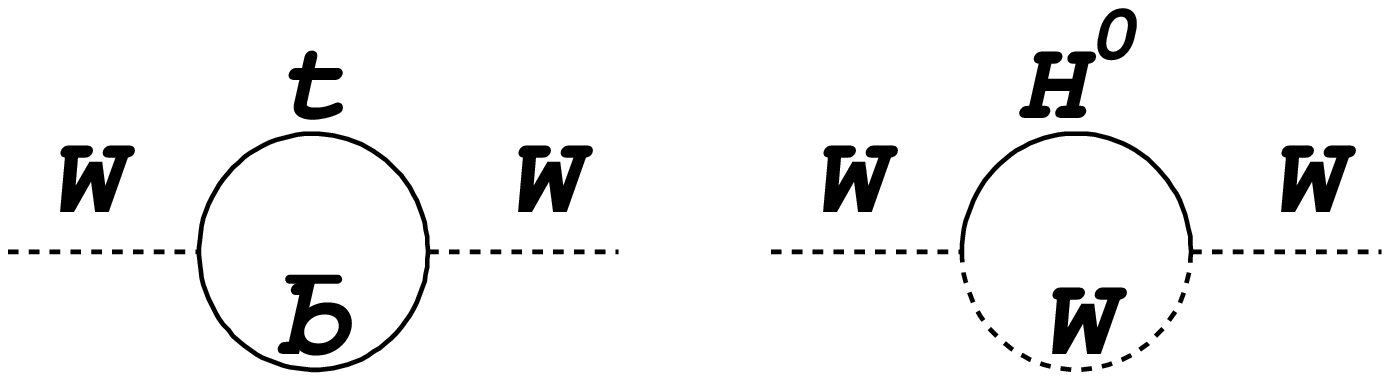,width=4.2in,height=1.9in}}
\vspace{-1cm}
\caption{ Loop diagrams contributing to the \wb\ boson mass.}
\label{fig:loop}
\end{figure}

The Standard Model links the \wb\ boson mass to other parameters,
\begin{equation}
\mw = \left({\pi\alpha\over \sqrt{2} G_F}\right)
^{1\over2}{1\over\sin\theta_W\sqrt{1-\Delta r}}.
\label{eq:mw1}
\end{equation}
In the ``on shell'' scheme \cite{onshell}
\begin{equation}
\cos\theta_W = {\mw\over\mz},
\label{eq:mw2}
\end{equation}
where $\theta_W$ is the weak mixing angle. Aside from the radiative corrections
$\Delta r$, the \wb\ boson mass is thus
determined by three precisely measured quantities, the mass of the \zb\ boson
\mz \cite{mz}, the Fermi constant $G_F$ \cite{PDG}
and the electromagnetic coupling constant $\alpha$ evaluated at $Q^2=\mz^2$
\cite{alphaZ}:
\begin {eqnarray}
\mz & = & 91.1865\pm0.0020\ \hbox{GeV}, \label{eq:mz} \\
G_F & = & (1.16639\pm0.00002)\times10^{-5}\ \hbox{GeV}^{-2}, \\
\alpha & = & (128.896\pm0.090)^{-1}.
\end{eqnarray}
From the measured \wb\ boson mass we can derive the size of
the radiative corrections $\Delta r$. Within
the framework of the Standard Model, these corrections are dominated
by loops involving the top quark and the Higgs boson (see
Fig.~\ref{fig:loop}). The correction from the $t\overline b$ loop is substantial
because of the large mass difference between the two quarks. It is proportional
to $m_t^2$ for large values of the top quark mass $m_t$. Since $m_t$ has
been measured \cite{mtop_lj}, this
contribution can be calculated within the Standard Model. For a large Higgs
boson mass, $m_H$, the correction from the Higgs loop is proportional to $\ln
m_H$. In extensions to the Standard Model new particles may give rise
to additional corrections to the value of \mw. In the Minimal Supersymmetric
extension of the Standard Model (MSSM), for example,
additional corrections can increase the predicted
\wb\ mass by up to 250 MeV~\cite{susy}.

A measurement of the \wb\ boson mass therefore constitutes a test of
the Standard Model.
In conjunction with a measurement of the top quark mass
the Standard Model predicts \mw\ up to a 200 MeV uncertainty due to the
unknown Higgs boson mass. 
By comparing with the measured value of the \wb\ boson mass
we can constrain the mass of the Higgs boson,
the agent of the electroweak symmetry breaking that has up to now eluded
experimental detection.
A discrepancy with the range allowed by the Standard Model could indicate new
physics.
The experimental challenge is thus to measure the \wb\
boson mass to sufficient precision, about 0.1\%, to be sensitive to these
corrections.

\section{Overview}
\label{sec-overview}
\subsection {Conventions}
We use a Cartesian coordinate system with the $z$-axis defined by
the direction of the proton beam, the $x$-axis pointing radially out of the
Tevatron ring and the $y$-axis pointing up. A vector $\vec p$ is then defined
in terms of its projections on these three axes, \px, \py, \pz.  Since
protons and antiprotons in the Tevatron are unpolarized,
all physical processes are invariant
with respect to rotations around the beam direction.
It is therefore
convenient to use a cylindrical coordinate system, in which the same
vector is given by the magnitude of its component transverse to the
beam direction, \pt, its azimuth $\phi$, and \pz. In \ppbar\ collisions
the center of mass frame of the parton-parton collisions is approximately at
rest in the plane transverse to the beam direction but has an undetermined
motion along the beam direction. Therefore the plane transverse to the beam
direction is of special importance and sometimes we work with two-dimensional
vectors defined in the $x$-$y$ plane. They are written with a subscript
$T$, \eg\ $\vec \pt$. We also use spherical coordinates by replacing
\pz\ with the colatitude
$\theta$ or the pseudorapidity $\eta=-\ln\tan\left(\theta/2\right)$. The origin
of the coordinate system is in general the reconstructed position of the \ppbar\
interaction when describing the interaction, and the geometrical
center of the detector when describing the detector.
For convenience, we use units in which $c=\hbar=1$.

\subsection {\wb\ and \zb\ Boson Production and Decay}

\begin{figure}[ht]
\vspace{ -2.9cm}
\centerline{\psfig{figure=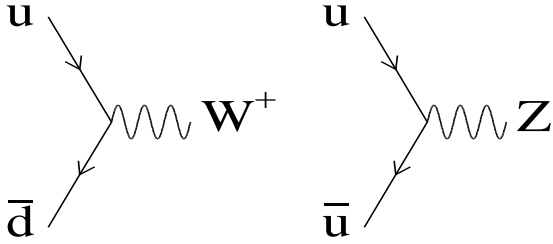,height=8cm,width=10cm}}
\vspace{-.8cm }
\caption{ Lowest order diagrams for \wb\ and \zb\ boson production.}
\label{fig:wzproduction}
\end{figure}

In \ppbar\ collisions at $\sqrt{s}=1.8$ TeV, \wb\ and \zb\ bosons are produced
predominantly through quark-antiquark annihilation. Figure
\ref{fig:wzproduction} shows the lowest-order diagrams. The quarks in the
initial state may radiate gluons which are usually
very soft but may sometimes be energetic enough to give rise to hadron jets in
the detector. In the reaction the initial proton and antiproton
break up and the fragments hadronize. We refer to everything except the
vector boson and its decay products collectively as the underlying event. Since
the initial proton and antiproton momentum vectors add to zero, the same must be
true for the vector sum of all final state momenta and therefore the vector
boson recoils against all particles in the underlying event. The sum of the
transverse momenta of the recoiling particles must balance the transverse
momentum of the boson, which is typically small compared to its mass
but has a long tail to large values.

We identify \wb\ and
\zb\ bosons by their leptonic decays. The \Dzero\ detector
(Sec.~\ref{sec-exper})
is best suited for a precision measurement of electrons and
positrons\footnote{In the following we use ``electron'' generically for
both electrons and positrons.}, and we therefore use the decay channel
\wev\ to measure the \wb\ boson mass. \zee\ decays serve as an important
calibration sample. About 11\% of the \wb\ bosons decay to $e\nu$ and about
3.3\% of the \zb\ bosons decay to $ee$. The leptons typically have
transverse momenta  of about half the mass of the decaying boson and are
well isolated from other large energy deposits in the calorimeter. Intermediate
vector boson
decays are the dominant source of isolated high-\pt\ leptons at the Tevatron,
and therefore these decays allow us to select a clean sample of \wb\ and \zb\
boson decays.

\subsection {Event Characteristics}

In events due to the process $\ppbar\to(\wev)+X$, where $X$ stands for the
underlying event, we detect the electron and all
particles recoiling against the \wb\ with pseudorapidity $-4<\eta<4$. The
neutrino escapes undetected.
In the calorimeter we cannot
resolve individual recoil particles, but we measure their energies summed over
detector segments. Recoil particles with $|\eta| \gt 4$ escape unmeasured
through
the beampipe, possibly carrying away substantial momentum along the beam
direction. This means that we cannot measure the sum of the $z$-components of
the recoil momenta, $u_z$, precisely. Since these particles escape at a very
small angle
with respect to the beam, their transverse momenta are typically small and can
be neglected in the sum of the transverse recoil momenta, \utv. We measure \utv\
by summing the observed energy flow vectorially over all detector segments.
Thus, we reduce the reconstruction of every candidate event to a measurement of
the electron momentum \pev\ and \utv.

Since the neutrino escapes undetected, the sum of all measured final state
transverse momenta does not add to zero. The missing transverse momentum \mptv,
required to balance the transverse momentum sum, is a measure of the transverse
momentum of the neutrino.
The neutrino momentum component along the beam direction cannot be determined,
because $u_z$ is not measured well.
The signature of a \wev\ decay is therefore an isolated
high-\pt\ electron and large missing transverse momentum.

In the case of \zee\ decays the signature consists of two isolated high-\pt\
electrons and we measure the momenta of both leptons, \peonev\ and \petwov, and
\utv\ in the detector.

\subsection {Mass Measurement Strategy}

Since \pnuz\ is unknown, we cannot reconstruct the $e\nu$ invariant mass for
\wev\ candidate events and therefore must resort to other kinematic variables
for the mass measurement.

For recent measurements \cite{UA2,CDF,D0} the transverse mass,
\begin{equation}
\mt = \sqrt{2 \pte \ptnu \left(1-\cos\left(\phie-\phinu\right)\right)}
\label{eq:mtdefn}
\end{equation}
was used. This variable has the advantage that its spectrum is relatively
insensitive to the production dynamics of the \wb. Corrections to \mt\
due to the motion
of the \wb\ are of order $\left(\qt/\mw\right)^2$,
where \qt\ is the transverse momentum of the \wb\ boson. It is also insensitive
to selection biases that prefer certain event topologies
(Sec.~\ref{sec-elec-upar}). However, it makes use of
the inferred neutrino \pt\ and is therefore sensitive to the response of the
detector to the recoil particles.

The electron \pt\ spectrum provides an alternative measurement of the \wb\ mass.
It is measured with better resolution than the neutrino \pt\ and is insensitive
to the recoil momentum measurement. However, its shape is sensitive to the
motion
of the \wb\ and receives corrections of order $\qt/\mw$. It thus requires a
better understanding of the \wb\ boson production dynamics than the \mt\
spectrum.

\begin{figure}[ht]
\vskip -0cm
\centerline{\psfig{figure=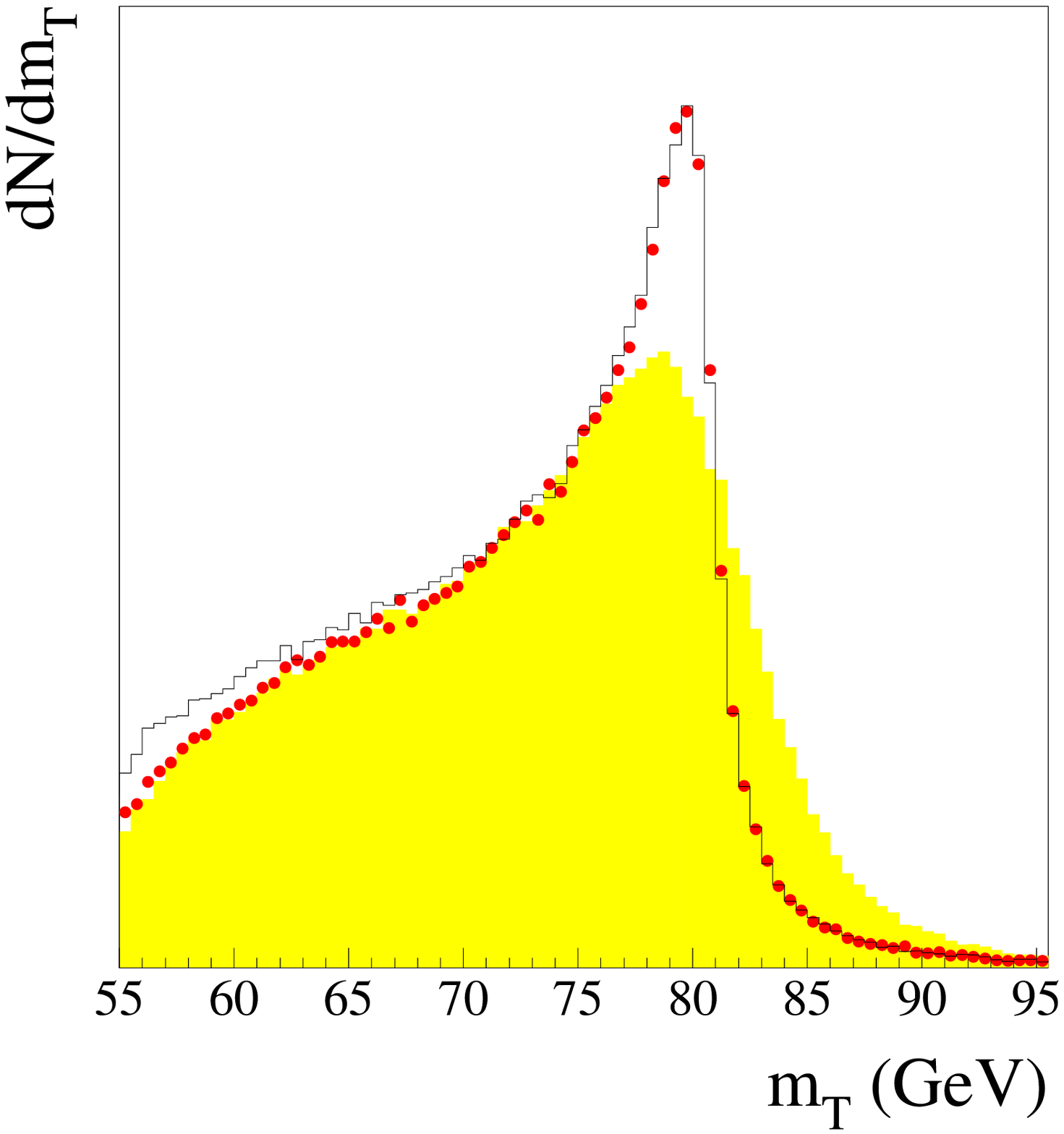,height=3.5in,width=3.5in}}
\caption{The \mt\ spectrum for \wb\ bosons with $q_T=0$ (------), with the
correct $q_T$ distribution ($\bullet$), and with detector resolutions (shaded).}
\label{fig:sensitivities1}
\end{figure}

\begin{figure}[ht]
\vskip -0cm
\centerline{\psfig{figure=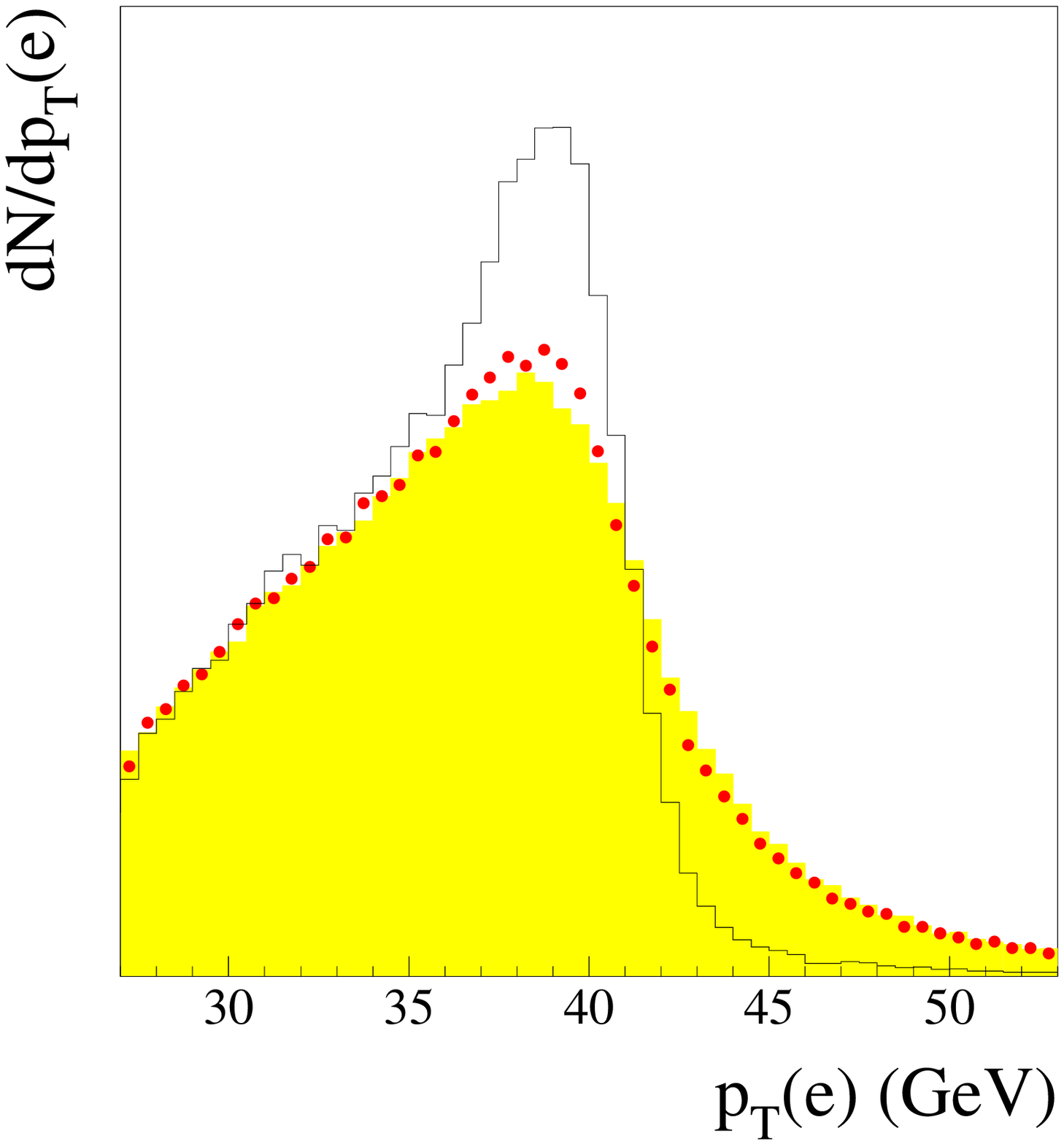,height=3.5in,width=3.5in}}
\caption{The \pte\ spectrum for \wb\ bosons with $q_T=0$ (------), with the
correct $q_T$ distribution ($\bullet$), and with detector resolutions (shaded).}
\label{fig:sensitivities2}
\end{figure}

The \mt\ and \pte\ spectra thus provide us with two complementary measurements.
This is illustrated in Figs.~\ref{fig:sensitivities1} and
\ref{fig:sensitivities2}, which show the effect of the motion of the
\wb\ bosons and the detector resolutions on the shape of each of the two
spectra.
The solid line shows the shape of the distribution before the detector
simulation and with \qt=0. The points show the shape after \qt\ is added to
the system, and the shaded histogram also includes the detector simulation.
We observe that the shape of the \mt\ spectrum is dominated by detector
resolutions and the
shape of the \pte\ spectrum by the motion of the \wb. By performing the
measurement using both spectra we provide a powerful cross-check with
complementary systematics.

Both spectra are equally sensitive to the electron energy response of the
detector. We calibrate this response by forcing the observed dielectron mass
peak in the \zee\ sample to agree with the known \zb\ mass\cite{mz}
(Sec.~\ref{sec-elec}). This means that we effectively measure the ratio
of \wb\ and \zb\ masses, which is equivalent to a measurement of the \wb\ mass
because the \zb\ mass is known precisely.

To carry out these measurements we perform a maximum likelihood fit to the
spectra. Since the shape of the spectra, including all
the experimental effects, cannot be computed analytically, we need a Monte Carlo
simulation program that can predict the shape of the spectra as a function of
the \wb\ mass.  To perform a measurement of the \wb\ mass to a
precision of order 100 MeV we have to estimate individual systematic effects
to 10 MeV. This requires a Monte Carlo sample of
2.5 million accepted \wb\ bosons for each such effect. The program
therefore must be capable of generating large samples in a reasonable time. We
achieve the required performance by employing a parameterized model of the
detector response.

We next summarize the aspects of the accelerator and detector
that are important for our measurement (Sec.~\ref{sec-exper}). Then we
describe the data selection (Sec.~\ref{sec-data}) and
the fast Monte Carlo model (Sec.~\ref{sec-mc}). Most parameters in
the model are determined from our data. We describe the determination of
the various components of the Monte Carlo model in
Secs.~\ref{sec-elec}-\ref{sec-back}.
After tuning the model we fit the
kinematic spectra (Sec.~\ref{sec-fit}),
perform some consistency checks (Sec.~\ref{sec-checks}),
and discuss the systematic uncertainties (Sec.~\ref{sec-syst}).
Section~\ref{sec-results} summarizes the results and presents
the conclusions.

\section {Experimental Setup}
\label{sec-exper}
\subsection{ Accelerator }
The Fermilab Tevatron\cite{Tevatron} collides proton and antiproton
beams at a center-of-mass energy of $\sqrt{s}=1.8$ TeV. Six bunches each
of protons and antiprotons circulate around the ring in opposite directions.
Bunches cross at the intersection regions every 3.5 $\mu$s. During the
1994--1995 running period, the accelerator reached a peak luminosity of
$2.5\times10^{31} \hbox{cm}^{-2} \hbox{s}^{-1}$ and
delivered an integrated luminosity of
about 100 pb$^{-1}$.

The Tevatron tunnel also houses a 150 GeV proton synchrotron, called the Main
Ring, which is used as an injector for the Tevatron. The Main Ring also serves
to accelerate protons for antiproton production during collider operation.
Since the Main Ring beampipe passes through the outer section of the \Dzero\
calorimeter, passing proton bunches give rise to backgrounds in the detector.
We eliminate this background using timing cuts based on the accelerator clock
signal.

\subsection{ Detector }
\label{sec-detect}

\subsubsection{Overview}
\begin{figure}[htpb!]
\vspace{0.5in}
\centerline{\psfig{figure=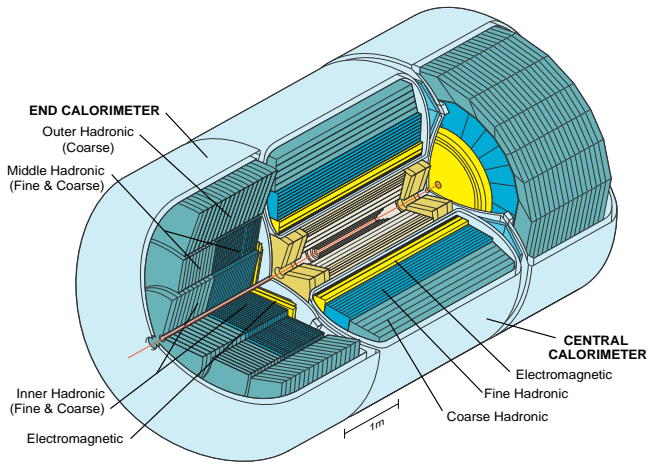,width=3.5in,height=2.5in}}
\vspace{-0.02in}
\caption{ A cutaway view of the \Dzero\ calorimeter and tracking system.}
\label{fig:d0cal}
\end{figure}

The \Dzero\ detector consists of three major subsystems: a central
detector, a calorimeter (Fig.~\ref{fig:d0cal}), and a muon spectrometer.
It is described in
detail in Ref.~\cite{d0nim}. We describe only the features that are most
important for this measurement.

\subsubsection{Central Detector}
The central detector is designed to measure the trajectories of charged
particles.
It consists of a vertex drift chamber, a transition
radiation detector, a central drift chamber (CDC), and two forward drift
chambers (FDC). There is no central magnetic field.
The CDC covers the region $|\eta|<1.0$. It is a jet-type drift chamber
with delay lines to give the hit coordinates in the $r$-$z$ plane.
The FDC covers the region $1.4<\abseta<3.0$.

\subsubsection{Calorimeter}
\label{sec-exper-cal}

The calorimeter is the most important part of the
detector for this measurement. It is a sampling calorimeter and uses uranium
absorber plates and liquid argon as the
active medium. It is divided into three parts: a central calorimeter
(CC) and two end calorimeters (EC), each housed in its own cryostat.
Each is segmented into an electromagnetic (EM) section, a fine
hadronic (FH) section, and a coarse hadronic (CH) section, with increasingly
coarser sampling. The CC-EM section is constructed of 32 azimuthal modules.
The entire calorimeter is divided into about 5000 pseudo-projective
towers, each covering 0.1$\times$0.1 in $\eta\times\phi$. The EM section
is segmented into four layers, 2, 2, 7, and 10 radiation lengths
thick. The third layer, in which electromagnetic showers typically reach their
maximum, is transversely segmented into cells covering 0.05$\times$0.05 in
$\eta\times\phi$. The hadronic section is segmented into four layers (CC) or
five layers (EC). The entire calorimeter is 7--9 nuclear interaction lengths
thick. There are no projective cracks in the calorimeter and it provides
hermetic and almost uniform coverage for particles with $\abseta<4$.
Figure~\ref{fig:d0cal} shows a view of the calorimeter and the central detector.

The signals from arrays of 2$\times$2 calorimeter towers, covering
0.2$\times$0.2 in $\eta\times\phi$, are added together
electronically for the EM section only and for all sections, and shaped with a
fast rise time for use in the Level 1 trigger. We refer to these arrays of
2$\times$2 calorimeter towers as ``trigger towers".

\begin{figure}[htpb!]
\vspace{-0.3in}
\centerline{\psfig{figure=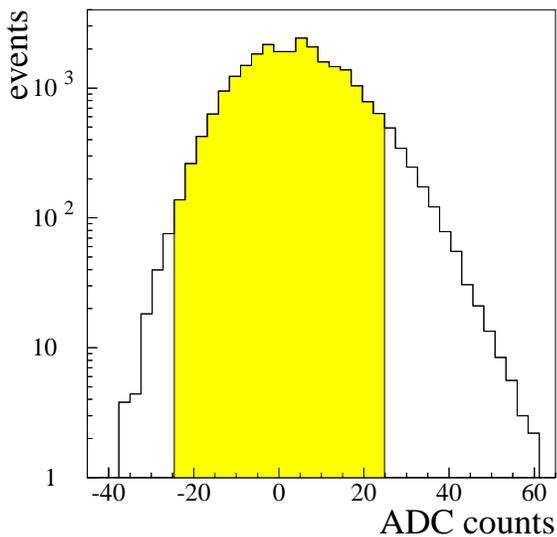,width=3.5in,height=3.5in}}
\vspace{-0.02in}
\caption{ The pedestal spectrum of a central calorimeter cell, where the mean
pedestal has been subtracted.   The shaded region are the events removed by the
zero-suppression. }
\label{fig:calped}
\end{figure}

Figure~\ref{fig:calped} shows the pedestal spectrum of a calorimeter cell. The
spectrum has an asymmetric tail from ionization caused by the intrinsic
radioactivity of the uranium absorber plates. The data are corrected such
that the mean pedestal is zero for each cell. To reduce the amount of data that
have to be stored, the calorimeter readout is zero-suppressed.
Only cells with a signal that deviates from zero by more than twice the
rms of the pedestal distribution are read out. This region of the pedestal
spectrum is indicated by the shaded region in Fig.~\ref{fig:calped}. Due to
its asymmetry, the spectrum does not average to zero after
zero-suppression. Thus the zero-suppression effectively causes a pedestal shift.

\begin{figure}[htpb!]
\vspace{-0.3in}
\centerline{\psfig{figure=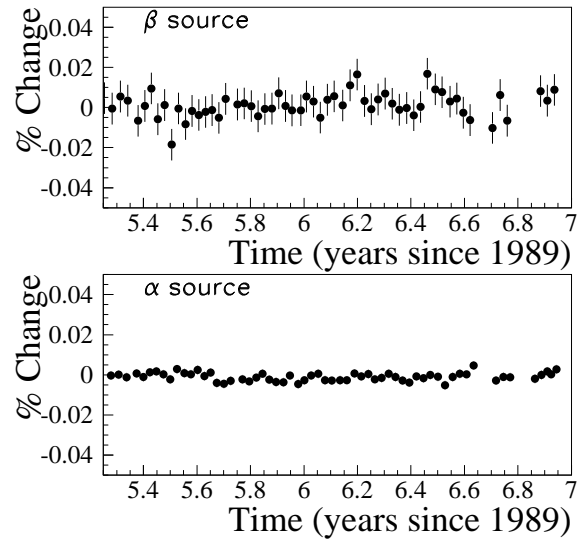,width=3.5in,height=3.5in}}
\vspace{-0.02in}
\caption{ The response of the liquid argon in the central calorimeter as
monitored by $\alpha$ and $\beta$ sources.}
\label{fig:calgain}
\end{figure}

\begin{figure}[htpb!]
\vspace{-0.in}
\begin{tabular}{c}
\epsfxsize = 6.cm \epsffile[0 0 400 400]{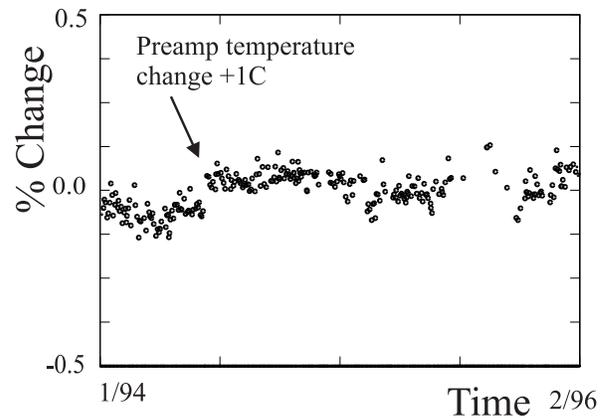}
\end{tabular}
\vspace{-0.in}
\caption{ The percentage change in the central calorimeter gains over the course
of the run.} 
\label{fig:calelec1}
\end{figure}

\begin{figure}[htpb!]
\vspace{-0.3in}
\begin{tabular}{c}
\epsfxsize = 6.0cm \epsffile[0 0 400 400]{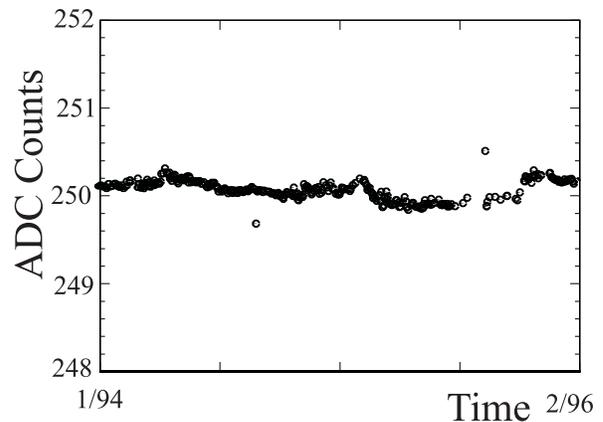}
\end{tabular}
\vspace{-0.in}
\caption{ The change in the central calorimeter pedestals over the course of the
run.} 
\label{fig:calelec2}
\end{figure}

The liquid argon has unit gain and therefore the calorimeter response was
extremely stable during the entire run. Figure~\ref{fig:calgain} shows the
response of the liquid argon as monitored with radioactive sources of
$\alpha$ and $\beta$ particles. Figures~\ref{fig:calelec1}
and \ref{fig:calelec2} show the gains and
pedestals of a typical readout channel throughout the run.

The EM calorimeter provides a measurement of energy and position of the
electrons from the \wb\ and \zb\ decays. Due to the fine segmentation of the
third layer, we can measure the position of the shower centroid with a
precision of 2.5 mm in the azimuthal direction and 1 cm in the $z$-direction.

\begin{figure}[htpb!]
\vspace{-0.in}
\centerline{\psfig{figure=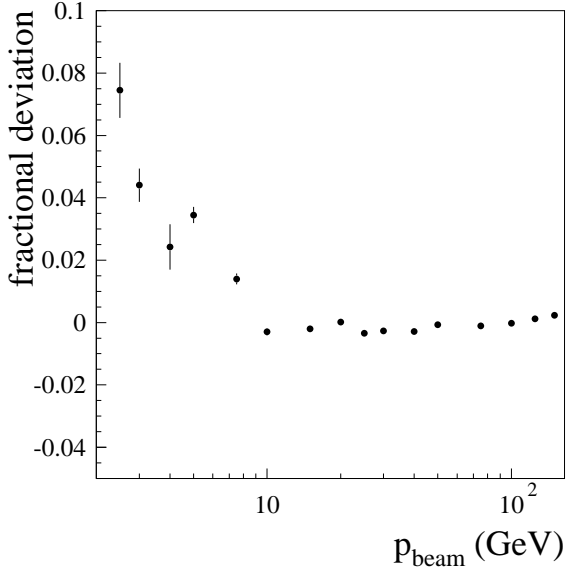,width=3.5in,height=3.5in}}
\vspace{-0.in}
\caption{ The fractional deviation of the reconstructed electron energy
from the beam momentum from beam tests of a CC-EM module.}
\label{fig:tbresponse}
\end{figure}

We study the response of the EM calorimeter to electrons in beam tests
\cite{testbeam}. To reconstruct the
electron energy we add the signals $a_i$ observed in each EM layer
($i=1\ldots4$) and the first FH layer ($i=5$) of an array of  5$\times$5
calorimeter towers, centered on the most energetic tower,
weighted by a layer dependent sampling weight $s_i$,
\begin{equation}
E = A \sum_{i=1}^5 s_i a_i - \deltaem.
\label{eq:emresponse}
\end{equation}
To determine the sampling weights we minimize
\begin{equation}
\chi^2 = \sum {(p_{\rm beam}-E)\over \sigma_{\rm EM}^2},
\end{equation}
where the sum runs over all events and $\sigma_{\rm EM}$ is the resolution given
in Eq. \ref{eq:emresolution}.
We obtain $A=2.96$ MeV/ADC count, $\deltaem =-347$ MeV, $s_1=1.31$, $s_2=0.85$,
$s_4=0.98$, and $s_5=1.84$. We arbitrarily fix $s_3=1$. The value of $\deltaem$
depends on the amount of dead material in front of the calorimeter.  The
parameters $s_1$ to $s_4$ weight
the four EM layers and $s_5$ the first FH layer.
Figure~\ref{fig:tbresponse} shows the fractional deviation of $E$ as a
function of the beam momentum $p_{\rm beam}$. Above 10 GeV they deviate by less
than 0.3\% from each other.

\begin{figure}[htpb!]
\vspace{-0.3in}
\centerline{\psfig{figure=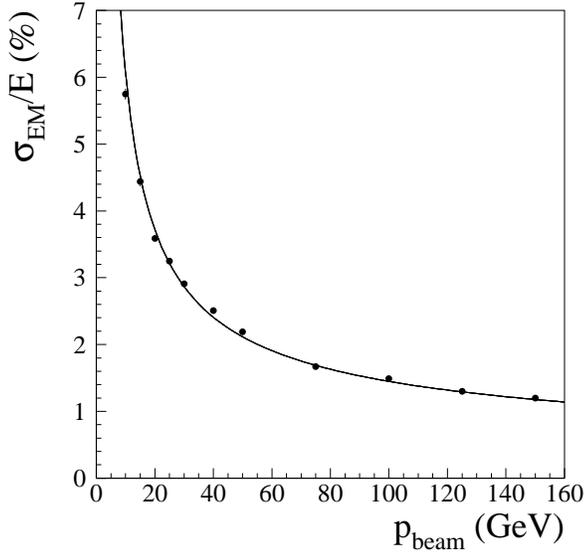,width=3.5in,height=3.5in}}
\vspace{-0.02in}
\caption{ The fractional electron energy resolution measured in beam tests of a
CC-EM module for the data ($\bullet$) and the parameterization (-----).}
\label{fig:tbresolution}
\end{figure}

The fractional energy resolution can
be parameterized as a function of electron energy using constant, sampling, and
noise terms as
\begin{equation}
\left({\sigma_{\rm EM}\over E}\right)^2 = \cem^2 +
\left({\sem \over\sqrt{E\sin\theta}}\right)^2 + \left({\nem \over E}\right)^2
\label{eq:emresolution}
\end{equation}
with $\cem =0.003$, $\sem =0.135$ GeV$^{1/2}$
\cite{Zhu_thesis,Heuring_thesis}, and $\nem =0.43$ GeV in the central
calorimeter. The angle $\theta$ is the colatitude of the electron.
Figure~\ref{fig:tbresolution} shows the fractional electron energy resolution
versus beam momentum for a CC-EM module. The line shows the parametrization
of the resolution from Eq.~\ref{eq:emresolution}.

\subsubsection {Luminosity Monitor}
Two arrays of scintillator hodoscopes, mounted in front of the EC cryostats,
register hits with a 220 ps time resolution. They serve to detect that
an inelastic \ppbar\ interaction has taken place. The particles from the
breakup of the proton give rise to hits in the hodoscopes on one side of the
detector that are tightly clustered in time. The detector has a 91\% acceptance
for inelastic \ppbar\ interactions. For events with a single interaction
the location of the interaction vertex can be determined with a resolution of 3
cm from the time difference between the hits on the two sides of the detector
for use in the Level 2 trigger.
This array is also called the Level 0 trigger because the detection of an
inelastic \ppbar\ interaction is a basic requirement of most trigger
conditions.

\subsubsection{Trigger}
Readout of the detector is controlled by a two-level trigger system.

Level 1 consists of an and-or network, that can be programmed to trigger on a
\ppbar\ crossing if a number of preselected conditions are true. The Level 1
trigger decision is taken within the 3.5 $\mu$s time interval between crossings.
As an extension to Level 1, a trigger processor (Level 1.5) may be invoked
to execute simple algorithms on the limited information available at the time
of a Level 1 accept. For electrons, the processor uses the energy deposits in
each trigger tower as inputs. The detector cannot accept any triggers until the
Level 1.5 processor completes execution and accepts or rejects the event.

Level 2 of the trigger consists of a farm of 48 VAXstation 4000's.
At this level the complete event is available.
More sophisticated algorithms refine the
trigger decisions and events are accepted based on preprogrammed conditions.
Events accepted by Level 2 are written to magnetic tape for offline
reconstruction.

\section{ Data Selection }
\label{sec-data}
\subsection{Trigger }
\label{sec:trigger}

The conditions required at trigger Level 1 for \wb\ and \zb\ candidates are:
\begin{itemize}
\item $\underline{\hbox{\ppbar\ interaction:}}$ Level 0 hodoscopes register
hits consistent with a \ppbar\ interaction. This condition accepts 98.6\%
of all \wb\ and \zb\ bosons produced.

\item $\underline{\hbox{Main Ring Veto:}}$ No Main Ring proton bunch passes
through the detector less than 800 ns before or after the   \ppbar\ crossing
and no protons were injected into the Main Ring less than 400 ms before the
\ppbar\ crossing.

\item $\underline{\hbox{EM trigger towers:}}$ There are one or more EM trigger
towers with $E\sin\theta>T$, where $E$ is the energy measured in the tower,
$\theta$ its angle with the beam measured from the center of the detector, and
$T$ a programmable threshold. This requirement is fully efficient for electrons
with $\pt>2 T$.
\end{itemize}

The Level 1.5 processor recomputes the transverse electron energy by adding the
adjacent EM trigger tower with the largest signal to the EM trigger tower that
exceeded the Level 1 threshold. In addition, the signal in the EM trigger tower
that exceeded the Level 1 threshold must constitute at least 85\% of the signal
registered in this tower if the hadronic layers are also included. This EM
fraction requirement is fully efficient for electron candidates that pass our
offline selection (Sec.~\ref{sec-data_samples}).

Level 2 uses the EM trigger tower that exceeded the Level 1 threshold as a
starting point.
The Level 2 algorithm finds the most energetic of the four calorimeter
towers that make up the trigger tower, and sums the energy in the EM
sections of a 3$\times$3 array of calorimeter towers around it.
It checks the longitudinal shower shape by applying cuts on the fraction of the
energy in the different EM layers. The transverse shower shape is characterized
by the energy deposition pattern in the third EM layer. The difference between
the energies in concentric regions covering 0.25$\times$0.25 and
0.15$\times$0.15 in
$\eta\times\phi$ must be consistent with an electron.
Level 2 also imposes an isolation
condition requiring
\begin{eqnarray}
 \frac{\sum_iE_i\sin\phi_i - \pt}{\pt} \lt 0.15
\end{eqnarray}
where the sum runs over
all cells within a cone of radius $R=\sqrt{\Delta\phi^2+\Delta\eta^2}=0.4$
around the electron direction and \pt\ is the transverse momentum of the
electron~\cite{McKinley_thesis}.

\begin{figure}[htpb!]
\vspace{-0.3in}
\centerline{\psfig{figure=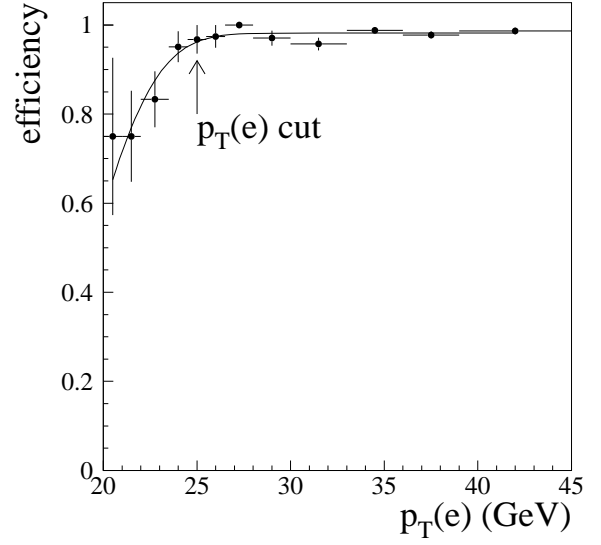,width=3.5in,height=3.5in}}
\vspace{-0.02in}
\caption{ The relative efficiency of the Level 2 electron filter
for a threshold of 20 GeV.  The arrow indicates the cut applied in the final
event selection.  }
\label{fig:ptetrig}
\end{figure}

The \pt\ of the electron computed at Level 2 is based
on its energy and the $z$-position of
the interaction vertex measured by the Level 0 hodoscopes. Level 2 accepts
events that have a minimum number of EM clusters that satisfy the shape cuts
and have \pt\ above a preprogrammed threshold. Figure
\ref{fig:ptetrig} shows the measured relative efficiency of the Level 2 electron
filter versus electron \pt\ for a Level 2 \pt\ threshold of 20 GeV.
We determine this efficiency using \zb\ data taken with a lower
threshold value (16 GeV).
The efficiency is the fraction of electrons above a Level 2 \pt\
threshold of 20 GeV.
The curve is the parameterization used in the fast Monte Carlo.

\begin{figure}[htpb!]
\vspace{-0.3in}
\centerline{\psfig{figure=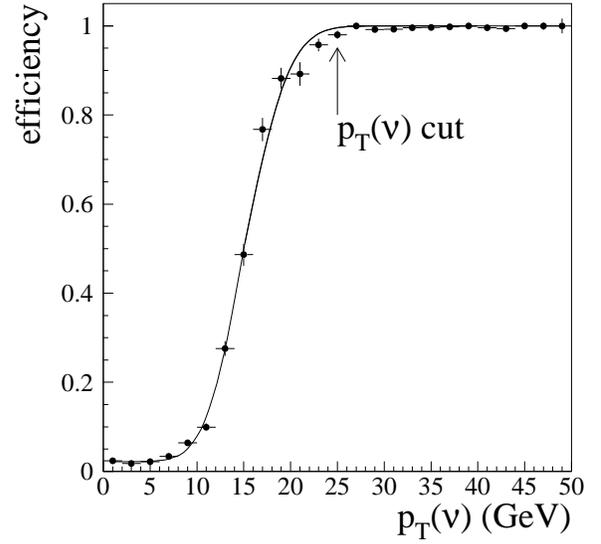,width=3.5in,height=3.5in}}
\vspace{-0.02in}
\caption{ The efficiency of a 15 GeV Level 2 \mpt\ requirement.
The arrow indicates the cut applied in the final
event selection.  }
\label{fig:ptmisstrig}
\end{figure}

Level 2 also computes the missing transverse momentum based on the energy
registered in each calorimeter cell and the vertex $z$-position.
We determine the efficiency curve for a 15 GeV Level 2 \mpt\ requirement
from data taken without the Level 2 \mpt\ condition. Figure
\ref{fig:ptmisstrig} shows the measured efficiency versus \ptnu.
The curve is the parameterization used in the fast Monte Carlo.

\subsection{ Reconstruction }

\subsubsection{ Electron }
\label{sec-data-elec}

We identify electrons as clusters of adjacent calorimeter cells with significant
energy deposits. Only clusters with at least 90\% of
their energy in the EM section and at least 60\% of their energy in the
most energetic calorimeter tower are considered as electron candidates.
For most electrons we also reconstruct a track in the CDC or FDC that points
towards the centroid of the cluster.

We compute the electron energy \Ee\ from the signals in all cells of the EM
layers and the first FH layer in a window covering 0.5$\times$0.5 in
$\eta\times\phi$ and centered on the tower which registered the highest fraction
of the electron energy. In the computation we use the sampling weights and
calibration constants determined using
the testbeam data (Sec.~\ref{sec-exper-cal}) except for the offset
$\deltaem$, which we take from an in situ calibration
(Sec.~\ref{sec-EMresponse}), \ie\ $\deltaem =-0.16$ GeV
for electrons in the CC.

The calorimeter shower centroid position (\xcal, \ycal, \zcal), the
center of gravity of the track (\xtrk, \ytrk, \ztrk) and the proton
beam trajectory define the electron direction. The shower
centroid algorithm is documented in Appendix~\ref{app:pos_cc}. The center
of gravity of the CDC track is defined by the mean hit coordinates of all the
delay line hits on the track. The calibration of the measured $z$-coordinates
contributes a significant systematic uncertainty to the \wb\ boson mass
measurement and is described in Appendices~\ref{app:pos_cdc} and
\ref{app:pos_cc}. Using tracks from many events reconstructed in the vertex
drift chamber, we measure the beam trajectory for every run.
The closest approach to the beam trajectory of the line through shower centroid
and track center of gravity defines the position of the interaction
vertex (\xvtx, \yvtx, \zvtx). In \zee\ events we may have
two electron candidates with tracks. In this case we take the point midway
between the vertex positions determined from each electron as the
interaction vertex. Using only the electron track to determine the position of
the interaction vertex, rather than all tracks in the event, makes the
resolution of this measurement less sensitive to the
luminosity and avoids confusion
between vertices in events with more than one \ppbar\ interaction.

We then define the azimuth \phie\ and the colatitude \te\ of the electron using
the vertex and the shower centroid positions,
\begin{eqnarray}
\tan\phie & = & {\ycal - \yvtx \over \xcal - \xvtx }, \\
\tan\te & = &   {\sqrt{\xcal^2+\ycal^2} - \sqrt{\xvtx^2+\yvtx^2}
                        \over \zcal - \zvtx }.
\end{eqnarray}
Neglecting the electron mass, the momentum of the electron is given by
\begin{equation}
\pev = \Ee \left(\begin{array}{l} \sin\te\cos\phie \\
                                  \sin\te\sin\phie \\
                                  \cos\te            \end{array} \right).
\end{equation}

\subsubsection{ Recoil }
We reconstruct the transverse momentum of all particles recoiling against the
\wb\ or \zb\ boson by taking the vector sum
\begin{equation}
\utv = \sum_i E_i \sin\theta_i \left(
\begin{array}{c} \cos\phi_i \\ \sin\phi_i \end{array} \right),
\end{equation}
where the sum runs over all calorimeter cells that were read out, except those
that belong to electron clusters. $E_i$ are the
cell energies, and $\phi_i$ and $\theta_i$ are the azimuth and colatitude of
the center of cell $i$ with respect to the interaction vertex.

\subsubsection{ Derived Quantities }
\begin{figure}[htb]
\vspace{-0.3in}
\centerline{\psfig{figure=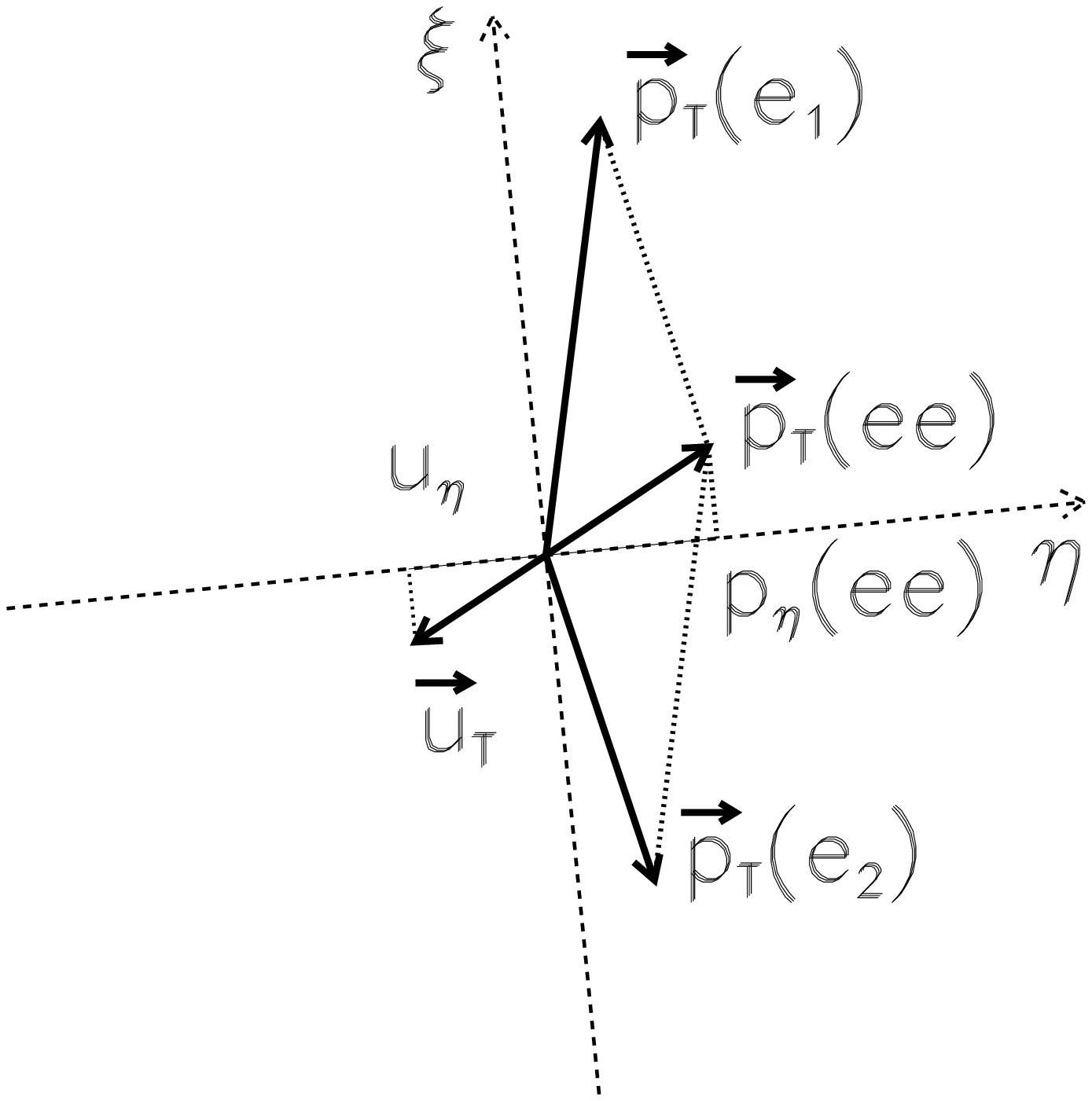,width=3.5in,height=3.5in}} 
\vspace{-0.2in}
\caption{ Illustration of momentum vectors in the transverse plane
for \zee\ candidates. The vectors drawn with thick lines are directly
measured.}
\label{fig:zdef}
\end{figure}

In the case of \zee\ decays we define the dielectron momentum
\begin{equation}
\peev = \peonev + \petwov
\end{equation}
and the dielectron invariant mass
\begin{equation}
\mee = \sqrt{2 E(e_1) E(e_2) (1-\cos\omega)},
\end{equation}
where $\omega$ is the opening angle between the two electrons. It is
useful to define a coordinate system in the plane transverse
to the beam that depends only on the electron directions. We follow the
conventions first introduced by UA2\cite{UA2} and call the axis
along the inner bisector of the two electrons the $\eta$-axis and the axis
perpendicular to that the $\xi$-axis. Projections on these axes are denoted
with subscripts $\eta$ or $\xi$. Figure~\ref{fig:zdef} illustrates these
definitions.

\begin{figure}[htb]
\vspace{-0.3in}
\centerline{\psfig{figure=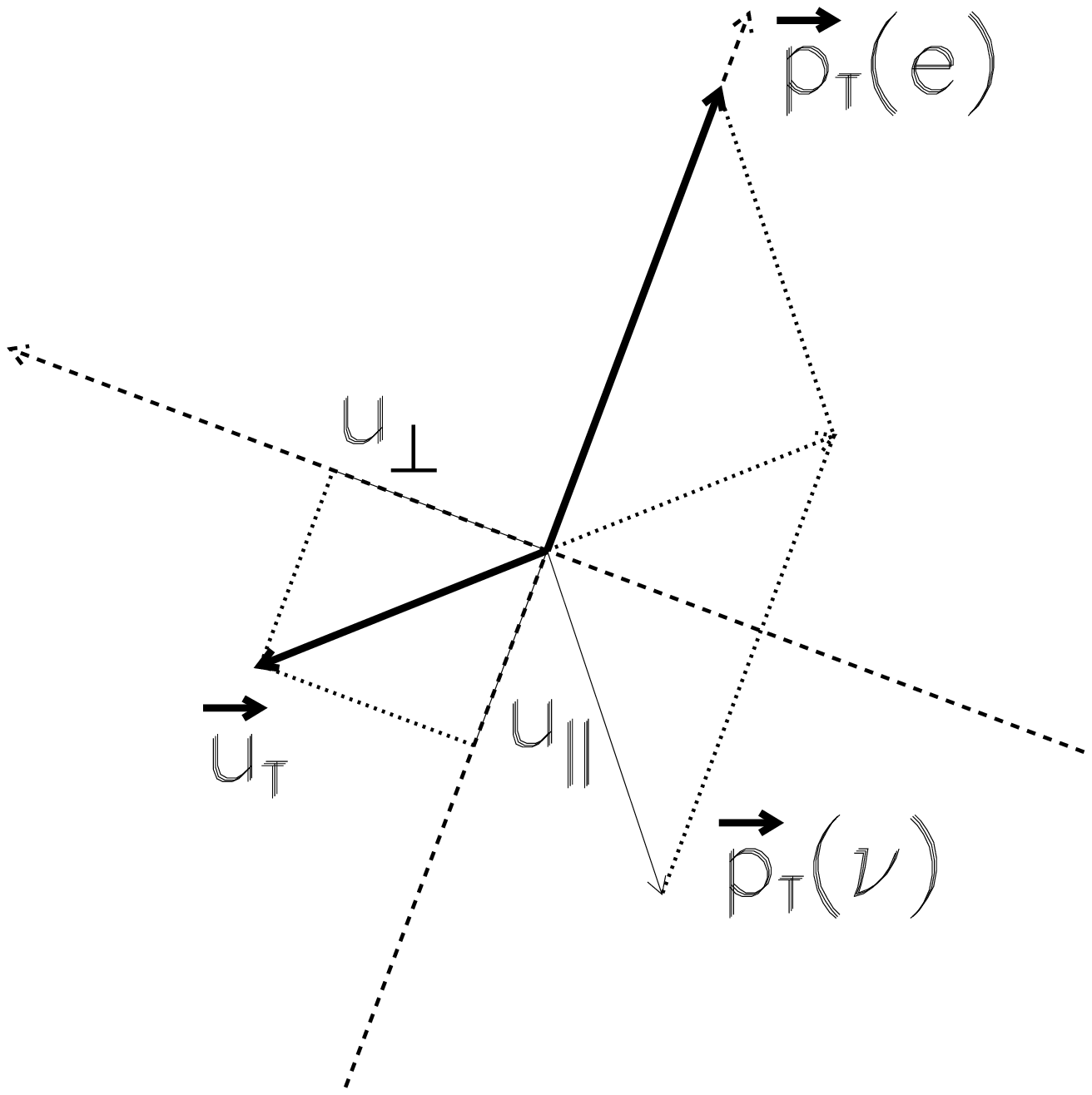,width=3.5in,height=3.5in}} 
\vspace{-0.2in}
\caption{ Illustration of momentum vectors in the transverse plane
for \wev\  candidates. The vectors drawn with thick lines are directly
measured.}
\label{fig:wdef}
\end{figure}

In case of \wev\ decays we define the transverse neutrino momentum
\begin{equation}
\ptnuv = -\ptev - \utv
\end{equation}
and the transverse mass (Eq.~\ref{eq:mtdefn}).
Useful quantities are the projection of the transverse recoil momentum on the
electron direction,
\begin{equation}
\upar = \utv\cdot\hat \pte,
\end{equation}
and the projection on the direction perpendicular to the electron direction,
\begin{equation}
\uper = \utv\cdot(\hat\pte\times\hat z).
\end{equation}
Figure~\ref{fig:wdef} illustrates these definitions.

\subsection{ Electron Identification }

\subsubsection{ Fiducial Cuts}
To ensure a uniform response we accept only electron candidates that are well
separated in azimuth ($\Delta\phi$) from the calorimeter module boundaries in
the CC-EM and from the edges of the calorimeter by cutting on $\Delta\phi$ and
\zcal.
We also remove electrons for which the $z$-position of the track center
of gravity is near the edge of the CDC.
For electrons in the EC-EM we cut on the index of the most energetic
tower, $i_\eta$. Tower 15 covers
$1.4<\eta<1.5$ with respect to the detector center and tower 25 covers
$2.4<\eta<2.5$.

\subsubsection{ Quality Variables }
We test how well the shape of a cluster agrees with that expected for an
electromagnetic shower by computing a
quality variable ($\chi^2$) for all cell energies using a
41-dimensional covariance matrix. The covariance matrix was determined from
\GEAN\ \cite{geant} based simulations \cite{top_prd}.

To determine how well a track matches a cluster we extrapolate the track to the
third EM layer in the calorimeter and compute the distance between the
extrapolated track and the cluster centroid in the azimuthal direction,
$\Delta s$, and in the $z$-direction, $\Delta z$. The variable
\begin{equation}
\sigm^2 = \left({\Delta s\over \delta s}\right)^2 + \left({\Delta z\over \delta
z}\right)^2,
\end{equation}
quantifies the quality of the match. In the
EC-EM $z$ is replaced by $r$, the radial distance from the
center of the detector.  The parameters $\delta s=0.25$ cm,
$\delta z=2.1$ cm, and $\delta r=1.0$ cm are the resolutions with which
$\Delta s$, $\Delta z$, and $\Delta r$ are measured, as
determined with the electrons from \wev\ decays.

In the EC, electrons must have a matched track in the forward drift chamber. In
the CC, we define ``tight'' and ``loose'' criteria. The tight criteria require a
matched track in the CDC. The loose criteria do not require a matched track and
help increase the electron finding efficiency for \zee\ decays.

The isolation fraction is defined as
\begin{equation}
    \fiso = {E_{\rm cone}-E_{\rm core}\over E_{\rm core}},
\end{equation}
where $E_{\rm cone}$ is the energy in a cone of
radius $R=\sqrt{\Delta\phi^2+\Delta\eta^2}=0.4$ around the direction of
the electron, summed over the entire depth of the calorimeter and
$E_{\rm core}$ is the energy in a cone of $R=0.2$, summed over the EM
calorimeter only.

\begin{figure}[htpb!]
\vspace{-0.3in}
\centerline{\psfig{figure=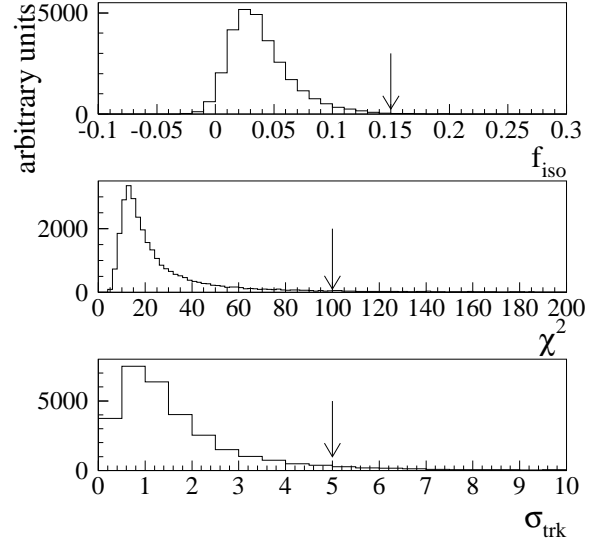,width=3.5in,height=3.5in}}
\vspace{-0.in}
\caption{ Distributions of the electron identification variables.
The arrows indicate the cut values. }
\label{fig:eid}
\end{figure}

Figure~\ref{fig:eid} shows the distributions of the three quality variables
for electrons in the CC with the arrow showing the cut values.
Table~\ref{tab:e_selection} summarizes the electron selection criteria.

\subsection{ Data Samples }
\label{sec-data_samples}
The data were taken during the 1994--1995 Tevatron run. After the removal of
runs in which parts of the detector were not operating adequately, they amount
to an integrated luminosity of about 82 pb$^{-1}$.
We select \wb\ decay candidates by requiring:

\begin{tabular}{ll}
Level 1: & \ppbar\ interaction \\
         & Main Ring Veto \\
	 & EM trigger tower above 10 GeV \\
Level 1.5: & $\geq 1$ EM cluster above 15 GeV \\
Level 2: & electron candidate with $\pt>20$ GeV \\
         & momentum imbalance $\mpt>15$ GeV \\
offline: & $\geq 1$ tight electron candidate in CC \\
         & $\pte > 25$ GeV \\
         & $\ptnu > 25$ GeV \\
         & $\ut < 15$ GeV \\
\end{tabular}

\noindent
We select \zb\ decay candidates by requiring:

\begin{tabular}{ll}
Level 1: & \ppbar\ interaction \\
         & $\geq 2$ EM trigger towers above 7 GeV \\
Level 1.5: & $\geq 1$ EM cluster above 10 GeV \\
Level 2: & $\geq 2$ electron candidates with $\pt>20$ GeV \\
offline: & $\geq 2$ electron candidates \\
	 & $\pte > 25$ GeV \\
         & $70<\mee<110$ GeV \\
\end{tabular}

\noindent We accept \zee\ decays with at least one electron candidate in the
CC and the other in the CC or the EC.
One CC candidate must pass the tight electron
selection criteria. If the other candidate is also in the
CC it may pass only the
loose criteria. We use the 2{,}179 events with both electrons in the CC
(CC/CC \zb\ sample) to calibrate the
calorimeter response to electrons (Sec.~\ref{sec-elec}). These events need
not pass the Main Ring Veto cut because Main Ring background does not affect
the EM calorimeter. The 2{,}341 events for which both electrons have tracks and
which pass the Main Ring Veto (CC/CC+EC $Z$ sample) serve to calibrate the
recoil
momentum response (Sec.~\ref{sec-recoil}). Table~\ref{tab:sample} summarizes
the data samples.

\begin{figure}[htpb!]
\vspace{-0.3in}
\centerline{\psfig{figure=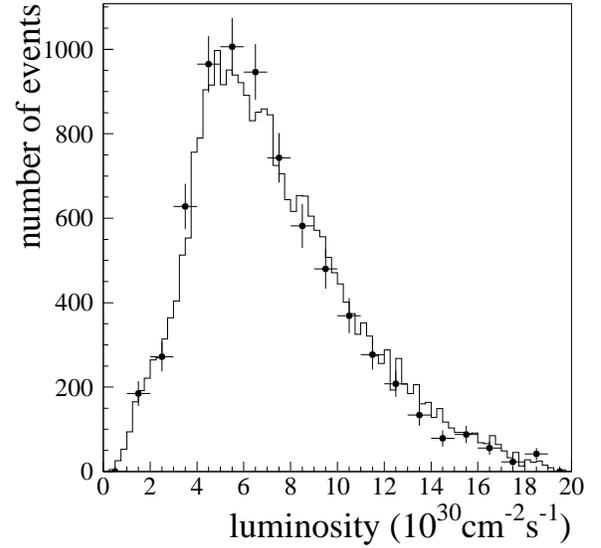,width=3.5in,height=3.5in}}
\vspace{-0.02in}
\caption{ The luminosity distribution of the \wb\ (------) and the \zb\
($\bullet$) samples.}
\label{fig:lum}
\end{figure}

Figure~\ref{fig:lum} shows the luminosity of the colliding beams
during the \wb\ and \zb\ data collection.

On several occasions we use a sample of 295{,}000 random \ppbar\ interaction
events for calibration purposes. We collected
these data concurrently with the \wb\ and
\zb\ signal data, requiring only a \ppbar\ interaction at
Level 1. We refer to these data as ``minimum bias events".

\section{ Fast Monte Carlo Model }
\label{sec-mc}
\subsection {Overview}

The fast Monte Carlo model consists of three parts. First we simulate the
production of the \wb\ or \zb\ boson by generating the boson four-momentum and
other characteristics of the event like the $z$-position of the interaction
vertex and the luminosity.  The event luminosity is required for luminosity
dependent parametrizations in the detector simulation.  Then we simulate the
decay of the boson. At this point
we know the true \pt\ of the boson and the momenta of its decay products. We
then apply a parameterized detector model to these momenta in order to simulate
the observed transverse recoil momentum and the observed electron momenta.

\subsection { Vector Boson Production }
\label{sec-vb_prod}

In order to specify completely the production dynamics of vector bosons
in \ppbar\ collisions we need to know the differential
production cross section
in mass $Q$, rapidity $y$, and transverse momentum \qt\ of the
produced \wb\ bosons. To speed up the event generation, we factorize
this into
\begin{equation}
{\d^3\sigma\over \d \qt^2  \d y \d Q } \approx
\left.{\d^2\sigma\over \d \qt^2 \d y}\right|_{Q^2=\mw^2} \times
{\d\sigma\over \d Q}
\label{eq:prod}
\end{equation}
to generate \qt, $y$, and $Q$ of the bosons.

For \ppbar\ collisions, the vector boson production
cross section is given by the parton cross section $\widetilde\sigma_{i,j}$
convoluted with the parton distribution functions  $f(x,Q^2)$ and summed
over parton flavors $i,j$:
\begin {eqnarray}
{\d^2\sigma\over \d \qt^2  \d y} &=&
     \sum_{i,j}\int dx_1 \int dx_2 f_i(x_1,Q^2) f_j(x_2,Q^2) \nonumber \\
 & & \ \ \delta(sx_1x_2-Q^2){\d^2\widetilde\sigma_{i,j}\over \d \qt^2  \d y}.
\label{eq:wprod}
\end {eqnarray}
Several authors \cite{LY,AK} have computed
$\left.{\d^2\sigma\over \d \qt^2 \d y}\right|_{Q^2=\mw^2}$ using a perturbative
calculation \cite{AR} for the high-\qt\ regime and the Collins-Soper
resummation formalism \cite{CSS,AEMG} for the low-\qt\ regime. We use the
code provided by the authors of Ref.~\cite{LY} and the MRSA$'$ parton
distribution functions \cite{mrsa} to compute the cross section.
We evaluate Eq.~\ref{eq:wprod} separately for
interactions involving at least one valence quark and for interactions involving
two sea quarks.

The parton cross section is given by
\begin{equation}
{\d^2\widetilde\sigma\over\d \qt^2 \d y} = {\widetilde\sigma_0\over4\pi \hat{s}}
\left\{\int \d^2 b e^{i\vec\qt\cdot\vec b} \cdot\widetilde{W}(b) \times e^{-S} +
Y\right\},
\label{eq:dydpt}
\end{equation}
where $\widetilde\sigma_0$ is the tree-level cross section,
$\hat{s}$ is the parton
center-of-mass energy, and $b$ is the impact parameter in transverse momentum
space. $\widetilde{W}$ and $Y$ are
perturbative  terms and $S$ parameterizes the non-perturbative physics. In the
notation of Ref.~\cite{LY}
\begin{equation}
S=\left[g_1+g_2\ln\left({Q\over2Q_0}\right)\right]b^2 + g_1g_3\ln(100x_1x_2)b
\label{eq:sud}
\end{equation}
where $Q_0$ is a cut-off parameter, $x_1$ and $x_2$ are the momentum fractions
of the initial state partons.
The parameters $g_1$, $g_2$, and $g_3$ have to be determined experimentally
(Sec.~\ref{sec-constraints}).

We use a Breit-Wigner curve with mass dependent width for the line shape of the
\wb\ boson. The intrinsic width of the \wb\ is $\wwidth=2.062 \pm 0.059 $ GeV
\cite{Wwidth}. The line shape is skewed due to the momentum
distribution of the quarks inside the proton and antiproton. The mass spectrum
is given by
\begin {eqnarray}
{\d\sigma\over\d Q}={\cal L}_{\qqbar}(Q)
{Q^2\over (Q^2-\mw^2)^2+{Q^4\Gamma_W^2\over\mw^2}}.
\end {eqnarray}
We call
\begin {eqnarray}
{\cal L}_{\qqbar}(Q) = {2Q\over s} \sum_{i,j}\int_{Q^2/s}^1 {\d x\over x}
f_i(x,Q^2)f_j(Q^2/sx,Q^2)
\end {eqnarray}
the parton luminosity. To evaluate it we generate \wev\ events using the \HERW\
Monte Carlo event generator \cite{herwig}, interfaced with \PDFL\ \cite{PDFLIB},
and select the events subject to the same kinematic and fiducial cuts as for the
\wb\ and \zb\ samples with all electrons in CC. We plot the mass spectrum
divided by the intrinsic line shape of the \wb\ boson. The result is
proportional to the parton luminosity and we parameterize
the spectrum with the function~\cite{D0}
\begin{equation}
{\cal L}_{\qqbar}(Q) = {e^{-\beta Q}\over Q}.
\label{eq:partlum}
\end{equation}
Table~\ref{tab:wprod} shows $\beta$ for \wb\ and \zb\ events for some modern
parton distribution functions. The value of $\beta$ depends on the rapidity
distribution of the \wb\ bosons, which is restricted by the kinematic and
fiducial cuts that we impose on the decay leptons. The values of $\beta$ given
in Table~\ref{tab:wprod} are for the rapidity distributions of \wb\ and \zb\
bosons that satisfy the kinematic and fiducial cuts given in
Sec.~\ref{sec-data}.
The uncertainty in $\beta$ is about 0.001, due to Monte Carlo
statistics and uncertainties in the acceptance.

To generate the boson four-momenta we treat $\d\sigma/\d Q$ and $\d^2\sigma/
\d \qt^2 \d y$ as probability density functions and pick $Q$ from the former
and a pair of $y$ and \qt\ values from the latter. For a fraction $f_{ss}$ we
use $\d^2\sigma/\d \qt^2\d y$ for interactions between two sea quarks.
Their helicity is $+1$ or $-1$ with equal probability. For the remaining \wb\
bosons
we use $\d^2\sigma/\d \qt^2\d y$ for interactions involving at least one
valence quark. They always have helicity $-1$. Finally, we pick the
$z$-position of the interaction vertex from a Gaussian
distribution centered at $z=0$ with a standard deviation of 25 cm and
a luminosity for each event from the histogram in Fig.~\ref{fig:lum}.

\subsection{ Vector Boson Decay }

\begin{figure}[htpb!]
\vspace{-0.3in}
\centerline{\psfig{figure=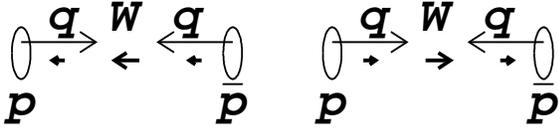,width=4.2in,height=1.58in}} 
\vspace{-0.02in}
\caption{ Polarization of the \wb\ produced in \ppbar\ collisions if the
quark comes from the proton (left) and if the antiquark comes from the proton
(right). The thick arrows indicate the orientation of the particle spins.}
\label{fig:wpol}
\end{figure}

At lowest order the \wb\ boson is fully polarized along the beam direction due
to the $V$--$A$ coupling of the charged current. The resulting angular
distribution of the charged lepton in the \wb\ rest frame is given by
\begin{equation}
\label{eq:stnd_angle}
{\d \sigma\over \d \cos\theta^*} \propto (1-\lambda q\cos\theta^*)^2,
\end{equation}
where $\lambda$ is the helicity of the \wb\ with respect to the proton
direction,
$q$ is the charge of the lepton, and $\theta^*$ is the angle between the charged
lepton and proton beam directions in the \wb\ rest frame.
The spin of the  \wb\ points along the direction of the incoming antiquark.
Most of the time the quark comes from the proton and the antiquark from the
antiproton, so that $\lambda=-1$. Only if both quark and antiquark come from
the sea of the proton and antiproton is there a 50\% chance that the quark
comes from the antiproton and the antiquark from the proton and in that case
$\lambda=1$ (Fig.~\ref{fig:wpol}). We determine the fraction of sea-sea
interactions, \fss, using the parameterizations of the parton distribution
functions given in \PDFL\ \cite{PDFLIB}.

\begin{figure}[htpb!]
\vspace{-0.3in}
\centerline{\psfig{figure=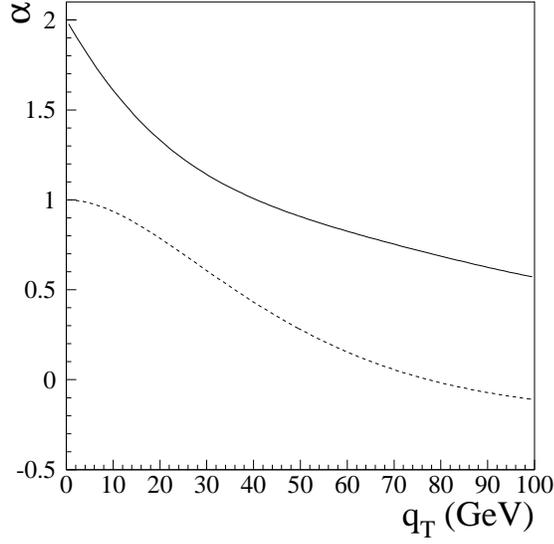,width=3.5in,height=3.5in}}
\vspace{-0.in}
\caption{The calculations of
$\alpha_1$\hbox{(-----)}  and $\alpha_2$\hbox{(- - -)} as
a function of the transverse momentum of the \wb\ boson.}
\label{fig:alpha}
\end{figure}

When ${\cal O}(\alpha_s)$ processes are included, the boson acquires finite
\pt\ and Eq.~\ref{eq:stnd_angle} is changed to~\cite{mirk}
\begin{equation}
\label{eq:cos_angle}
{\d\sigma\over \d\cos\theta_{\rm CS}} \propto
    \left( 1+\alpha_1(\qt)\cos\theta_{\rm CS}+\alpha_2(\qt)\cos^2\theta_{\rm CS}
    \right)
\end{equation}
for $W^+$ bosons with $\lambda=-1$ and after integration over $\phi$. The angle
$\theta_{\rm CS}$ in Eq.~\ref{eq:cos_angle} is now defined in the Collins-Soper
frame \cite{csframe}. The values of $\alpha_1$ and $\alpha_2$ as a function of
transverse boson momentum have been calculated at ${\cal O}(\alpha_s^2)$
\cite{mirk} and are shown
in Fig.~\ref{fig:alpha}. We have implemented the angular distribution given
in Eq.~\ref{eq:cos_angle} in the fast Monte Carlo. The effect
is smaller if the \wb\ bosons are selected with $\ut<15$ GeV than for $\ut<30$
GeV. The angular distribution of the leptons from \zee\ decays is also
generated according to Eq.~\ref{eq:cos_angle}, but with $\alpha_1$ and
$\alpha_2$ computed for \zee\ decays~\cite{mirk}.

To check whether neglecting the correlations between
the mass and the other parameters in Eq.~\ref{eq:prod} introduces an
uncertainty, we use the \HERW\ program to generate \wev\ decays including
the correlations neglected in our model. We apply our parameterized
detector model to them and fit them with probability density functions
that were generated without the correlations. The fitted
\wb\ mass values agree with the \wb\ mass used in the Monte Carlo generation
within the statistical uncertainties of 25 MeV.

Radiation from the decay electron or the \wb\ boson biases the mass
measurement. If the decay electron radiates a photon and the photon is well
enough separated from the electron so that its energy is not included in the
electron energy, or if an on-shell \wb\ boson radiates a photon and therefore is
off-shell when it decays, the measured mass is biased low. We use the
calculation of Ref.~\cite{rad_decays_th} to generate \wegam\
decays. The calculation gives the fraction of events in which a photon with
energy $\Eg>E_0$ is radiated, and the angular distribution and energy
spectrum of the photons. Only radiation from the decay electron and the \wb\
boson, if the final state \wb\ is off-shell, is included to order $\alpha$.
Radiation by the initial quarks or the \wb, if the final \wb\ is on-shell,
does not affect the mass of the $e\nu$ pair from the \wb\ decay. We use a
minimum photon energy $E_0=50$ MeV, which means that in 30.6\% of all \wb\
decays a photon with $\Eg>50$ MeV is radiated. Most of these photons are emitted
close to the electron direction and cannot be separated from the electron in
the calorimeter. For \zee\ decays there is a 66\% probability that any one of
the electrons radiates a photon with $\Eg>50$ MeV.

\begin{figure}[htpb!]
\vspace{-0.3in}
\centerline{\psfig{figure=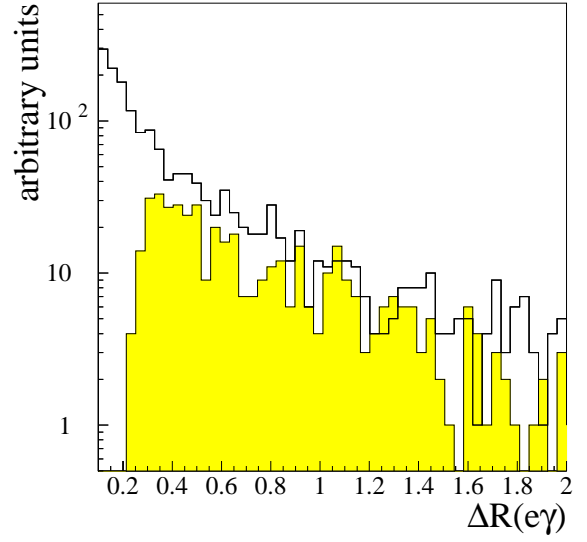,width=3.5in,height=3.5in}}
\vspace{-0.in}
\caption{ The distribution of \regam\ of photons from \wegam\ decays that
are reconstructed as separate objects (shaded) and those that are not, either
because they are too close to the electron or too low in energy (------).}
\label{fig:radiation}
\end{figure}

The separation of the electron and photon in the lab frame is
\begin{equation}
    \regam = \sqrt{(\phie-\phig)^2 + (\etae-\etag)^2}.
\end{equation}
Figure~\ref{fig:radiation} shows the calculated distribution of photons
as a function of \regam. The shaded histogram in the figure
shows the photons that are reconstructed as separate objects.
If the photon and electron are close
together they cannot be separated in the calorimeter.
The momentum of a photon
with $\regam\lt\rzero$ is therefore added to the electron momentum,
while for $\regam\geq\rzero$ a photon is considered separated
from the electron and its momentum is added to the recoil momentum. We use
$\rzero =0.3$, which is the approximate size of the window in which the electron
energy is measured. This procedure has been verified to give the same results as
an explicit \GEAN\ simulation of radiative \wb\ decays.
In only about 3.5\% of the \wev\ decays does the photon separate far
enough from the electron, \ie\ $\regam\gt\rzero$, to cause a
mismeasurement of the transverse mass.

\wb\ boson decays through the channel $W\to\tau\nu\to e\nu\overline\nu\nu$ are
topologically indistinguishable from  \wev\ decays. We therefore include
these decays in the \wb\ decay model, properly accounting for the polarization
of the tau leptons in the decay angular distributions. The fraction of \wb\
bosons that decay in this way is $B(\tev)/\left(1+B(\tev)\right) = 0.151$.

We let the generated \wb\ bosons decay
with an angular distribution corresponding to their helicity. For 15.1\%
of the \wb\ bosons the decay is to $\tau\nu\to e\nu\overline\nu\nu$. For 30.6\%
of the remaining \wb\ bosons a photon is radiated.
For 66\% of the \zb\ bosons
the decay is to $\ee\gamma$ and for the remainder to \ee.

\subsection{ Detector Model }

The detector simulation uses a parameterized model for response and resolution
to obtain a prediction for the distribution of the observed electron and recoil
momenta.

When simulating the detector response to an electron of energy $E_0$, we
compute the observed electron energy as
\begin{equation}
E(e) = \alphaem E_0 + \Delta E({\cal L},u_{||}) +
\sigma_{\rm EM}\cdot X,
\end{equation}
where \alphaem\ is the response of the electromagnetic calorimeter,
$\Delta E$ is the energy due to particles from the underlying event within the
electron window (parameterized as a function of luminosity ${\cal L}$ and
$u_{||}$), $\sigma_{\rm EM}$ is the energy resolution of the electromagnetic
calorimeter, and $X$ is a random variable from a normal parent distribution
with zero mean and unit width.

The transverse energy measurement depends on the measurement of the electron
direction as well. We determine the shower centroid position by intersecting the
line defined by the event vertex and the electron direction with a cylinder
coaxial with the beam
and 91.6 cm in radius (the radial center of the EM3 layer).  We
then smear the azimuthal and $z$-coordinates of the intersection point by their
resolutions.  We determine the $z$-coordinate of the center of gravity of the
CDC track by intersecting the same line with a cylinder of 62 cm radius, the
mean radial position of all delay lines in the CDC,
and smearing by the resolution. The measured angles are then obtained from the
smeared points as described in Section~\ref{sec-data-elec}.

The model for the particles recoiling against the \wb\ has two components: a
``hard'' component that models the $p_T$ of the \wb,
and a ``soft'' component that models detector noise and pile-up. Pile-up refers
to the effects of additional \ppbar\ interactions in the same or previous beam
crossings. For the soft component we use the transverse momentum balance
\mptv\ from a minimum bias event recorded in the detector.
The observed recoil \pt\ is then given by
\begin{eqnarray}
\label{eq:ut}
\vec u_T &=& -\bigl(\rrec\qt+\sigrec\cdot X \bigr)\hat \qt \nonumber \\
         & & -\dupar({\cal L},\upar) \hat\pte \\
         & & +\alphamb \mptv, \nonumber
\end{eqnarray}
where \qt\ is the generated value of the boson transverse momentum,
\rrec\ is the (in general momentum dependent) response, \sigrec\ is the 
resolution of the
calorimeter, \dupar\ is the transverse energy flow into the electron window
(parameterized as a function of luminosity ${\cal L}$ and \upar), and
\alphamb\ is a correction factor that allows us to adjust the resolution to
the data. The quantity \dupar\ is different from the energy added to the
electron, $\Delta E$, because of the zero-suppression in the calorimeter
readout.

We simulate selection biases due to the trigger requirements and the electron
isolation by accepting events with the estimated efficiencies.
Finally, we compute all the derived quantities from these observables
and apply fiducial and kinematic cuts.

\section{ Electron Measurement }
\label{sec-elec}
\subsection { Angular Resolutions }

\begin{figure}[htpb!]
\vspace{-0.3in}
\centerline{\psfig{figure=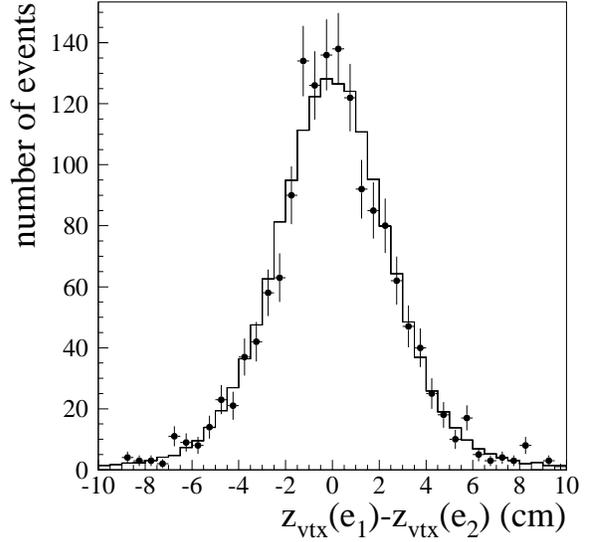,width=3.5in,height=3.5in}}
\vspace{-0.02in}
\caption{ The distribution of $\zvtx(e_1) - \zvtx(e_2)$ for the
\zee\ sample ($\bullet$) and the fast Monte Carlo simulation~\hbox{(------)}.}
\label{fig:ztrk}
\end{figure}

The resolution for the $z$-coordinate of the track center of gravity, \ztrk, is
determined from the \zee\ sample. Both electrons originate from the same
interaction vertex and therefore the difference between the interaction vertices
reconstructed from the two electrons separately, $\zvtx(e_1) -\zvtx(e_2)$, is a
measure of the resolution with which the electrons point back to the vertex. The
points in Fig.~\ref{fig:ztrk} show the distribution
of $\zvtx(e_1) -\zvtx(e_2)$ observed in the CC/CC \zb\ sample
with tracks required for both electrons.

A Monte Carlo study based on single electrons generated with a \GEAN\
simulation shows that the resolution of the
shower centroid algorithm can be parameterized as
\begin{equation}
\sigma(\zcal) = (a + b\, \lambda(e)) + (c + d\, \lambda(e)) \zcal,
\end{equation}
where $\lambda(e)=|\te-90^\circ|$, $a=0.33$ cm, $b=5.2\times10^{-3}$ cm,
$c= 4.2\times10^{-4}$, and $d=7.5\times10^{-5}$. We then tune the
resolution function for \ztrk\ in the fast Monte Carlo so that it
reproduces the shape of the $\zvtx(e_1) - \zvtx(e_2)$ distribution observed
in the data. We find that a resolution function consisting of two Gaussians
0.31 cm and 1.56 cm wide, with 6\% of the area under the wider Gaussian, fits
the data well. The histogram in Fig.~\ref{fig:ztrk} shows the Monte
Carlo prediction for the best fit, normalized to the same number of events as
the data. The \wb\ mass measurement is very insensitive to these resolutions.
The uncertainties in the resolution parameters cause less than 5 MeV uncertainty
in the fitted \wb\ mass.

The calibration of the $z$-position measurements from the CDC and calorimeter
is described in Appendix~\ref{app:pos_cdc}.
We quantify the calibration uncertainty in terms of
scale factors $\alphacdc=0.988\pm0.001$ and $\alphacc=0.9980\pm0.0005$ for the
$z$-coordinate. The
uncertainties in these scale factors lead to a finite uncertainty in the \wb\
mass measurement.

\subsection { Underlying Event Energy }
\label{sec-ue}

The energy in an array of 5$\times$5 towers in the four EM layers and the
first FH layer around the most energetic
tower of an electron cluster is assigned to
the electron. This array contains the entire energy deposited by the electron
shower plus some energy from other particles.
The energy in the window is  excluded from the computation of \utv. This causes
a bias in \upar, the component of \utv\ along the direction of the electron.
For $\ptw\ll\mw$
\begin{equation}
\mt \approx 2 \pte + \upar,
\label{eq:mt_upar}
\end{equation}
so that this bias propagates directly into a bias in the transverse mass.
We call this bias \dupar. It is equal to the momentum flow observed
in the EM and first FH sections of a 5$\times$5 array of calorimeter towers.

\begin{figure}[htpb!]
\vspace{-0.3in}
\centerline{\psfig{figure=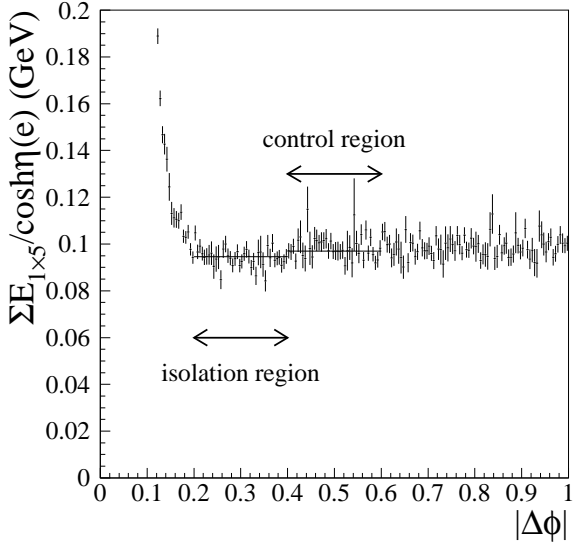,width=3.5in,height=3.5in}}
\vspace{-0.02in}
\caption{  The transverse energy flow into 1$\times$5 tower segments as a
function of azimuthal separation from the electron in the \wb\ sample.}
\label{fig:wuelin}
\end{figure}

\begin{figure}[htpb!]
\vspace{-0.3in}
\centerline{\psfig{figure=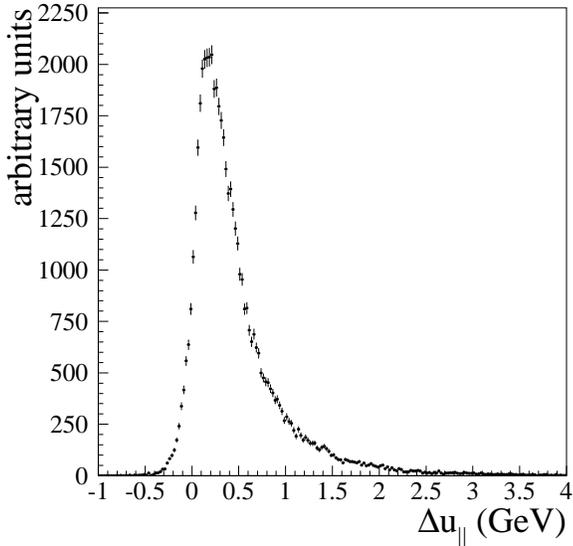,width=3.5in,height=3.5in}}
\vspace{-0.2in}
\caption{ The distribution of \dupar\ in
the \wb\ signal sample.}
\label{fig:deltaupar}
\end{figure}

We use the \wb\ and \zb\ data samples to measure \dupar. For every
electron in the \wb\ and \zb\ samples we compute the energy flow into an
azimuthal ring of calorimeter towers, 5 towers wide in $\eta$ and centered on
the tower with the largest fraction of the electron energy. For every electron
we plot the transverse energy flow into one-tower-wide azimuthal segments of
this ring as a function of the azimuthal separation $|\Delta\phi|$ between the
center of the segment and the electron shower centroid. The energy flow
$\sum E_{1\times5}$ is computed as the sum of all energy deposits in the four EM
layers and the first FH layer in the 1$\times$5 tower segment. Figure
\ref{fig:wuelin} shows the transverse energy flow $\sum
E_{1\times5}/\cosh\eta(e)$ versus $|\Delta\phi|$ for the electrons in the \wb\
sample with $\ut<15$ GeV. For small $|\Delta\phi|$ we
see the substantial energy flow from the electron shower and for larger
$|\Delta\phi|$ the constant noise level.
The electron shower is contained in a window of
$|\Delta\phi|<0.2$. We estimate the energy flow into the 5$\times$5 tower window
around the electron from the energy flow into segments of the azimuthal ring
with $|\Delta\phi|>0.2$. The level of energy flow is sensitive to the isolation
cut. The region $0.2<|\Delta\phi|<0.4$, which
is used for the isolation variable, is maximally biased by the cut; the
region, $0.4<|\Delta\phi|<0.6$, which is close to the electron but outside the
isolation region, is minimally biased. We expect the energy flow under the
electron to lie somewhere in between the energy flow into these two regions. We
therefore compute \dupar\ based on the average transverse energy
flow into both regions and assign a systematic error equal to half the
difference between the two regions. We repeat the same analysis for the
electrons in the CC/CC \zb\ sample. The results are
tabulated in Table~\ref{tab:ue}. We find $\dupar=479\pm2(stat)\pm6(syst)$
MeV for \wb\ events with $\ut<15$ GeV. For the \zb\ sample \dupar\ is
$11\pm7$ MeV lower. Figure~\ref{fig:deltaupar} shows
the spectrum of \dupar.

\begin{figure}[htpb!]
\vspace{-0.5in}
\centerline{\psfig{figure=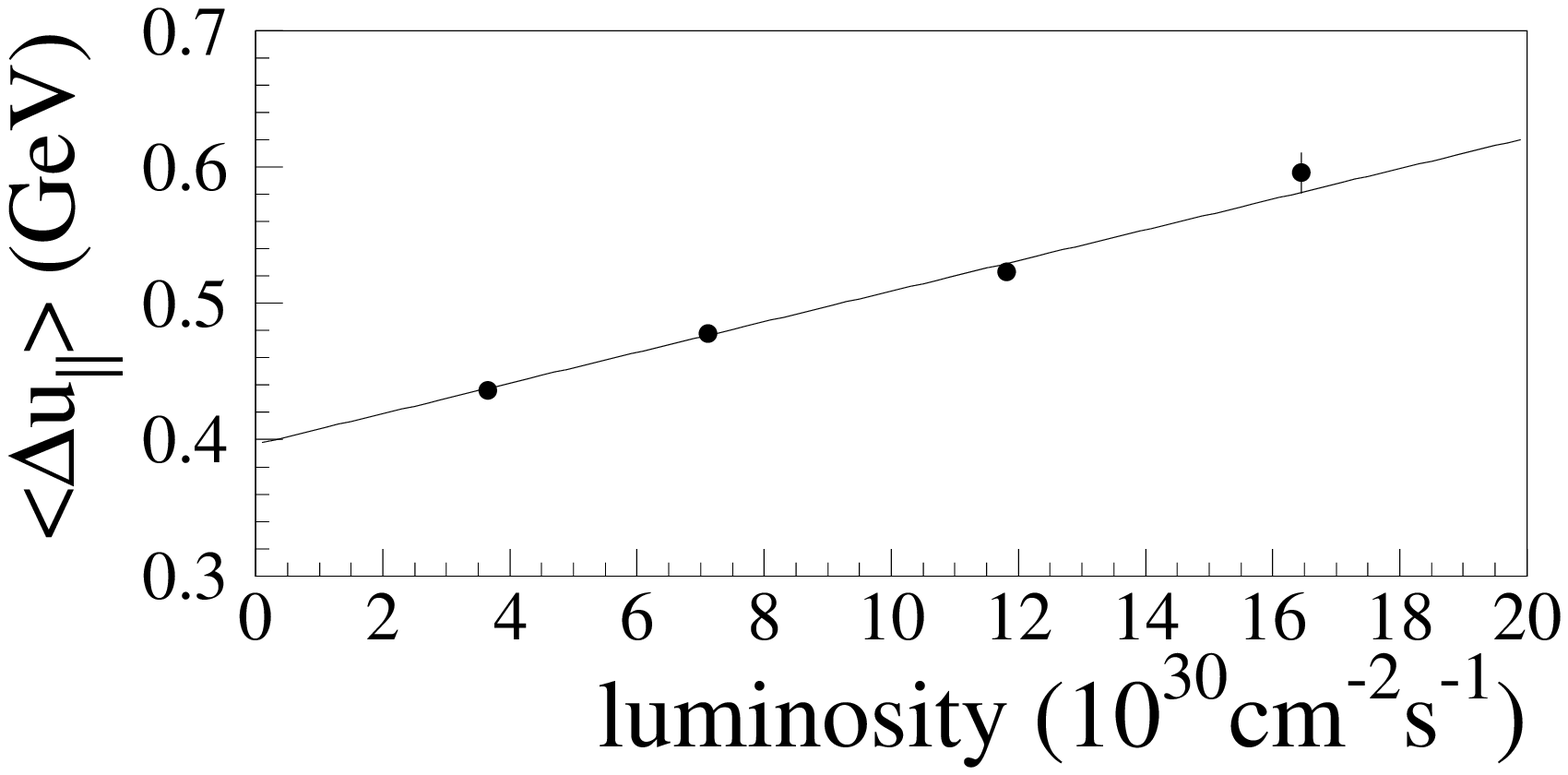,width=3.5in,height=3.5in}}
\vspace{-1.6in}
\caption{ The luminosity dependence of \mdupar.}
\label{fig:duparlum}
\vspace{-0.4in}
\centerline{\psfig{figure=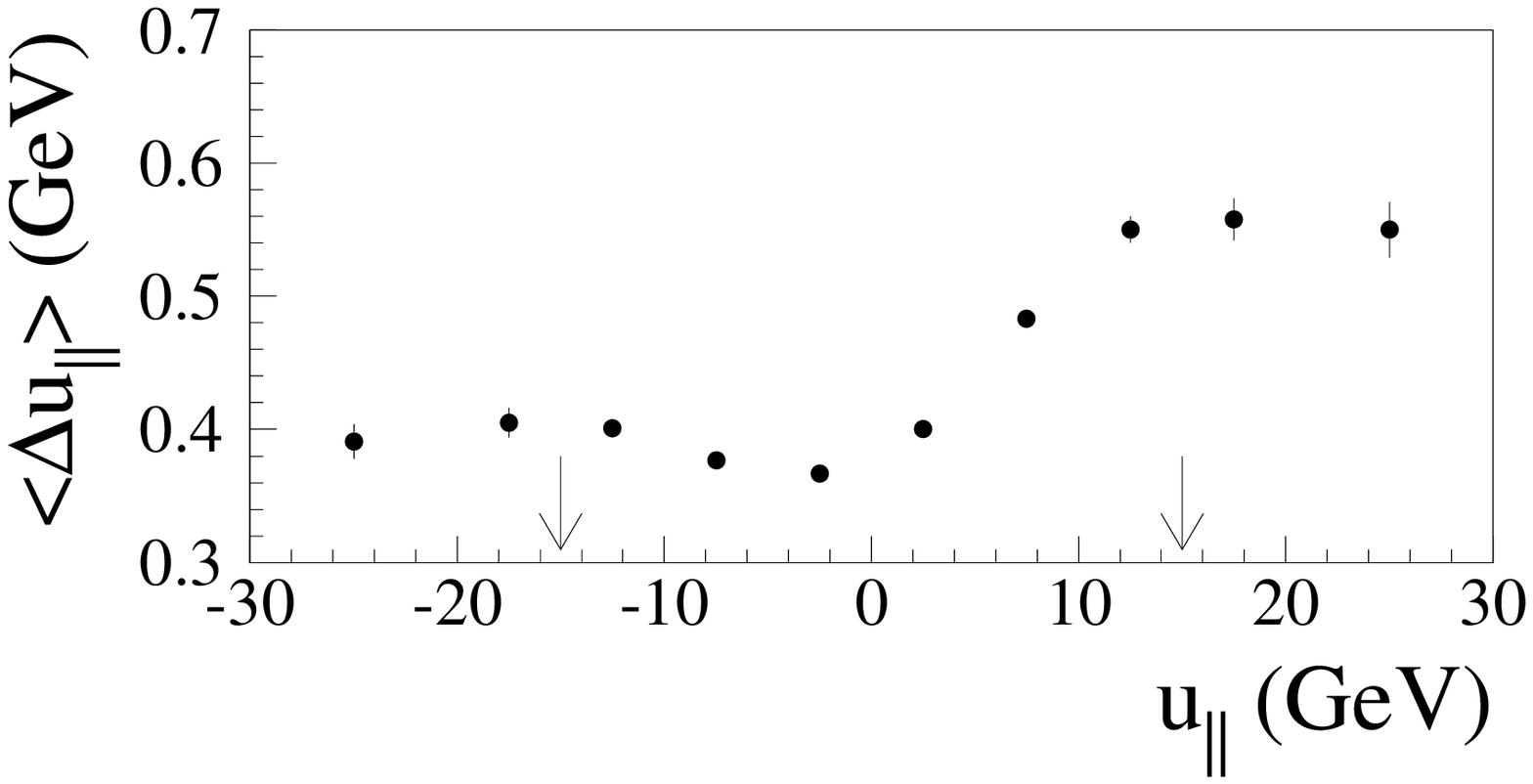,width=3.5in,height=3.5in}}
\vspace{-1.5in}
\caption{ The variation of  \mdupar\ as a function of \upar.
The region between the arrows is populated by the \wb\ sample.  }
\label{fig:duupar}
\end{figure}

At higher luminosity the average number of interactions per event increases and
therefore \dupar\ increases. This is shown in Fig.~\ref{fig:duparlum}.
The mean value of \dupar\ increases by 11.2 MeV per
10$^{30}$cm$^{-2}$s$^{-1}$. The underlying event energy flow into the electron
window also depends on \upar. Figure~\ref{fig:duupar} shows 
$\langle \dupar(0, \upar )\rangle$, the mean value for
\dupar\ corrected back to zero luminosity, as
a function of \upar. In the fast Monte Carlo model a value \dupar\ is picked
from the distribution shown in Fig.~\ref{fig:deltaupar} for every event and then
corrected for \upar\ and luminosity dependences.

The measured electron energy is biased upwards  by the additional energy $\Delta
E$ in
the window from the underlying event. $\Delta E$ is not equal to \dupar\ because
the additional energy deposited by the electron may lift some cells that would
have been zero-suppressed in the calorimeter readout above the zero-suppression
threshold. Therefore
\begin{equation}
\Delta E = \dupar - \Delta_{\rm ped}
\end{equation}
where $\Delta_{\rm ped}=212\pm25$ MeV is a correction for the pedestal shift
introduced by the zero-suppression in the calorimeter readout. This is
determined by superimposing single electrons simulated with a \GEAN\ simulation
on minimum bias events that were recorded without zero-suppression in the
calorimeter readout. Most of this bias cancels in the
\wb\ to \zb\ mass ratio so that the \wb\ mass measurement is not sensitive to
$\Delta_{\rm ped}$.

\subsection { \upar\ Efficiency }
\label{sec-elec-upar}

The efficiency for electron identification depends on their environment.
Well-isolated electrons are identified correctly more often than electrons near
other particles. Therefore \wb\ decays in which the electron is emitted in the
same direction as the particles recoiling against the \wb\ are selected
less often than \wb\ decays in which the electron is emitted in the direction
opposite the recoiling particles. This causes a bias in the lepton \pt\
distributions, shifting \pte\ to larger values and
\ptnu\ to lower values, whereas the \mt\ distribution is only slightly
affected.

\begin{figure}[htpb!]
\vspace{-0.3in}
\centerline{\psfig{figure=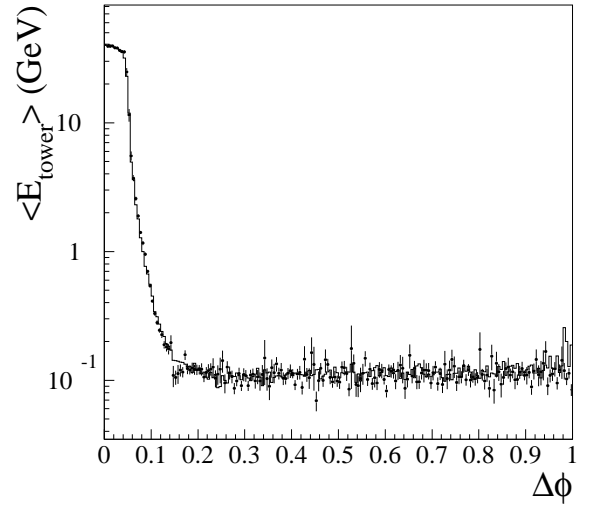,width=3.5in,height=3.5in}}
\vspace{-0.02in}
\caption{ The transverse energy flow into 1$\times$5 tower segments as a
function of the azimuthal separation from the electron for the electrons from
\wev\ decays ($\bullet$) and the superimposed Monte Carlo electron sample
(------).} 
\label{fig:etrans}
\end{figure}

\begin{figure}[htpb!]
\vspace{-0.3in}
\centerline{\psfig{figure=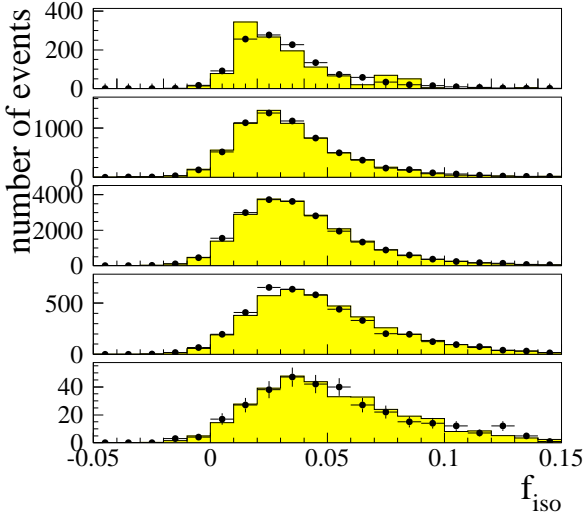,width=3.5in,height=3.5in}}
\vspace{0.2in}
\caption{ The isolation spectrum for five different \upar\ regions,
$\upar<-15$, $-15<\upar<-5$, $-5<\upar<5$, $5<\upar<15$, $\upar>15$ GeV
(from top to bottom), for the electrons from \wev\ decays ($\bullet$) and
the superimposed electron sample (shaded).}
\label{fig:MCiso}
\end{figure}

\begin{figure}[htpb!]
\vspace{-0.3in}
\centerline{\psfig{figure=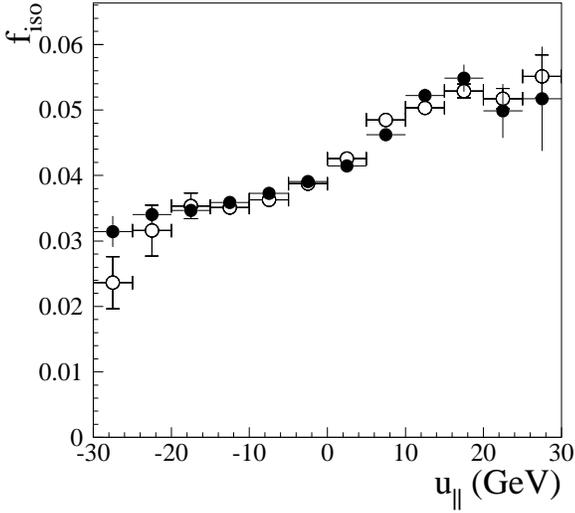,width=3.5in,height=3.5in}}
\vspace{0.2in}
\caption{ The mean isolation versus \upar\ for the \wb\ electron sample
($\circ$) and the superimposed Monte Carlo electron sample ($\bullet$).}
\label{fig:MCisoprof}
\end{figure}

We estimate the electron finding efficiency as a function of \upar\
by superimposing Monte Carlo electrons, simulated using the \GEAN\ program, 
onto the events
from our \wb\ signal sample.  We use the \wb\ sample in order to ensure that
the underlying event is correctly modeled.
The sample of superimposed electrons, which are spatially separated from the
electron that is already in the event, matches the data well. It is
important that the superimposed sample model the transverse shower shape and
isolation well, because these are the dominant effects that cause the efficiency
to vary with \upar. Figure~\ref{fig:etrans} shows the transverse shower profile
of the superimposed electron sample and the electron sample from \wb\ decays.
Figure~\ref{fig:MCiso} shows the distribution of the isolation for the two
electron samples in five \upar\ regions. Figure~\ref{fig:MCisoprof} compares the
mean isolation versus \upar\ for the two samples.

\begin{figure}[htpb!]
\vspace{-0.3in}
\centerline{\psfig{figure=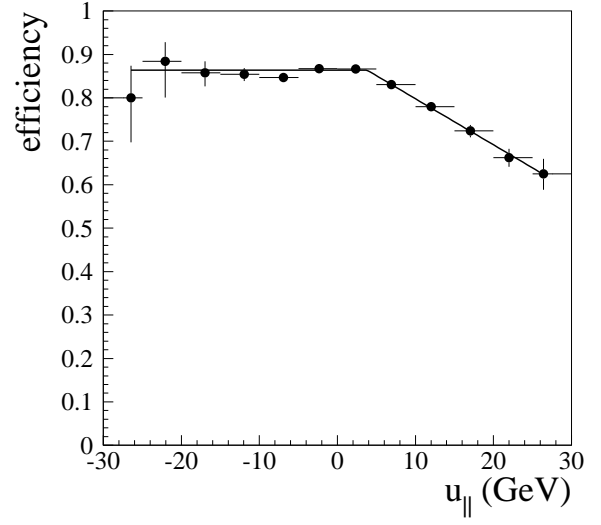,width=3.5in,height=3.5in}} 
\vspace{0.2in}
\caption{ The electron selection efficiency as a function of \upar.}
\label{fig:upareff}
\end{figure}

We then apply the shower shape and isolation cuts
used to select the \wb\ signal sample and determine the fraction of the
electrons in the superimposed samples that pass all requirements
as a function of \upar. This efficiency
is shown in Fig.~\ref{fig:upareff}. The line is a fit to a function
of the form
\begin{equation}
\varepsilon(\upar)
= \varepsilon_0 \left\{ \begin{array}{ll} 1 & \hbox{for $\upar<u_0$} \\
                     1-s(\upar-u_0) & \hbox{otherwise.} \end{array} \right.
\end{equation}
The parameter $\varepsilon_0$ is an overall efficiency which is
inconsequential for the \wb\ mass measurement, $u_0$ is the value of \upar\
at which the efficiency starts to decrease as a function of \upar, and $s$ is
the rate of decrease. We obtain the best fit for $u_0=3.85\pm0.55$ GeV and
$s=0.013\pm0.001$ GeV$^{-1}$. These two values are strongly correlated. The
errors account for the finite number of superimposed Monte Carlo electrons.

\subsection { Electron Energy Response }
\label{sec-EMresponse}

Equation \ref{eq:emresponse} relates the reconstructed electron energy to the
recorded calorimeter signals. Since the values for the constants were
determined in a different setup, we determine the offset \deltaem\ and a
scale \alphaem, which essentially modifies $A$, in situ with collider
data for resonances that decay to electromagnetically showering particles:
$\pi^0\to\gamma\gamma$, $J/\psi\to\ee$, and \zee. We use $\pi^0$ and $J/\psi$
signals from an integrated luminosity of approximately 150 nb$^{-1}$, 
accumulated
during dedicated runs with low \pt\ thresholds for EM clusters in the trigger.

The fast Monte Carlo predicts the reconstructed electron energy
\begin {equation}
E(e) = \alphaem E_0 = A \sum_{i=1}^5 s_i a_i - \deltaem
\end{equation}
where $E_0$ is the generated electron energy.
To determine \deltaem\ and \alphaem, we compare the observed
resonances and Monte Carlo predictions as a function of \alphaem\ and
\deltaem.

\begin{figure}[htpb!]
\vspace{-0.3in}
\centerline{\psfig{figure=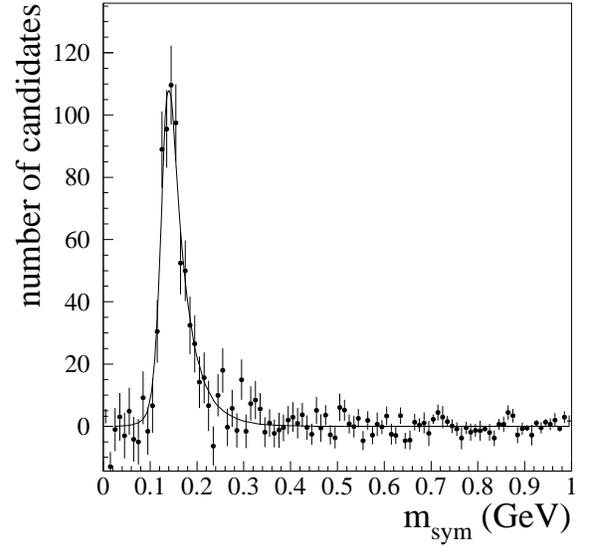,width=3.5in,height=3.5in}}
\vspace{.0in}
\caption{ The background-subtracted $m_{\rm sym}$ distribution. The superimposed
curve shows the Monte Carlo simulation.}
\label{fig:pizero}
\end{figure}

The photons from the decay of $\pi^0$s with $p_T>1$ GeV cannot be separated in
the calorimeter. There is about a 10\% probability for each photon to convert
to an \ee\ pair in the material in front of the CDC. If both photons
convert we can identify $\pi^0$ decays as EM clusters in the calorimeter with
two doubly-ionizing tracks in the CDC. We measure the $\pi^0$ energy $E(\pi^0)$
in the calorimeter and the opening angle $\omega$ between the two photons using
the two tracks. This allows us to compute the ``symmetric mass"
\begin{equation}
m_{\rm sym} = E(\pi^0) \sqrt{{1-\cos\omega\over2}},
\end{equation}
which is equal to the invariant mass if both
photons have the same energy, and is larger for asymmetric decays.
Figure~\ref{fig:pizero} shows the background
subtracted spectrum of $m_{\rm sym}$ for
$\pi^0$ candidates in the CC-EM superimposed with a Monte Carlo prediction of
the line shape.

\begin{figure}[htpb!]
\vspace{1in}
\begin{tabular}{c}
\epsfxsize = 6.0cm \epsffile[20 20 400 400]{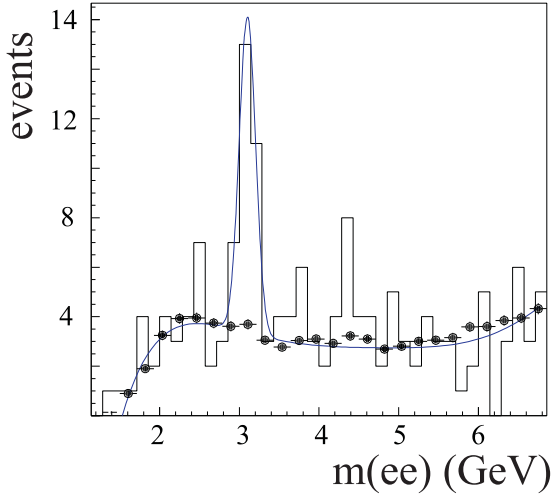}
\end{tabular}
\vspace{-.6in}
\caption{ The dielectron invariant mass spectrum for the $J/\psi\to ee$ sample
(histogram) and the background sample ($\bullet$). The smooth curve is a fit to
the data.}
\label{fig:jpsi}
\end{figure}

\begin{figure}[htpb!]
\vspace{-0.3in}
\centerline{
\psfig{figure=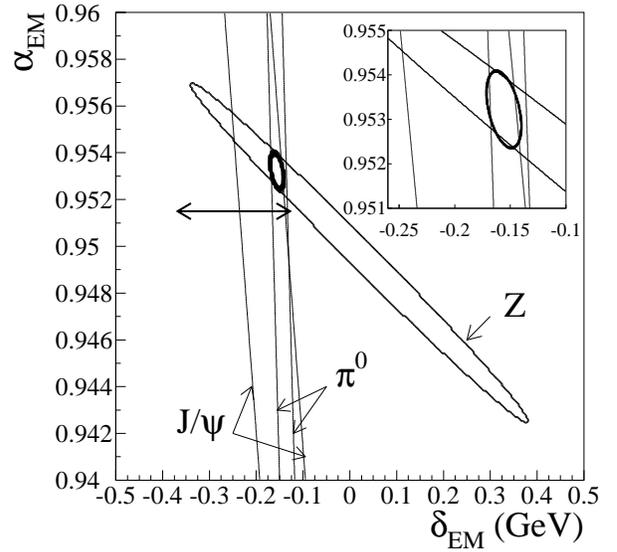,width=3.5in,height=3.5in}}
\vspace{-0.02in}
\caption{ The 68\% confidence level contours in \alphaem\ and
\deltaem\ from the $J/\psi$, $\pi^0$, and \zb\ data.
The inset shows an expanded view of the region where the
$\chi^2$ is minimized.}
\label{fig:EMcalib}
\end{figure}

Figure~\ref{fig:jpsi} shows the invariant mass spectrum of dielectron pairs
in the $J/\psi$ mass region. The smooth curve is the fit to a
Gaussian line shape above the background predicted using a sample of EM
clusters without CDC tracks. After correction for underlying event effects we
measure a mass of $3.03\pm0.04(stat)\pm0.19(syst)$ GeV. A Monte Carlo simulation
of $\ppbar\to\bbbar+X$, $b\to J/\psi+X$ tells us that we expect to observe a
mass
\begin{equation}
m_{\rm obs} = \alphaem m_{J/\psi} + 0.56\, \deltaem.
\end{equation}
Together with our measurement of $m_{\rm obs}$, this restricts the allowed
parameter space for \alphaem\ and \deltaem. The $\pi^0$ and $J/\psi$
analyses are described in detail in Ref.~\cite{D0}. Figure~\ref{fig:EMcalib}
shows the 68\% confidence level contours in \alphaem\ and \deltaem\
obtained from these data.

\begin{figure}[htpb!]
\vspace{-0.3in}
\centerline{\psfig{figure=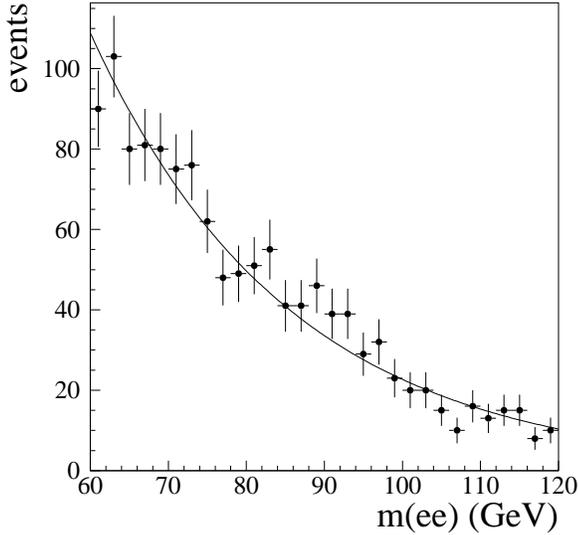,width=3.5in,height=3.5in}}
\vspace{-0.02in}
\caption{ The dielectron mass spectrum for the background data sample
to the CC/CC \zb\ sample.  The fit is an exponential.  }
\label{fig:zbkg}
\end{figure}

Fixing the observed \zb\ boson mass to the measured value
(Eq.~\ref{eq:mz})
correlates the values allowed for \alphaem\ and \deltaem. For a given
\deltaem\ we determine \alphaem\ so that the position of the \zb\ peak
predicted by the fast Monte Carlo agrees with the data. To determine the scale
factor that best fits the data, we perform a maximum likelihood fit to the \mee\
spectrum between 70 GeV and 110 GeV. In the resolution function we allow for an
exponential background shape whose slope is fixed to $-0.037\pm0.002\
\hbox{GeV}^{-1}$, the value obtained from a sample of events with two EM
clusters that fail the electron quality cuts (Fig.~\ref{fig:zbkg}).
The background normalization is allowed to float in the fit.
This is sufficient, together with the $\pi^0$ and $J/\psi$ data, to determine
both \alphaem\ and \deltaem.

\begin{figure}[htpb!]
\vspace{-0.3in}
\centerline{\psfig{figure=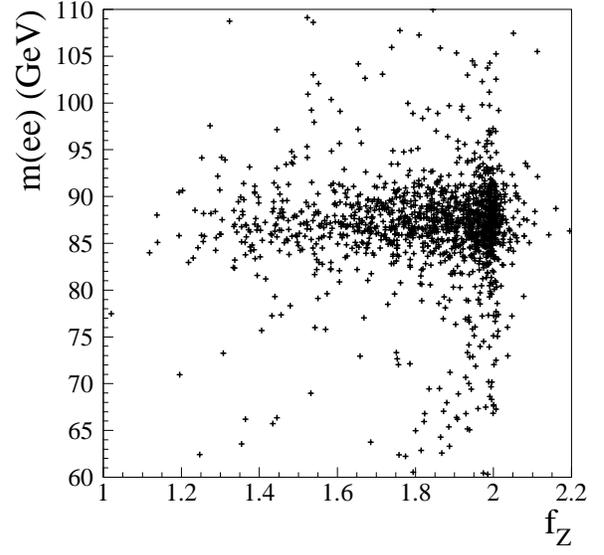,width=3.5in,height=3.5in}}
\vspace{-0.02in}
\caption{ The distribution of \mee\ versus $f_Z$ for the CC-CC \zee\ sample.}
\label{fig:binnedZ}
\end{figure}

Without relying on
the low energy data at all we can extract \alphaem\ and \deltaem\ from the \zb\
data alone. 
The electrons from \zb\ decays are not monochromatic and therefore we can make
use of their energy spread to constrain \alphaem\ and \deltaem\
simultaneously. For $\deltaem \ll E(e_1)+E(e_2)$ we can write
\begin{equation}
\mee = \alphaem \mz + f_Z \deltaem\ ,
\end {equation}
where $f_Z=(E(e_1)+E(e_2)) (1-\cos\omega)/\mee$ and $\omega$ is the opening
angle between the two electrons. We plot \mee\ versus $f_Z$
(Fig.~\ref{fig:binnedZ}) and compare it with the Monte Carlo predictions for
the allowed values of \alphaem\ and \deltaem\ using a binned maximum
likelihood fit.

Using the constraints on \alphaem\ and \deltaem\ from the \zb\ data
alone we obtain the contour labeled ``$Z$'' in Fig.~\ref{fig:EMcalib}
and $\deltaem =0.02\pm0.36$ GeV. The uncertainty in this measurement of
\deltaem\ is dominated by the statistical uncertainty due to the finite size
of the \zb\ sample.

The combined constraint from all three resonances
is shown by the thick contour in
Fig.~\ref{fig:EMcalib}. The $\pi^0$ and $J/\psi$ contours essentially fix
\deltaem, independent of \alphaem. The requirement that the \zb\ peak
position agree with the known \zb\ boson mass correlates
\alphaem\ and \deltaem. The contours in Fig.~\ref{fig:EMcalib} reflect
only statistical uncertainties. The uncertainty in the $\pi^0$ and $J/\psi$
contours is dominated by systematic effects in the underlying event corrections
and the deviation of the test beam data from the assumed response at low
energies.  The double arrow in Fig.~\ref{fig:EMcalib} represents the systematic
uncertainty in \deltaem. We determine
\begin{equation}
\deltaem = - 0.16 ^{+0.03}_{-0.21}\ \hbox{GeV}.
\label{eq:offset}
\end{equation}

\begin{figure}[htpb!]
\vspace{-0.3in}
\centerline{\psfig{figure=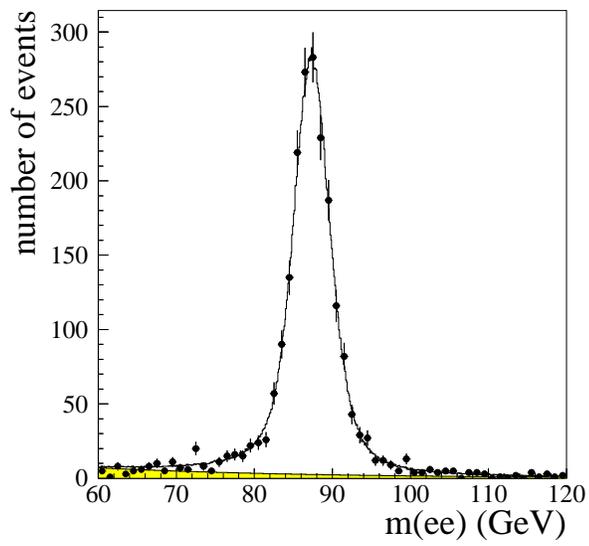,width=3.5in,height=3.5in}}
\vspace{-0.02in}
\caption{ The dielectron mass spectrum from the CC-CC \zb\ sample.
The superimposed curve shows the maximum likelihood fit and the shaded region
the fitted background.}
\label{fig:lzee}
\end{figure}

Figure~\ref{fig:lzee} shows the \mee\ spectrum for the CC/CC \zb\ sample and the
Monte Carlo spectrum that best fits the data for $\deltaem = - 0.16$ GeV.
The $\chi^2$ for the best fit to the CC/CC \mee\ spectrum is 33.5 for 39
degrees of freedom. For
\begin{equation}
\alphaem = 0.9533\pm0.0008
\label{eq:EMscale}
\end{equation}
the \zb\ peak position is consistent with the known \zb\ boson mass.
The error reflects the statistical uncertainty and the uncertainty in the
background normalization. The background slope has no measurable effect on the
result.

\begin{figure}[htpb!]
\vspace{-0.3in}
\centerline{
\psfig{figure=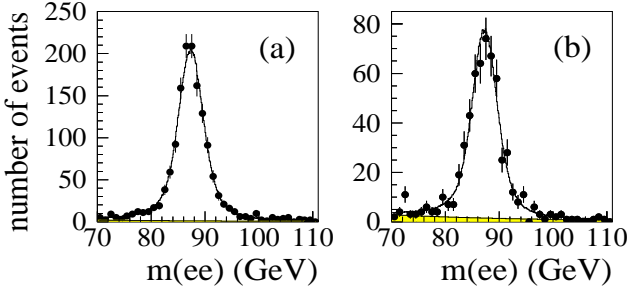,width=3.85in,height=2.2in}}
\vspace{-0.02in}
\caption{ The dielectron mass
spectra from (a) the tight/tight and (b) the tight/loose CC-CC
\zb\ samples. The curves show the fitted Monte Carlo spectra.}
\label{fig:zcc}
\end{figure}

If we split the CC/CC \zb\ sample into events with two tight electrons and
events with a tight and a loose electron and fit them separately
using the value of \alphaem\ given in Eq.~\ref{eq:EMscale} we obtain
\begin{eqnarray}
\mz & = & 91.206\pm0.086\ \hbox{GeV}\ \hbox{(tight/tight sample);} \\
\mz & = & 91.145\pm0.148\ \hbox{GeV}\ \hbox{(tight/loose sample).}
\end{eqnarray}
Figures~\ref{fig:zcc} (a) and (b) show the corresponding spectra and fits.

\subsection { Electron Energy Resolution }
\label{sec-elec-res}

\begin{figure}[htpb!]
\vspace{-0.3in}
\centerline{\psfig{figure=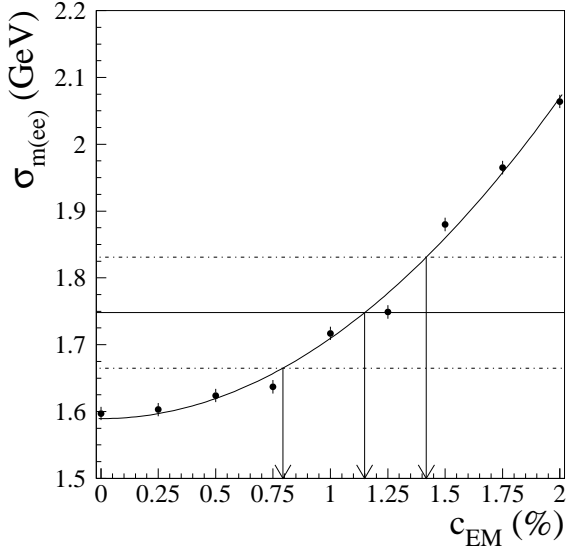,width=3.5in,height=3.5in}}
\vspace{-0.02in}
\caption{ The dielectron mass resolution versus
the constant term \cem.  }
\label{fig:cem}
\end{figure}

Equation~\ref{eq:emresolution} gives the functional form of the electron energy
resolution. We take the intrinsic resolution of the calorimeter, which is given
by the sampling term \sem, from the test beam measurements. The noise term
\nem\ is represented by the width of the $\Delta E$ distribution
(Fig.~\ref{fig:deltaupar}).
We measure the constant term \cem\ from the \zb\ line
shape of the data. We fit a
Breit-Wigner convoluted with a Gaussian, whose width characterizes the
dielectron mass resolution, to the \zb\ peak. Figure~\ref{fig:cem} shows the
width $\sigma_{\mee}$ of the Gaussian fitted to the \zb\ peak predicted by the
fast Monte Carlo as a function of \cem. The horizontal lines indicate the
width of the Gaussian fitted to the CC/CC \zb\ sample and its uncertainties,
$1.75\pm0.08$ GeV. We find that Monte Carlo and data agree if
$\cem = 0.0115^{+0.0027}_{-0.0036}$, as indicated by the arrows in
Fig.~\ref{fig:cem}. The measured \zb\ mass does not depend on \cem.

\section{ Recoil Measurement }
\label{sec-recoil}
\subsection { Recoil Momentum Response }

The detector response and resolution for particles recoiling against a \wb\
boson should be the same as for particles recoiling against a \zb\ boson. For
\zee\ events, we can measure the transverse momentum of the \zb\ from the
$e^+e^-$ pair, \ptee,  into
which it decays and from the recoil momentum \ut\ in the same way as for \wev\
events. By comparing \ptee\ and \ut\ we calibrate the recoil response relative
to the electron response.

\begin{figure}[htpb!]
\vspace{-0.3in}
\centerline{\psfig{figure=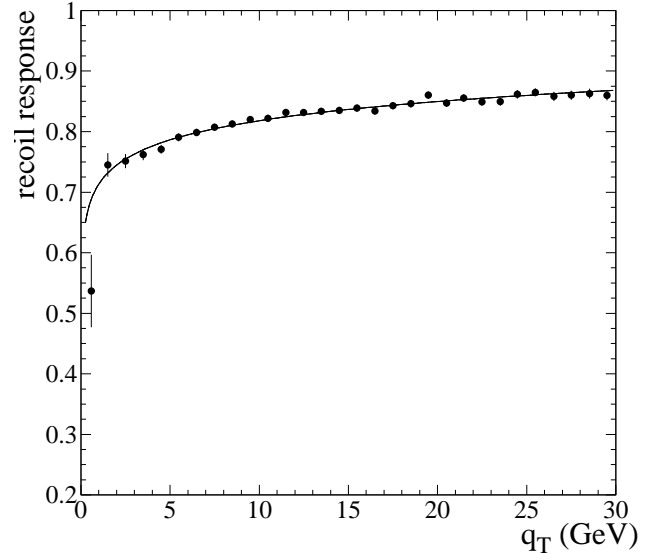,width=3.5in,height=3.5in}}
\vspace{-0.in}
\caption{ The recoil momentum response in the Monte Carlo \zb\ sample as a
function of $q_T$.}
\label{fig:MC_rec_response}
\end{figure}

The recoil momentum is carried by many particles, mostly hadrons, with a wide
momentum spectrum. Since the response of calorimeters to hadrons tends to be
nonlinear and the recoil particles are distributed all over the calorimeter,
including module boundaries with reduced response, we expect a momentum
dependent response function with values below unity.  
In order to fix the functional form of the recoil momentum response, we study
the response predicted by a Monte Carlo \zee\ sample obtained using the \HERW\
program and a \GEAN-based detector simulation. We project the
reconstructed transverse recoil momentum onto the direction of motion of the
\zb\ and define the response as
\begin{eqnarray}
    \rrec = \frac{\left| \utv\cdot\hat{q}_T \right|}{\left| \qt \right|},
\end{eqnarray}
where \qt\ is the generated transverse momentum of the \zb\ boson.
Figure~\ref{fig:MC_rec_response} shows this
response as a function of \qt. A response function of the form
\begin{equation}
\rrec = \alpharec + \betarec \log \left(\qt/\hbox{GeV}\right)
\end {equation}
fits the response predicted by \GEAN\ with $\alpharec =0.713\pm0.006$ and
$\betarec =0.046\pm0.002$.
This functional form also describes the jet energy response of the \Dzero\
calorimeter.

To measure the recoil response from the collider data we use the CC/CC+EC \zb\
sample. We allow one of the
leptons from the \zee\ decay to be in the CC or the EC, so that the rapidity
distribution of the \zb\ bosons approximates that of the \wb\ bosons. We
require both leptons to satisfy the tight electron criteria. This reduces the
background for the topology with one lepton in the EC. We also require the Main
Ring Veto as for the \wb\ sample (Sec.~\ref{sec-data}).

\begin{figure}[htpb!]
\vspace{-0.3in}
\centerline{\psfig{figure=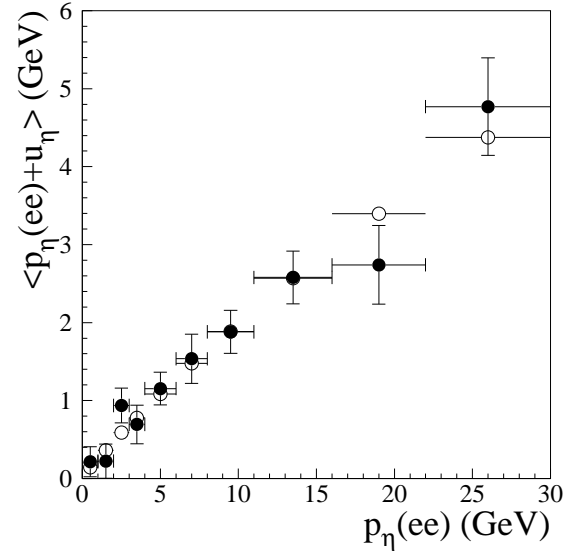,width=3.5in,height=3.5in}}
\vspace{-0.02in}
\caption{ The average $p_\eta(ee)+u_\eta$ versus $p_\eta(ee)$ for
the \zb\ data ($\bullet$) and the fast Monte Carlo simulation ($\circ$) .}
\label{fig:zbal}
\end{figure}

\begin{figure}[htpb!]
\vspace{-0.3in}
\centerline{\psfig{figure=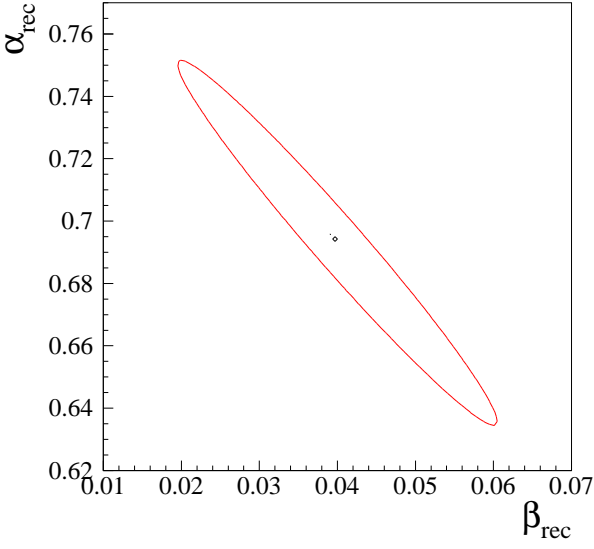,width=3.5in,height=3.5in}}
\vspace{-0.02in}
\caption{ The $\chi^2_0 + 1$ contour for the recoil momentum response
parameters.} 
\label{fig:zbal_chisq}
\end{figure}

We project the
transverse momenta of the recoil, \ut, and the \zb\ as measured by the two
electrons, \ptee, on the inner bisector of the electron directions
($\eta$-axis), as shown in Fig.~\ref{fig:zdef}. By projecting
the momenta on an axis that is independent of any energy
measurement, noise contributions to the momenta average to zero and do not bias
the result. We bin the data in $p_\eta(ee)$ and
plot the mean of the sum of the two projections, $u_\eta+p_\eta(ee)$, versus the
mean of $p_\eta(ee)$ (Fig.~\ref{fig:zbal}). We perform a
two-dimensional $\chi^2$ fit for the two parameters by comparing the data to
predictions of the fast Monte Carlo for different values of \alpharec\ and
\betarec. Figure~\ref{fig:zbal} also shows the prediction of the Monte
Carlo for the values of the parameters that give the best fit.
Figure~\ref{fig:zbal_chisq} shows the contour for $\chi^2=\chi^2_0 + 1$. The
best fit ($\chi^2_0=5$ for 8 degrees of freedom) is achieved for $\alpharec =
0.693\pm 0.060$ and $\betarec =  0.040 \pm 0.021$. The two parameters are
strongly correlated with a correlation coefficient $\rho=-0.979$.

\subsection { Recoil Momentum Resolution }
We parameterize the resolution for the hard component of the recoil as
\begin{equation}
\sigrec = \srec \sqrt{\ut},
\end{equation}
where \srec\ is a tunable parameter.

The soft component of the recoil is modeled by the transverse momentum balance
\mpt\ from minimum bias events, multiplied by a 
correction factor \alphamb\ (Eq.~\ref{eq:ut}).
This automatically models the effects of detector resolution and pile-up. To
model the pile-up
correctly as a function of luminosity, we need to take the minimum bias events
at the same luminosity as the \wb\ events.
At a given luminosity the mean number of interactions in
minimum bias events is always smaller than the mean number
of interactions in \wb\ events.  To model the detector
resolution correctly, the minimum bias events must have the
same interaction multiplicity spectrum as the \wb\ events. We therefore weight
the minimum bias events so that their interaction multiplicity approximates
that of the \wb\ events. As a measure of the interaction multiplicity on an
event-by-event basis, we use the multiplicity of vertices reconstructed from the
tracks in the CDC and the timing structure of the Level 0 hodoscope
signals~\cite{tracy}.

\begin{figure}[htpb!]
\vspace{-0.3in}
\centerline{\psfig{figure=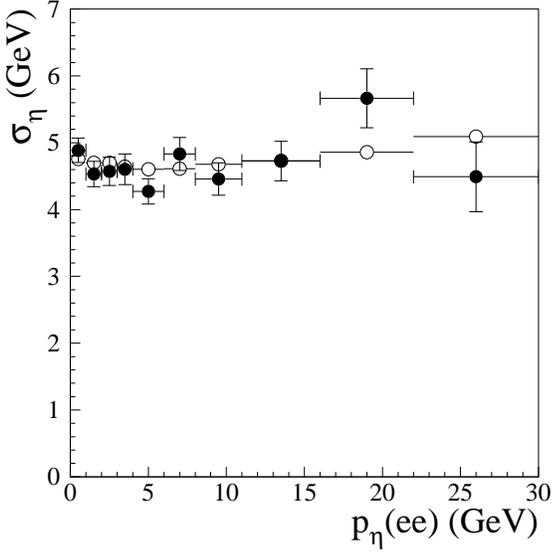,width=3.5in,height=3.5in}}
\vspace{-0.02in}
\caption{ The width of the $\eta$-balance distribution versus $p_\eta(ee)$ for
the \zb\ data ($\bullet$) and the fast Monte Carlo simulation ($\circ$).}
\label{fig:eta_res}
\end{figure}

\begin{figure}[htpb!]
\vspace{-0.3in}
\centerline{\psfig{figure=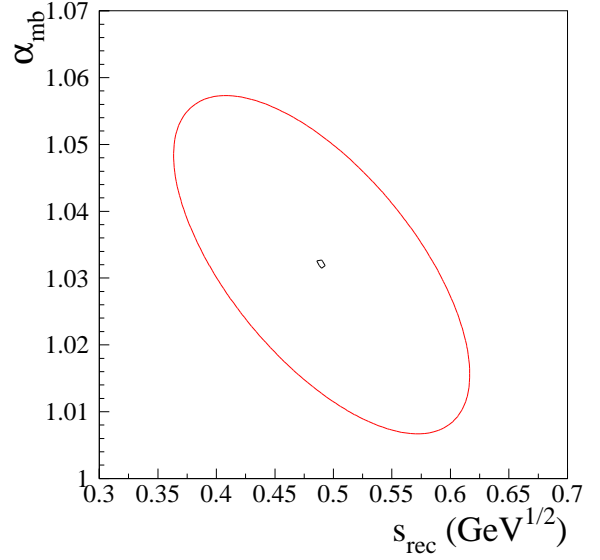,width=3.5in,height=3.5in}}
\vspace{-0.02in}
\caption{ The $\chi_0^2+1$ contour for the recoil resolution parameters
\alphamb\ and \srec.}
\label{fig:srec_vs_amb}
\end{figure}

We tune the two parameters \srec\ and \alphamb\ using the CC/CC+EC \zb\
sample. The width of the spectrum of the $\eta$-balance,
$u_\eta/\rrec+p_\eta(ee)$, is a measure of the recoil momentum
resolution. Figure~\ref{fig:eta_res} shows this width $\sigma_\eta$ 
as a function of $p_\eta(ee)$. The contribution of the
electron momentum resolution to the width of the $\eta$-balance is negligibly
small. The contribution of the recoil momentum resolution grows with
$p_\eta(ee)$ while the contribution from the minimum bias \mpt\ is
independent of $p_\eta(ee)$. This allows us to determine \srec\ and
\alphamb\ simultaneously and without sensitivity to the electron resolution
by comparing the width of the $\eta$-balance predicted by the Monte Carlo model
with that observed in the data in bins of $p_\eta(ee)$. We perform a $\chi^2$
fit comparing Monte Carlo and collider data. Figure~\ref{fig:srec_vs_amb} shows
contours of constant $\chi^2$ in the \alphamb-\srec\ plane. The best
agreement ($\chi^2_0=10.3$ for 8 degrees of freedom) occurs for
$\srec =0.49\pm0.14\ \hbox{GeV}^{1/2}$ and $\alphamb =1.032\pm0.028$
with a correlation coefficient $\rho=-0.60$ for the two parameters.
The $\xi$-balance, $u_\xi/\rrec+p_\xi(ee)$, is more sensitive to the
electron momentum resolution and is affected by changes in \srec\ and
\alphamb\ in the same way. We use it as a cross check only.

\begin{figure}[htpb!]
\vspace{-0.3in}
\centerline{\psfig{figure=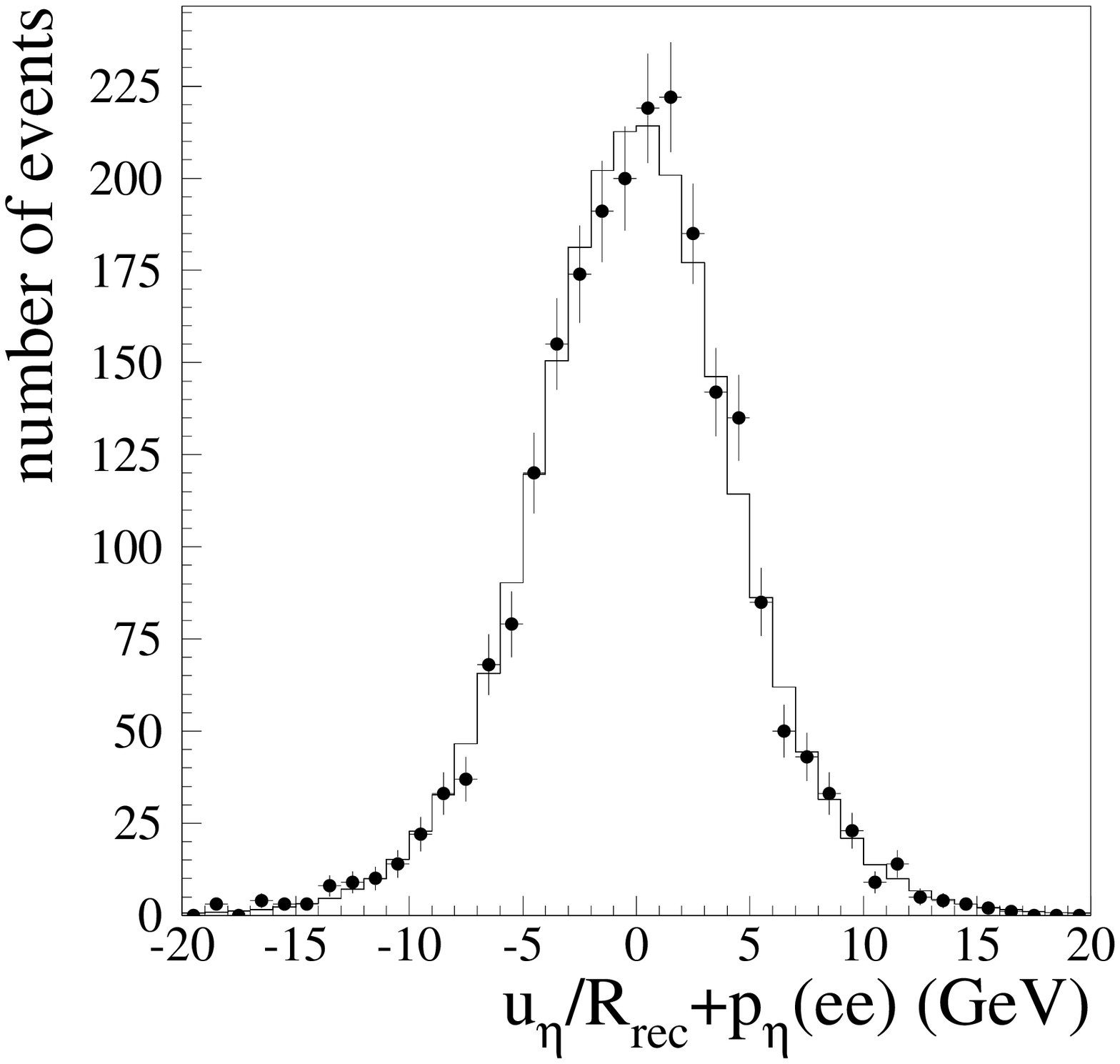,width=3.5in,height=3.5in}}
\vspace{-0.02in}
\caption{ The $\eta$-balance distribution for the \zb\ data ($\bullet$) and the
fast Monte Carlo simulation (-----).}
\label{fig:eta_balance}
\end{figure}

\begin{figure}[htpb!]
\vspace{-0.3in}
\centerline{\psfig{figure=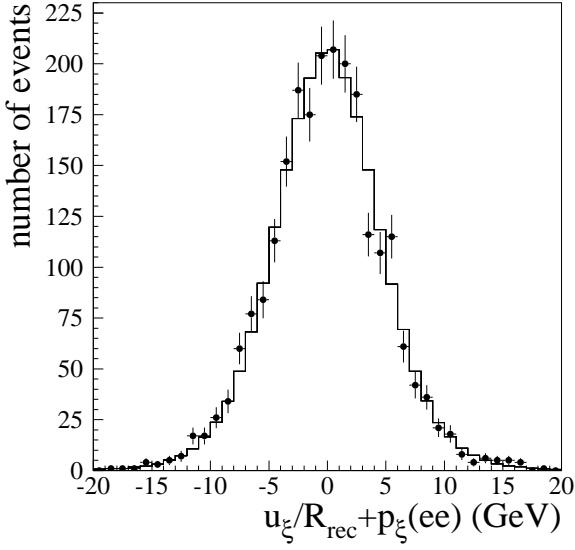,width=3.5in,height=3.5in}}
\vspace{-0.02in}
\caption{ The $\xi$-balance  distribution for the \zb\ data ($\bullet$) and
the fast Monte Carlo simulation (-----).}
\label{fig:xi_balance}
\end{figure}

Figure~\ref{fig:eta_balance} shows the spectrum of
$u_\eta/\rrec+p_\eta(ee)$ from the CC/CC+EC \zb\ data sample and from the fast
Monte Carlo with the tuned recoil resolution and response parameters. Figure
\ref{fig:xi_balance} shows the corresponding distributions for
$u_\xi/\rrec +p_\xi(ee)$. In both cases the agreement between data and Monte
Carlo simulation is good. A Kolmogorov-Smirnov test~\cite{KStest} gives 
confidence levels of $\kappa=0.33$ and 0.37
that the Monte Carlo and data spectra derive from the same parent
distribution. A $\chi^2$ test gives $\chi^2=25$ and 37, respectively, for 40
bins.

\begin{figure}[htpb!]
\vspace{-0.3in}
\centerline{\psfig{figure=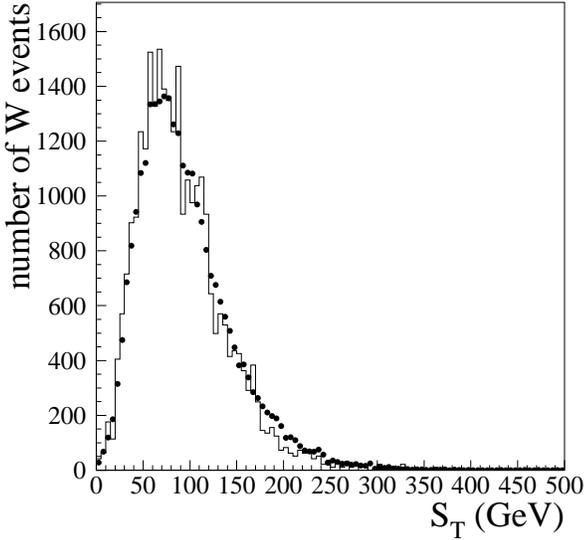,width=3.5in,height=3.5in}}
\vspace{-0.02in}
\caption{ The transverse energy flow in the \wb\ ($\bullet$)
and \zb\ (-----) data.}
\label{fig:wz_set}
\end{figure}

\begin{figure}[htpb!]
\vspace{-0.3in}
\centerline{
\psfig{figure=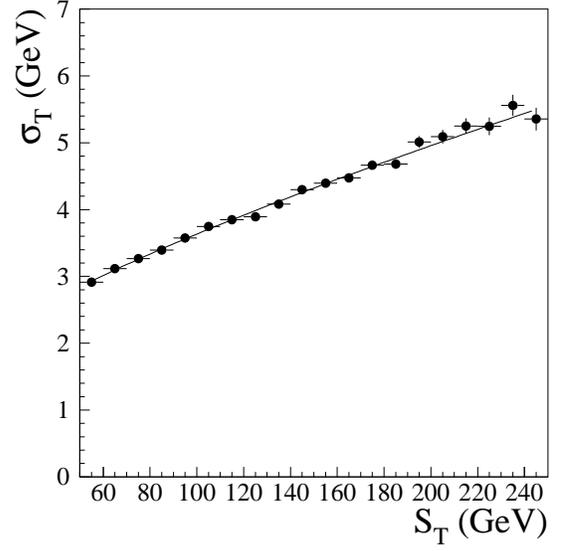,width=3.5in,height=3.5in}}
\vspace{-0.02in}
\caption{ The resolution for transverse momentum balance, $\sigma_T$, versus
the transverse energy flow, $S_T$, for minimum bias events ($\bullet$). The
smooth curve is a fit (Eq.~\protect\ref{eq:res_t}).}
\label{fig:mb_res}
\end{figure}

Figure~\ref{fig:wz_set} shows the overall energy flow transverse to the beam
direction measured by the sum $S_T = \sum_i E_i \sin\theta_i$ over all
calorimeter cells except cells belonging to an electron cluster. For
\wb\ events $\langle S_T \rangle=98.7\pm0.3$ GeV and for \zb\ events $\langle
S_T \rangle=91.0\pm0.9$ GeV. Increased transverse energy flow leads to a worse
recoil momentum resolution and therefore we need to correct the value of
\alphamb\ for the \wb\ sample to account for this difference. Figure
\ref{fig:mb_res} relates transverse energy flow $S_T$ to resolution
$\sigma_T$ for a minimum bias event sample. The resolution for measuring
transverse momentum balance along any direction is
\begin{eqnarray}
\sigma_T(S_T) = 1.42\ \hbox{GeV} + 0.15 \sqrt{S_T\,\hbox{GeV}} + 0.007 S_T
\label{eq:res_t}
\end{eqnarray}
for minimum bias events. The different energy flows in \wb\ and \zb\ events
lead to a correction to \alphamb\ of $\sigma_T(98.7\ \mbox{GeV})/
\sigma_T(91.0\ \mbox{GeV})=1.03\pm0.01$. The uncertainty reflects the
uncertainties in the determination of $\langle S_T \rangle$. This uncertainty
does not correlate with \srec.

\zb\ bosons are not intrinsically produced with less energy flow in the
underlying event than \wb\ bosons. Rather, the requirement of two reconstructed
isolated electrons biases the event selection in the \zb\ sample towards events
with lower energy flow compared to the events in the \wb\ sample which have only
one electron. We demonstrate this by
loosening the electron identification requirements for one of the electrons
in the \zb\ sample. We use events that were collected using less restrictive
trigger conditions for which at Level 2 only one of the electron candidates
must satisfy the shape and isolation requirements. We find that if all electron
quality cuts are removed for one electron $S_T$ increases by 7\%,
consistent with the ratio of the $S_T$ values in the \wb\ and \zb\ samples.

\subsection {Comparison with \wb\ Data}

\begin{figure}[htpb!]
\vspace{-0.3in}
\centerline{\psfig{figure=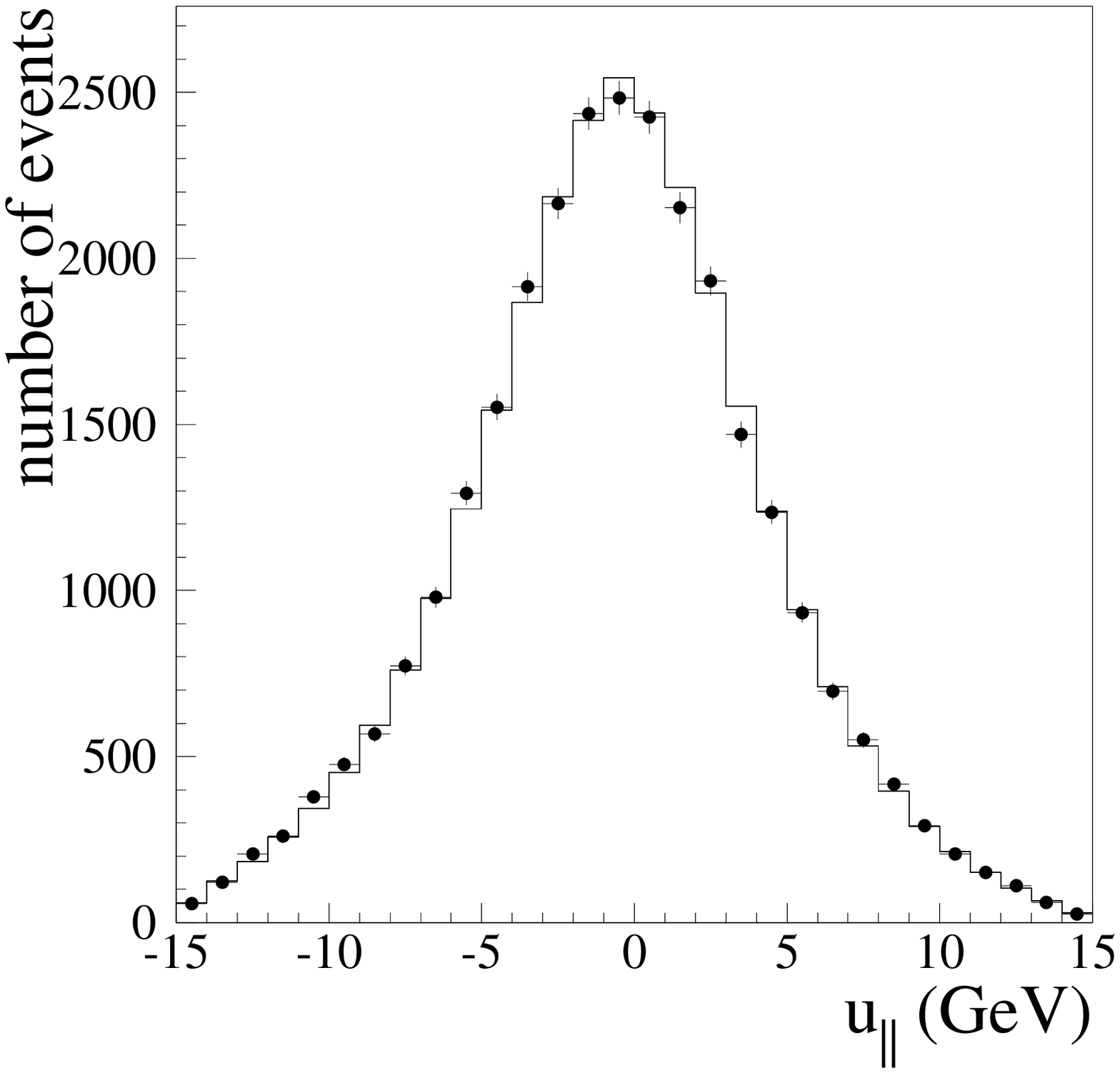,width=3.5in,height=3.5in}}
\vspace{-0.02in}
\caption{ The \upar\ spectrum for the \wb\ data
($\bullet$) and the Monte Carlo simulation (-----).}
\label{fig:uparproj}
\end{figure}

\begin{figure}[htpb!]
\vspace{-0.3in}
\centerline{\psfig{figure=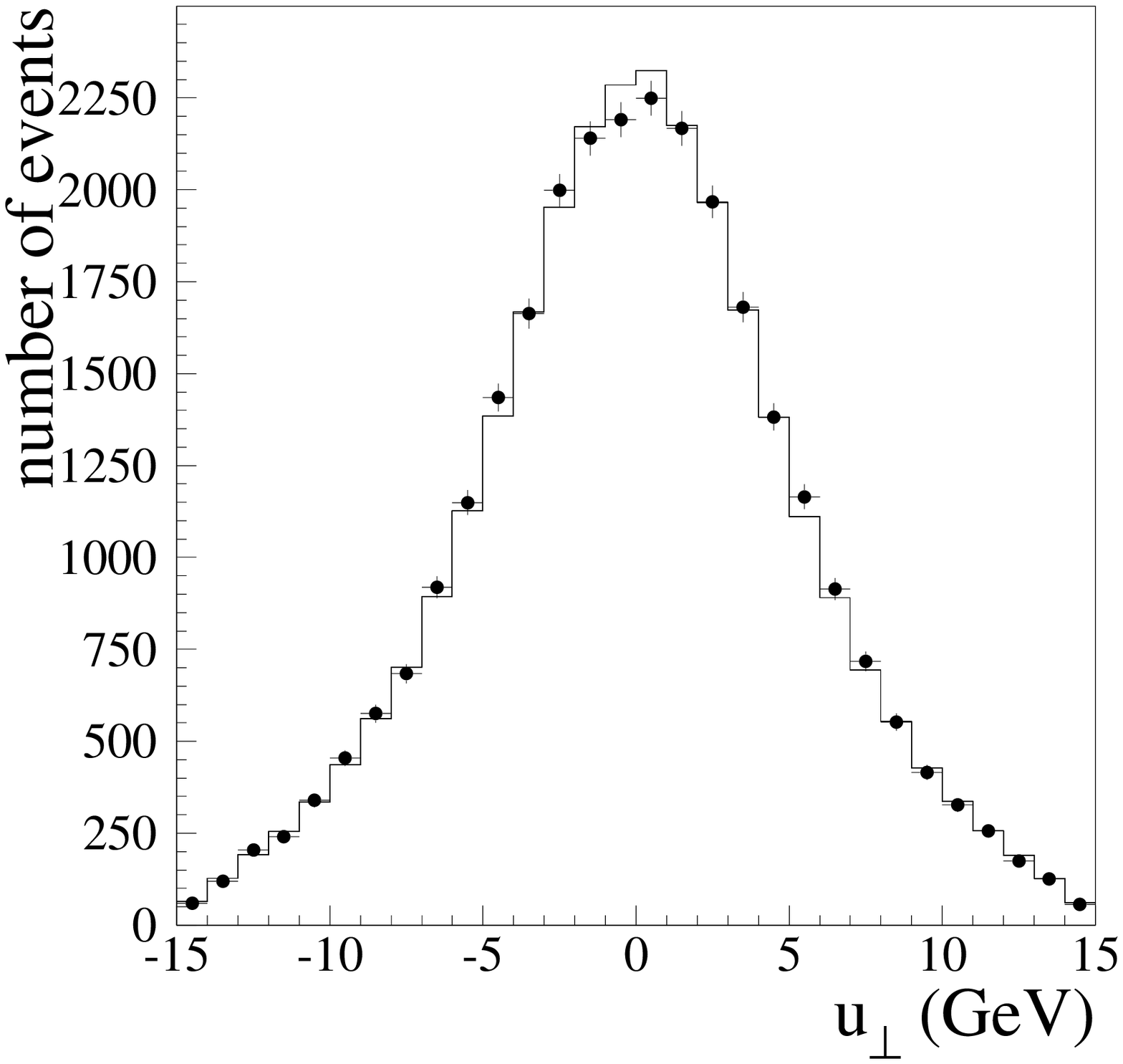,width=3.5in,height=3.5in}}
\vspace{-0.02in}
\caption{ The \uper\ spectrum for the \wb\ data
($\bullet$) and the Monte Carlo simulation (-----).}
\label{fig:uperproj}
\end{figure}

\begin{figure}[htpb!]
\vspace{-0.3in}
\centerline{\psfig{figure=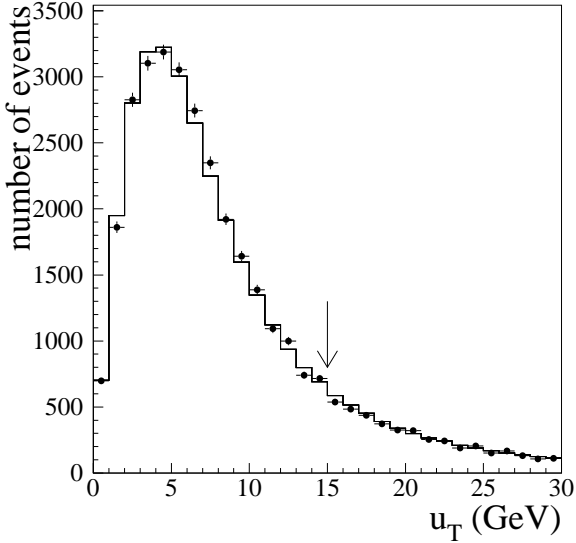,width=3.5in,height=3.5in}}
\vspace{-0.02in}
\caption{ The recoil momentum (\ut) spectrum for the \wb\ data
($\bullet$) and the Monte Carlo simulation (-----).  The arrow shows the
location of the cut.  }
\label{fig:ut}
\end{figure}

\begin{figure}[htpb!]
\vspace{-0.1in}
\centerline{\psfig{figure=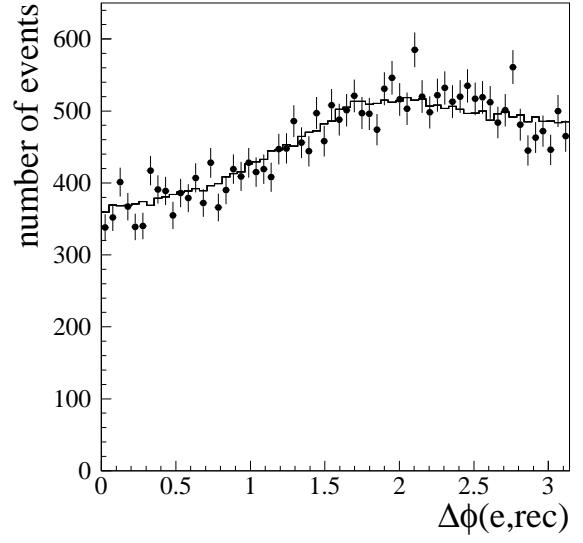,width=3.5in,height=3.5in}}
\vspace{-0.02in}
\caption{ The azimuthal difference between electron and recoil directions
for the \wb\ data ($\bullet$) and the Monte Carlo simulation (-----).}
\label{fig:deltaphi}
\end{figure}

We compare the recoil momentum
distribution in the \wb\ data to the predictions of the fast Monte Carlo,
which includes the parameters determined in this section and
Sec.~\ref{sec-elec}.
Figure~\ref{fig:uparproj} compares the \upar\ spectra from Monte Carlo and \wb\
data. The mean \upar\ for the \wb\ data is $-0.64\pm0.03$ GeV
and for the Monte Carlo prediction including backgrounds
it is $-0.61\pm0.01$ GeV, in very good agreement. This is important because a
bias in \upar\ would translate into a bias in the determination of \mt\
(Eq.~\ref{eq:mt_upar}). The agreement means that recoil momentum
response and resolution and the \upar\ efficiency parameterization
describe the data well. Figures \ref{fig:uperproj}--\ref{fig:deltaphi} show
\uper, \ut, and the azimuthal difference between electron and recoil directions
from Monte Carlo and \wb\ data.  The Kolmogorov-Smirnov probabilities for
Figs.~\ref{fig:uparproj}--\ref{fig:deltaphi} are
$\kappa=0.15$, 0.38, 0.16, and 0.11, respectively.

\section{ Constraints on the W Boson \pt\ Spectrum}
\label{sec-constraints}
\subsection {Parameters}

Since we cannot reconstruct a Lorentz invariant mass for \wev\ decays, knowledge
of the transverse momentum distribution of the \wb\ bosons is necessary to
measure the mass from the kinematic distributions. Theoretical calculations
provide a formalism to describe the boson \pt\ spectrum, but it includes
phenomenological parameters $g_1$, $g_2$, and $g_3$, which need to be determined
experimentally (Sec.~\ref{sec-vb_prod}). In addition, the boson \pt\ spectrum
also depends on the choice of parton distribution functions and \lqcd.

We can measure the \wb\ boson \pt\ spectrum only indirectly by measuring \utv,
the \pt\ of all particles that recoil against the \wb\ boson. Momentum
conservation requires the \wb\ boson \pt\ to be equal and opposite to \utv. 
The precision of the \utv\ measurement is
insufficient, especially for small \ut, to constrain the \wb\ spectrum as
tightly as is necessary for a precise \wb\ mass measurement.

We therefore have to find other data sets to constrain the model.
The formalism that describes the \pt\ spectrum of the \wb\ bosons has to
simultaneously describe the \pt\ spectrum of \zb\ bosons and the dilepton \pt\
spectrum from Drell-Yan production with the same model parameter values.
The authors of Ref.~\cite{LY} find
\begin{eqnarray}
g_1 & = & 0.11^{+0.04}_{-0.03}\ \GeV^2; \nonumber \\
g_2 & = & 0.58^{+0.1}_{-0.2}\ \GeV^2; \label{eq:g} \\
g_3 & = & -1.5^{+0.1}_{-0.1}\ \GeV^{-1} \nonumber
\end{eqnarray}
for mass cut-off $Q_0=1.6$ GeV in Eq.~\ref{eq:sud} and CTEQ2M parton
distribution functions, by fitting Drell-Yan and \zb\ data at different values
of $Q^2$. We further constrain these parameters using our much larger \zb\ data
sample.

\subsection {Determination of $g_2$ from \zee\ Data}

\begin{figure}[htpb!]
\vspace{-0.3in}
\centerline{\psfig{figure=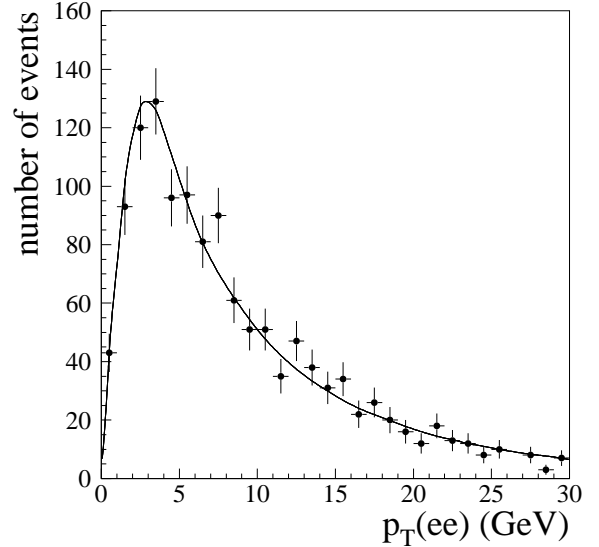,width=3.5in,height=3.5in}}
\vspace{-0.02in}
\caption{ Comparison of the \ptee\ data ($\bullet$) and simulation (-----)
for the best fit $g_2$ using MRSA$'$ parton distribution functions.}
\label{fig:ptz_data_vs_simulation}
\end{figure}

The \pt\ of \zb\ bosons can be measured more precisely than the \pt\ of \wb\
bosons by using the \ee\ pairs from their decays.
Figure~\ref{fig:ptz_data_vs_simulation} shows the \ptee\ spectrum observed in
the data.

To reduce the background contamination of the sample, the invariant
mass of \zb\ candidates must be within 10.5 GeV of the \zb\ peak position.
This mass window requirement reduces the background fraction to 2.5\%,
as determined from the dielectron invariant mass spectrum. As such it includes a
contribution from Drell-Yan \ee\ production, which has a \pt\ spectrum similar
to the signal and should not be counted as background in this case. To account
for this uncertainty we assign an error to the background fraction of \PM2.5\%.

\begin{figure}[htpb!]
\vspace{-0.3in}
\centerline{\psfig{figure=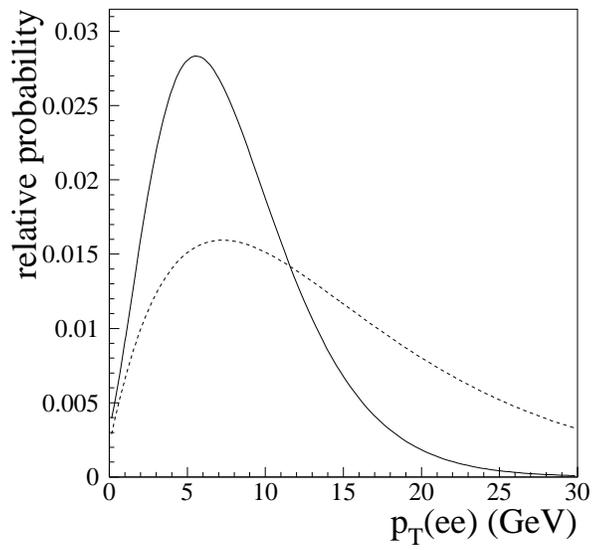,width=3.5in,height=3.5in}}
\vspace{-0.02in}
\caption{ The background parameterizations for the \ptee\ spectrum.}
\label{fig:pteebckgnd}
\end{figure}

The shape of the background is fixed by a sample of events with two
electromagnetic clusters which pass the same kinematic requirements as our
\zee\ sample, but fail the electron identification cuts\cite{dylanthesis}
(sample 1). As a cross-check we also use events with two jets,
each with more than 70\% of its energy in the EM
calorimeter (sample 2). Parameterizations of the two background shapes are shown
in Fig.~\ref{fig:pteebckgnd}. Their difference is taken to be the uncertainty
in background shape.

\begin{figure}[htpb!]
\vspace{-0.3in}
\centerline{
\psfig{figure=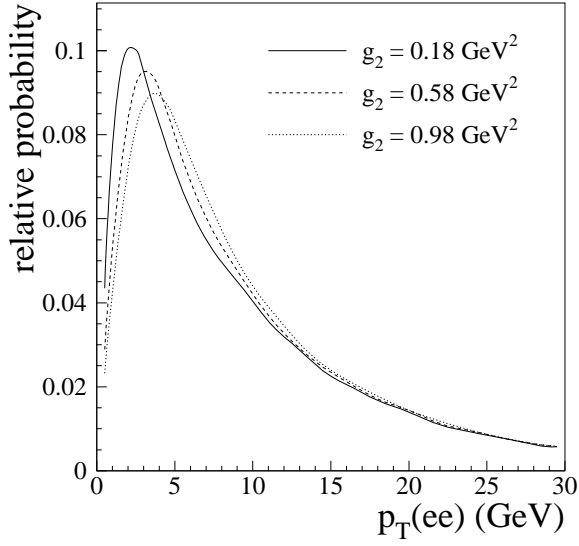,width=3.5in,height=3.5in}}
\vspace{-0.in}
\caption{The predicted \ptee\ spectra after detector simulation using MRSA$'$
parton distribution functions and $g_2=0.18$, 0.58, and 0.98 GeV$^2$.}
\label{fig:ptzsmearedvsg2}
\end{figure}

We use the fast Monte Carlo model to predict the \ptee\ spectrum from \zee\
decays for different sets of parameter values. The fast Monte Carlo simulates
the detector acceptance and resolution as discussed in the previous sections.
Figure~\ref{fig:ptzsmearedvsg2} shows the \ptee\ spectra predicted by the fast
Monte Carlo for MRSA$'$ parton distribution functions and three values of
$g_2$, with $g_1$ and $g_3$ fixed at the values given in Eq.~\ref{eq:g}.

The dominant effect of varying $g_2$ is to change the mean boson \pt.
Properly normalized and with the background contribution added, we use these
distributions as probability density functions to perform a maximum
likelihood fit for $g_2$. For a set of discrete values of $g_2$ we compute the
joint likelihood $L$ of the observed \ptee\ spectrum. We then fit $\log L$ as a
function of $g_2$ with a third order polynomial. The maximum of the polynomial
gives the fitted value of $g_2$. The value of $g_2$ has to be fit independently
for each parton distribution function choice.
We perform fits for four choices of parton distribution functions: MRSA$'$,
MRSD$-'$, CTEQ2M, and CTEQ3M. We fit the spectrum over the range $\ptee<15$
GeV, which corresponds to the range accepted by the \wb\ selection cuts.
The fits describe the data well.
Table~\ref{tab:g2fit_values} lists the fitted values for $g_2$ for the different
parton distribution function choices.
The result of the CTEQ2M fit is in good agreement with the value in
Eq.~\ref{eq:g}.

We estimate systematic uncertainties in the $g_2$ fit by running the fast
Monte Carlo with different parameter values and refitting the predicted \ptee\
spectrum with the nominal probability density functions. The uncertainties in
electron momentum response and resolution, \upar\ efficiency parametrization,
fiducial cuts, model of radiative decays, and background translate into a
systematic uncertainty in $g_2$ of 0.05 GeV$^2$.

As a cross-check we also fit the spectrum of the azimuthal separation \dphiee\
between the two electrons to constrain $g_2$.
The \dphiee\ spectrum has smaller systematic uncertainties but less
statistical sensitivity to $g_2$ than the \dphiee\ spectrum.
In Table~\ref{tab:g2fit_values} we also quote the results for $g_2$ from a fit
to the \dphiee\ spectrum.

The Monte Carlo prediction for the fitted
$g_2$ value using MRSA$'$ parton distribution functions is superimposed as a
smooth curve 
on Fig.~\ref{fig:ptz_data_vs_simulation}. The Kolmogorov-Smirnov probability
that the two distributions are from the same parent distribution is 
$\kappa=0.72$ and 
the $\chi^2$ is 25.5 for 29 degrees of freedom. Both of these tests indicate a
good fit. We use this model to compute the probability density functions for the
final fits to the kinematic spectra from the \wb\ sample.

\section{ Backgrounds }
\label{sec-back}
\subsection {\wte}

The decay \wte\ is topologically indistinguishable from \wev. It is included in
the fast Monte Carlo simulation (Sec.~\ref{sec-mc}). This decay is suppressed 
by the branching fraction for 
\tev, $\left( 17.83\pm0.08 \right)$\% \cite{PDG}, and by the
lepton \pt\ cuts. It accounts for 1.6\% of events in the \wb\ sample. 

\subsection {Hadronic Background}

QCD processes can fake the signature of a \wev\ decay if a hadronic jet fakes
the electron signature and the transverse momentum balance is mismeasured. 

We estimate this background from the \mpt\ spectrum of events with an
electromagnetic cluster. Electromagnetic clusters in events with
low \mpt\ are almost all due to jets. A fraction satisfy our electron
selection criteria and fake an electron. From the shape of the \mpt\ spectrum
for these events we determine how likely it is for these events to have
sufficient \mpt\ to enter our \wb\ sample. 

\begin{figure}[htpb!]
\centerline{\psfig{figure=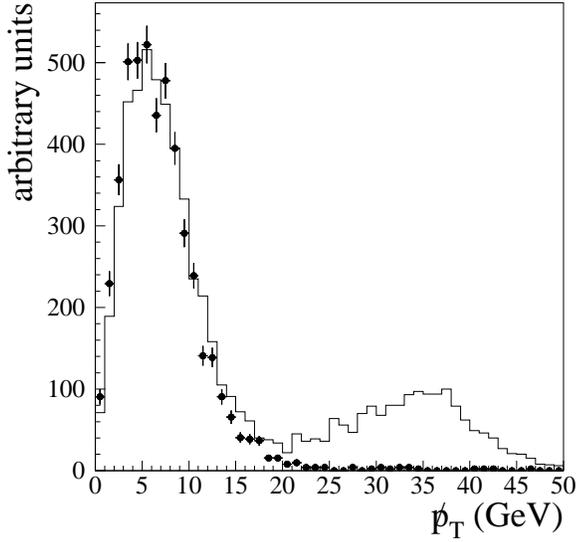,width=3.5in,height=3.5in}}
\vspace{-0.02in}
\caption{The \mpt\ spectra of a sample of events passing
electron identification cuts (------) and a sample of events failing
the cuts ($\bullet$).}
\label{fig:bkgmet}
\end{figure}

We determine this shape by selecting isolated 
electromagnetic clusters that  have
$\chi^2>200$ and $\sigm >10$. Almost all electrons fail this cut, so that
the remaining sample consists almost entirely of hadrons. We use data taken by
a trigger without the \mpt\ requirement to study the efficiencies of this cut
for jets. For $\mpt<10$ GeV we find 1973 such events, while in the same sample
3674 satisfy our electron selection criteria. If we normalize the background
spectrum to the electron sample we
obtain an estimate of the hadronic background in an electron candidate sample.
Figure~\ref{fig:bkgmet} shows the \mpt\ spectra of both  samples,
normalized for $\mpt<10$ GeV.

\begin{figure}[htpb!]
\centerline{\psfig{figure=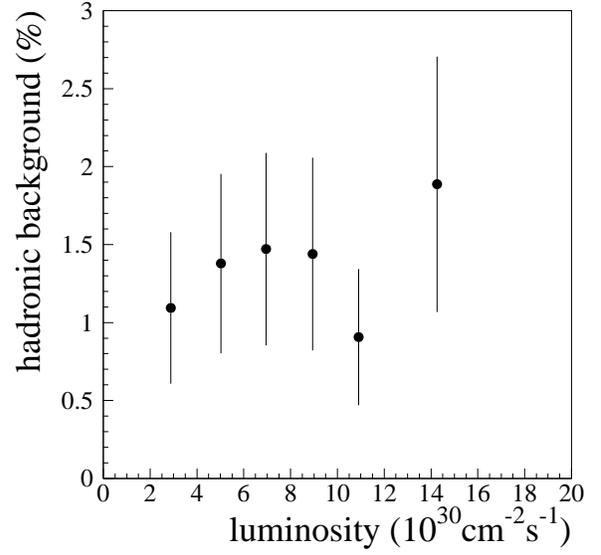,width=3.5in,height=3.5in}}
\vspace{-0.02in}
\caption{The fraction of hadron background as a function of luminosity.}
\label{fig:bkglum}
\end{figure}

\begin{figure}[htpb!]
\vspace{-0.3in}
\centerline{\psfig{figure=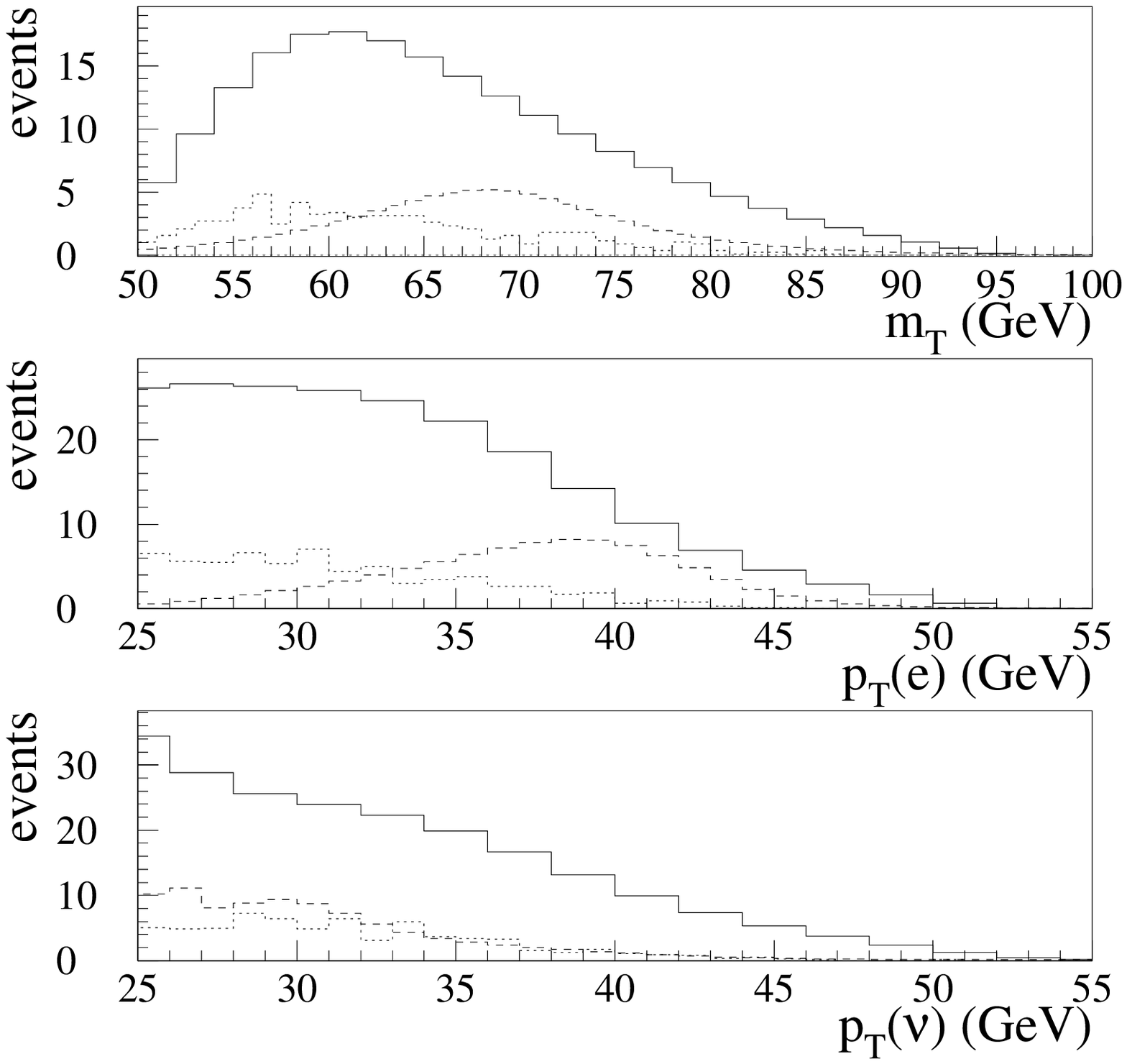,width=3.5in,height=3.5in}}
\vspace{-0.02in}
\caption{ Shape of \mt, \pte, and \ptnu\ spectra from hadron
(------), \zb\ (- - -), and $\tau\to$hadron backgrounds
($\cdot\cdot\cdot\cdot\cdot$) with the
proper relative normalization.}
\label{fig:bkg}
\end{figure}

In the data collected with the \wb\ trigger we find 204 events that satisfy
all the fiducial  and kinematic cuts, listed in Sec.~\ref{sec-data} for the
\wb\ sample, and have $\chi^2>200$ and $\sigm >10$. We therefore estimate
that 374 background  events entered the signal sample. This corresponds to a
fraction of the total \wb\ sample after all cuts 
of $f_{\rm had}  =  \left( 1.3 \pm 0.2 \right)$ \%.
For a looser cut on the recoil \pt, $\ut<30$ GeV, we find 
$f_{\rm had} = \left( 1.6 \pm 0.3 \right)$ \%. The error is dominated by
uncertainty in the relative normalization of the two samples 
at low \mpt. Figure~\ref{fig:bkglum} shows the background
fraction as a function of luminosity. There is no evidence for a significant
luminosity dependence. We use the background events with 
$\ptnu>25$ GeV to estimate the shape of the background contributions to the
\pte, \ptnu, and \mt\ spectra (Fig.~\ref{fig:bkg}).

\subsection{\zee}

To estimate the fraction of \zee\ events which satisfy the \wb\ selection, we
use a Monte Carlo sample of approximately 100{,}000 \zee\ events generated with
the \HERW\ program and a detector simulation based on \GEAN. The boson \pt\
spectrum generated by \HERW\ agrees reasonably well with the calculation in 
Ref.~\cite{LY}. \zee\ decays typically enter the \wb\ sample when one electron
satisfies the \wb\ cuts and the second electron is lost or mismeasured,
causing the event to have large \mpt.

Approximately 1.1\% of the \zee\ events have an electron with pseudorapidity
$|\eta|>4.0$, which is the acceptance limit of the end calorimeters.  
The fraction of \zee\ events which contain one electron with
$|\eta(e_1)|<1.0$ and $\pte>25$ GeV, and another with $|\eta(e_2)|>4.0$ is
approximately 0.04\%.  The contribution from the case of an electron lost
through the beampipe is therefore relatively small.

An electron is most frequently mismeasured when it goes into the regions
between the CC and one of the ECs, which are covered only by the hadronic
section of the calorimeter. These electrons therefore can 
not be identified and their
energy is measured in the hadronic calorimeter. Large \mpt\ is more likely for
these events than when both electrons hit the EM calorimeters. The mismeasured
electron contributes to the recoil when the event is treated as a \wb.  The
fraction of \zb\  events in the \wb\ sample therefore depends on the \ut\ cut.

We find that 10{,}987 Monte 
Carlo events pass the CC-CC \zee\ selection, and 758 (1{,}318) pass the \wb\
selection with a recoil cut of 15 (30) GeV. The fraction of 
\zb\ events in the \wb\ sample is therefore  
$f_Z = \left( 0.42 \pm 0.08 \right)\%$ for $\ut<15$ GeV 
and $\left( 0.62 \pm 0.08 \right)\%$ for $\ut<30$ GeV. The uncertainties quoted
include systematic uncertainties in the matching of momentum scales between
Monte Carlo and collider data. Figure~\ref{fig:bkg} shows the distributions of
\pte, \ptnu, and \mt\ for the events that satisfy
the \wb\ selection.

\subsection{\wth}

We estimate the background due to \wtv\ followed by a hadronic tau decay
based on two Monte Carlo samples. In a sample of \wth\ simulated using \GEAN, 
65 out of 4{,}514 events pass the fiducial and kinematic cuts of the \wb\
sample. We use a sample  of \wth\ simulated by replacing
the electron shower in \wev\ decays from collider data with the  hadrons from a
tau decay, generated by a Monte Carlo simulation, to estimate the
probability of the tau  decay products to fake an electron. Of 552 events 
that pass the fiducial and kinematic cuts 145 
pass the electron identification criteria. 
With the hadronic branching fraction for taus,
$B(\tau\to\hbox{hadrons})=64$\% we estimate a contamination of
the \wb\ sample of 0.24\% from hadronic tau decays. The expected background
shapes are plotted in Fig.~\ref{fig:bkg}. 

\subsection{ Cosmic Rays }

Cosmic ray muons can cause backgrounds when they coincide with a beam crossing
and radiate a photon of sufficient energy to mimic the signature of the
electron from \wev\ decays. We measure this background by searching for muons
near the electrons in the \wb\ signal sample. The muons have to be within
$10^\circ$ of the electron in azimuth. 
Using muon selection criteria similar to those in Ref.~\cite{top_xsect}
we observe 18 events with such muons in the \wb\ sample. 
We estimate the fraction of cosmic ray
events in the \wb\ sample to be $0.2\pm 0.1$\%. 
The effect of this background on the \wb\ mass measurement is negligible.

\section{ Mass Fits}
\label{sec-fit}
\subsection { Maximum Likelihood Fitting Procedure }

We use a binned maximum likelihood fit to extract the \wb\ mass.
Using the fast Monte Carlo program we compute the \mt, \pte, and \ptnu\
spectra for 200 hypothesized values of the \wb\ mass between 79.4 and 81.4 GeV.
For the \mt\
spectrum we use 100 MeV bins and for the lepton \pt\ spectra we use 50 MeV
bins. The statistical precision of the spectra for the \wb\ mass fit corresponds
to about 4 million \wb\ decays. When fitting
the collider data spectra we add the background contributions with the shapes
and normalizations described in Sec.~\ref{sec-back} to the signal spectra. We
normalize the spectra within the fit interval and interpret them as probability
density functions to compute the likelihood
\begin{equation}
L(m) = \prod_{i=1}^N p_i(m)^{n_i},
\end{equation}
where $p_i(m)$ is the probability density for bin $i$, assuming $\mw=m$, and
$n_i$ is the number of data entries in bin $i$. The product runs over all $N$
bins inside the fit interval. We fit $-\ln(L(m))$ with a quadratic function of
$m$. The value of $m$ at which the function assumes its minimum is the fitted
value of the \wb\ mass and the 68\% confidence level interval is the
interval in $m$ for which $-\ln(L(m))$ is within half a unit of its minimum.

As a consistency check of the fitting procedure we generate 105
Monte Carlo ensembles of 28{,}323 events each with \mw=80.4 GeV.
We then fit these ensembles with the same probability density functions as the
collider data spectra, except that we do not include the background
contributions. Table~\ref{tab:MCens} lists the mean, rms, and correlation
matrix of the fitted values.

\subsection { Electron \pt\ Spectrum }

\begin{figure}[htpb!]
\vspace{-0.3in}
\centerline{\psfig{figure=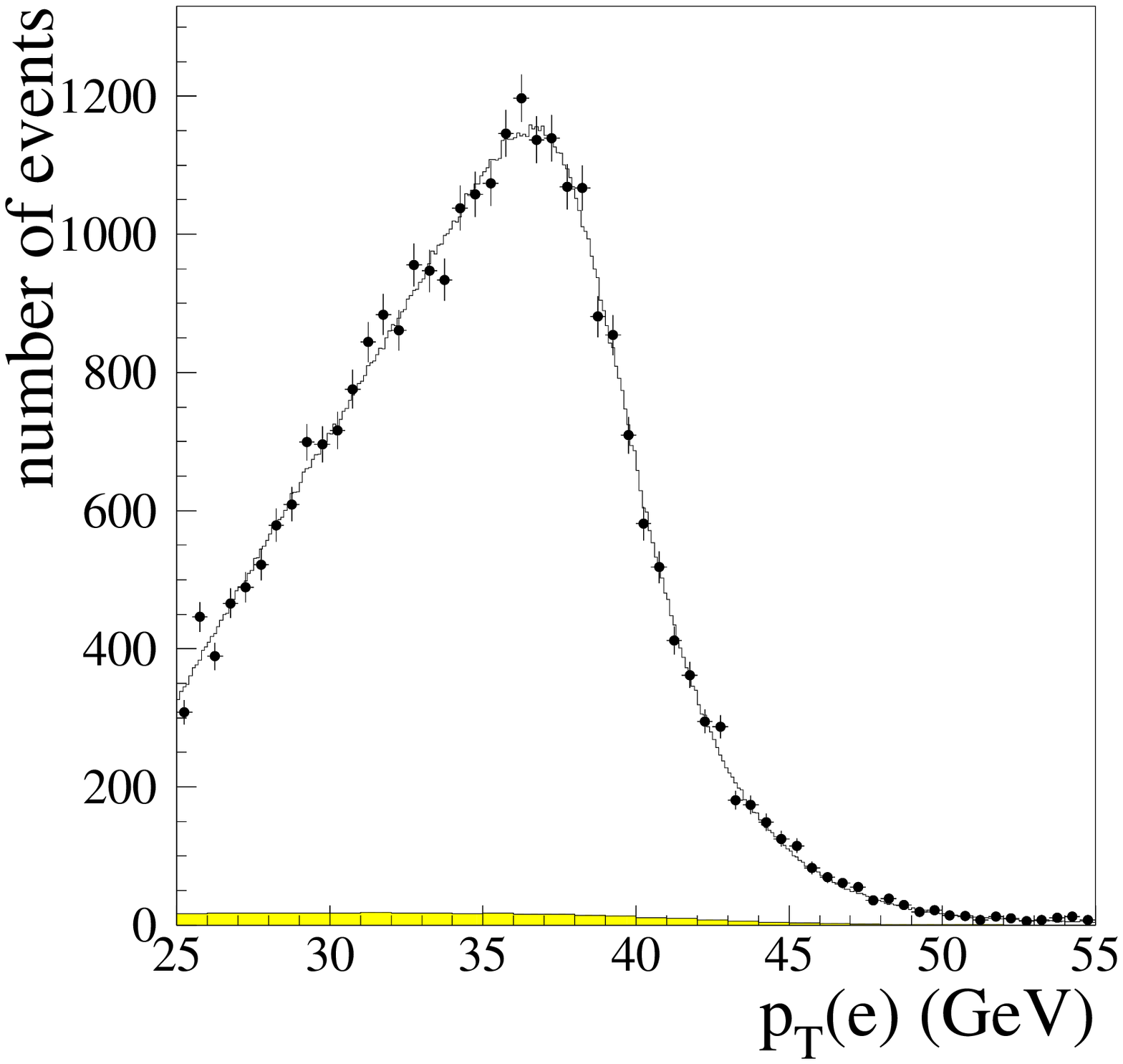,width=3.5in,height=3.5in}}
\vspace{-0.02in}
\caption{ Spectrum of \pte\ from the \wb\ data. The superimposed curve shows the
maximum likelihood fit and the shaded region the estimated background.}
\label{fig:eet}
\end{figure}

We fit the \pte\ spectrum in the region $30<\pte<50$ GeV. There are 22{,}898
events in this interval. The data points in Fig.~\ref{fig:eet} represent the
\pte\ spectrum from the \wb\ sample. The solid line shows the sum of the
simulated \wb\ signal and the estimated background for the best fit, and the
shaded region indicates the
sum of the estimated hadronic, \zee, and \wth\ backgrounds.
The maximum likelihood fit gives
\begin{equation}
\mw = 80.475 \PM 0.087\ \hbox{GeV}
\label{eq:mwpte}
\end{equation}
for the \wb\ mass.

\begin{figure}[htpb!]
\vspace{-0.3in}
\centerline{\psfig{figure=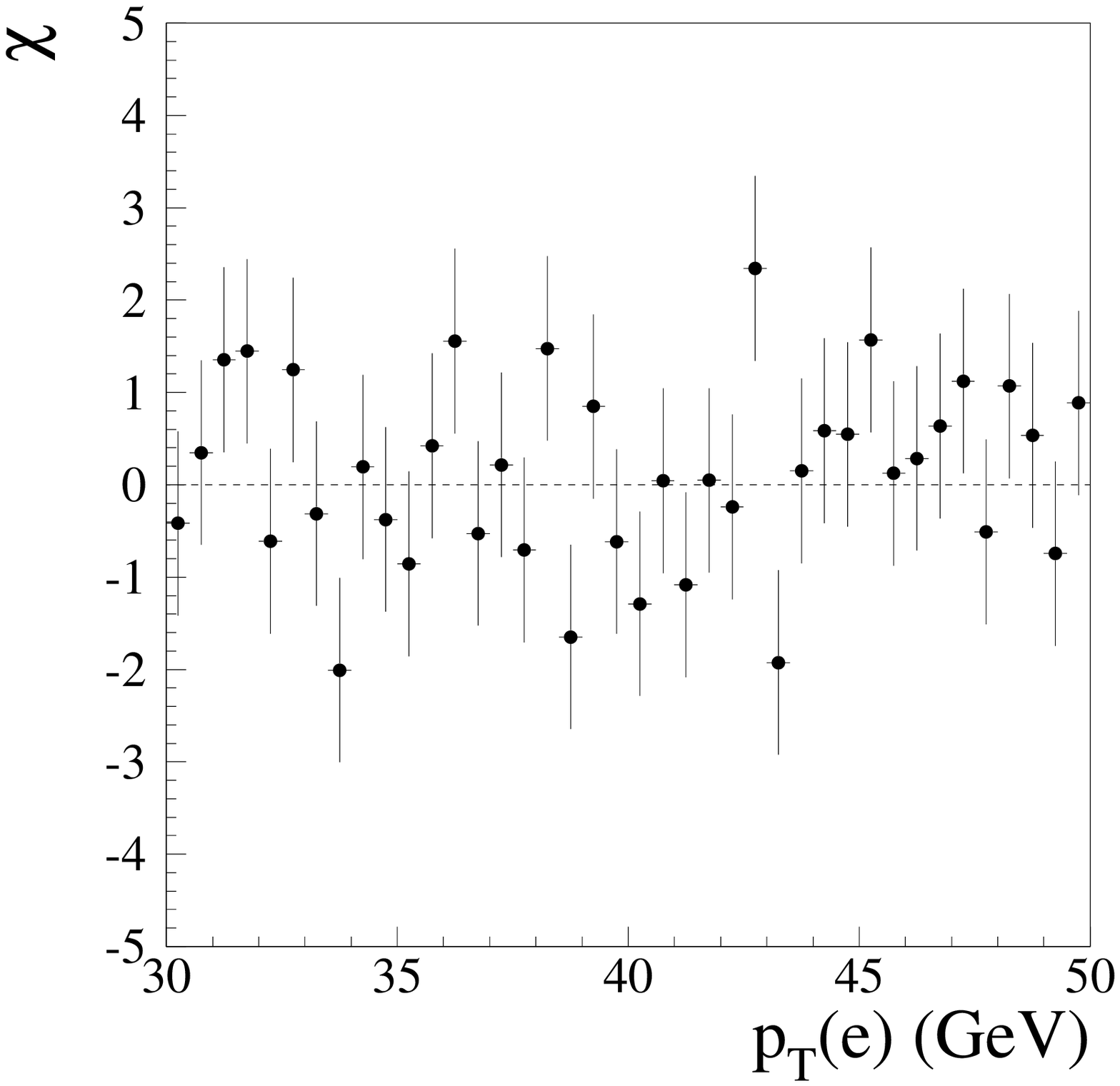,width=3.5in,height=3.5in}}
\vspace{-0.02in}
\caption{ The $\chi$ distribution for the fit to the \pte\ spectrum.}
\label{fig:chieet}
\end{figure}

As a goodness-of-fit test, we divide the fit interval into 0.5 GeV bins,
normalize the integral of the probability density function 
to the number of events in the fit interval, and compute
$\chi^2=\sum_{i=1}^N(y_i-P_i)^2/y_i$. The sum runs over all $N$ bins,
$y_i$ is the observed number of events in bin $i$, and $P_i$ is the
integral of the normalized probability density function over bin $i$. The parent
distribution is the $\chi^2$ distribution for $N-2$ degrees of
freedom. For the spectra in Fig.~\ref{fig:eet} we compute $\chi^2=40.6$. For
40 bins there is a 35\% probability for $\chi^2\ge40.6$. Figure~\ref{fig:chieet}
shows the contributions
$\chi_i=(y_i-P_i)/\sqrt{y_i}$ to $\chi^2$ for the 40 bins in the fit
interval.

We also compare the observed spectrum to the probability density function
using the Kolmogorov-Smirnov test.  For a comparison within the fit window
we obtain  $\kappa=0.81$ and for the entire histogram $\kappa=0.83$.

\begin{figure}[htpb!]
\vspace{-0.3in}
\centerline{\psfig{figure=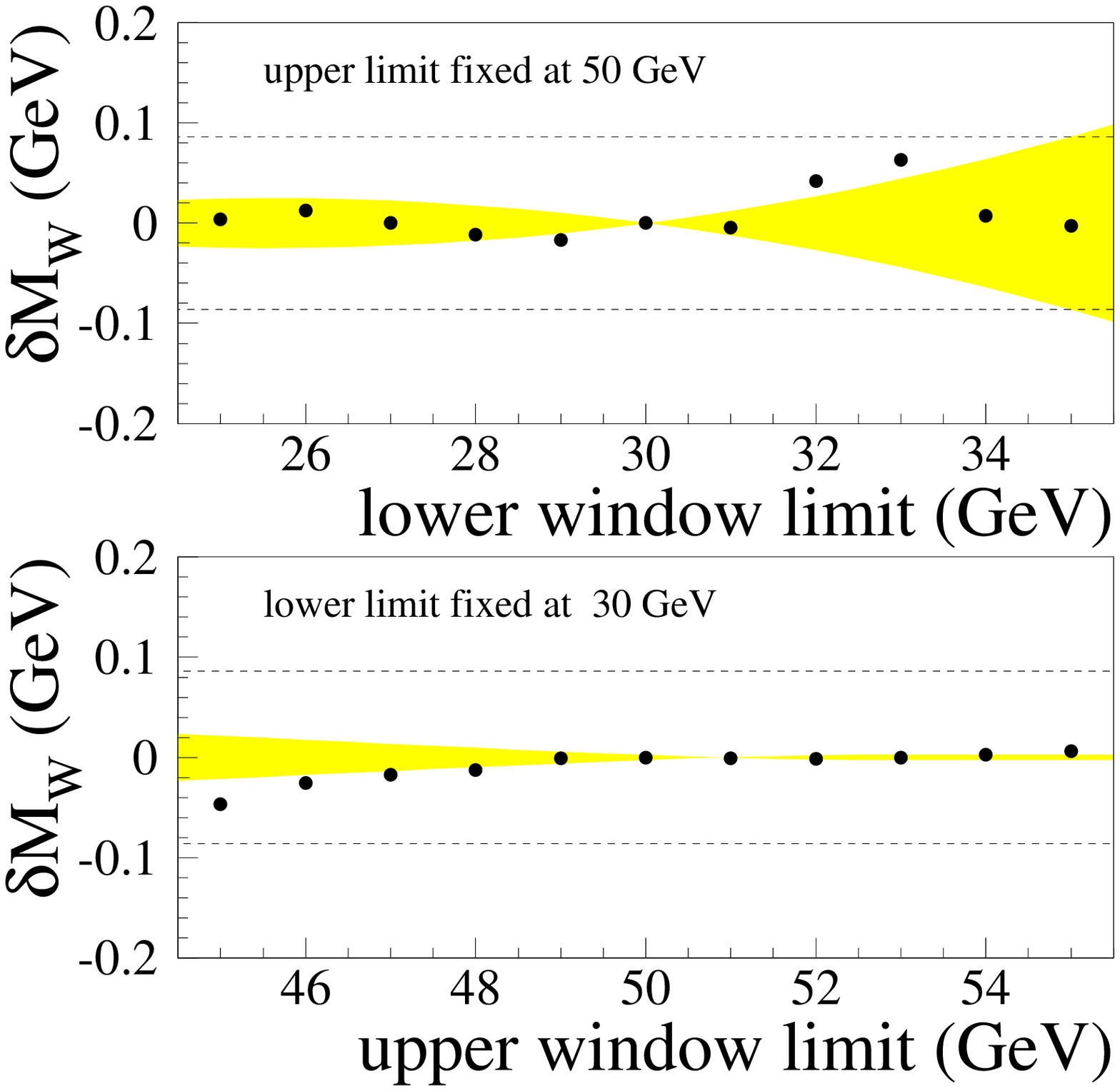,width=3.5in,height=3.5in}}
\vspace{-0.02in}
\caption{ Variation of the fitted mass with the \pte\ fit window limits. See
text for details. }
\label{fig:vareet}
\end{figure}

Figure~\ref{fig:vareet} shows the sensitivity of the
fitted mass value to the choice of fit interval. The points in the two plots
indicate the observed deviation of the fitted mass from the value given in
Eq.~\ref{eq:mwpte}. We expect some variation due to statistical
fluctuations in the spectrum and systematic uncertainties in the
probability density functions. We estimate the effect due to statistical
fluctuations using the Monte Carlo  ensembles described above. We expect the
fitted values
to be inside the shaded regions indicated in the two plots with 68\%
probability. The dashed lines indicate the statistical error for the nominal
fit.

All tests show that the probability density function provides a good description
of the observed spectrum.

\subsection { Transverse Mass Spectrum }

\begin{figure}[htpb!]
\vspace{-0.3in}
\centerline{\psfig{figure=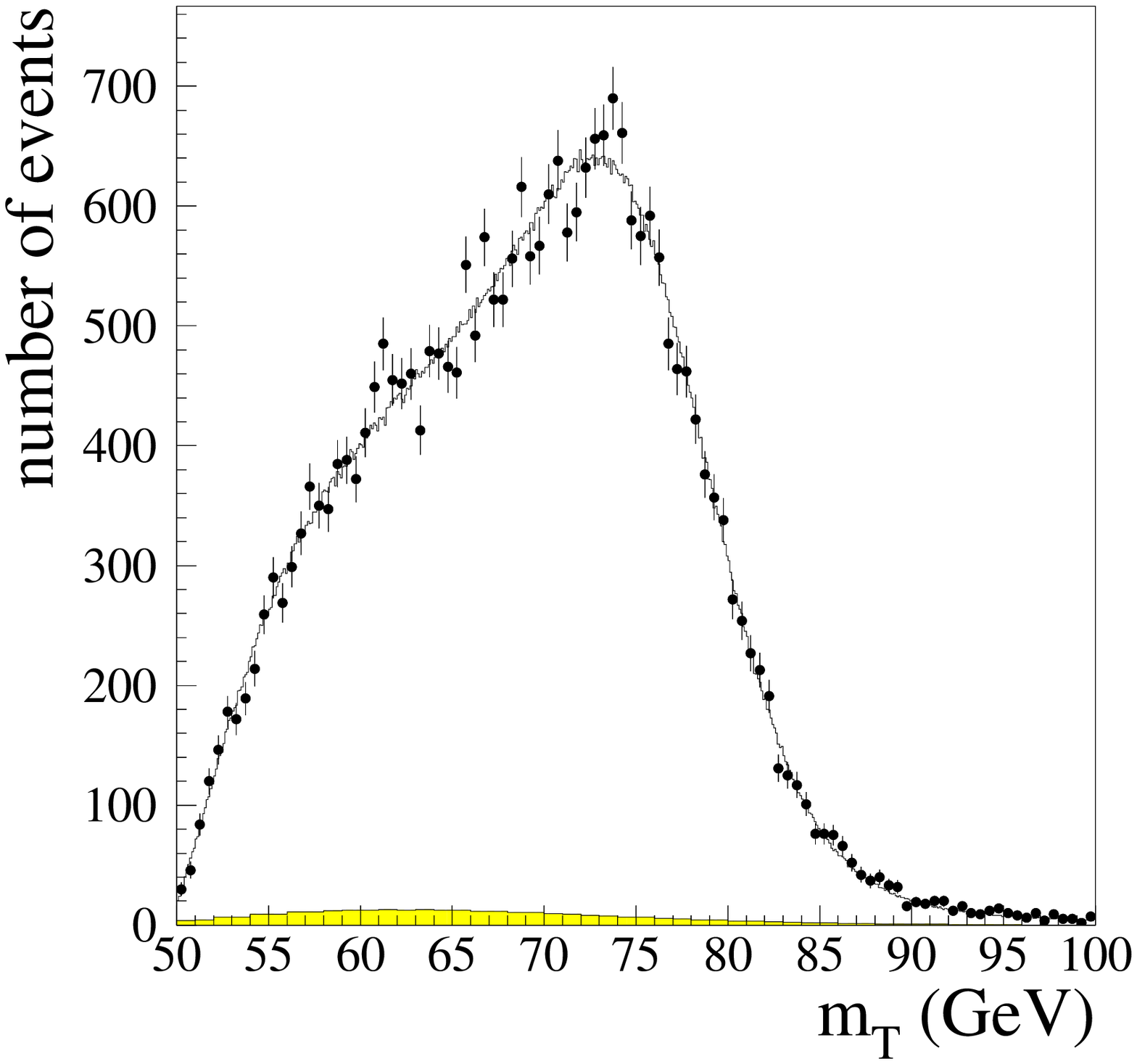,width=3.5in,height=3.5in}}
\vspace{-0.02in}
\caption{ Spectrum of \mt\ from the \wb\ data. The superimposed curve shows the
maximum likelihood fit and the shaded region shows the estimated background.}
\label{fig:mtw}
\end{figure}

\begin{figure}[htpb!]
\vspace{-0.3in}
\centerline{\psfig{figure=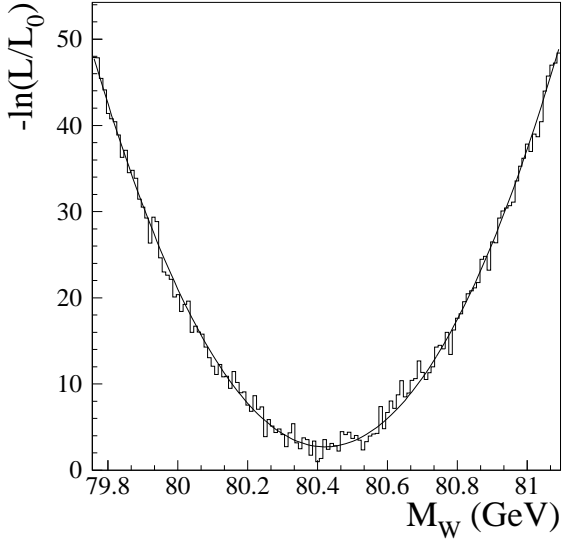,width=3.5in,height=3.5in}}
\vspace{-0.02in}
\caption{ The likelihood function for the \mt\ fit.}
\label{fig:like}
\end{figure}

Figure~\ref{fig:mtw} shows the \mt\ spectrum. The points are the observed
spectrum, the solid line shows signal plus background for the best fit, and
the shaded region indicates the estimated background contamination. We fit in
the interval
$60<\mt<90$ GeV. There are 23{,}068 events in this interval.
Figure~\ref{fig:like} shows
$-\ln(L(m)/L_0)$ for this fit where $L_0$ is an arbitrary number.
The best fit occurs for
\begin{equation}
\mw = 80.438 \PM 0.070\ \hbox{GeV}.
\label{eq:mw}
\end{equation}

\begin{figure}[htpb!]
\vspace{-0.3in}
\centerline{\psfig{figure=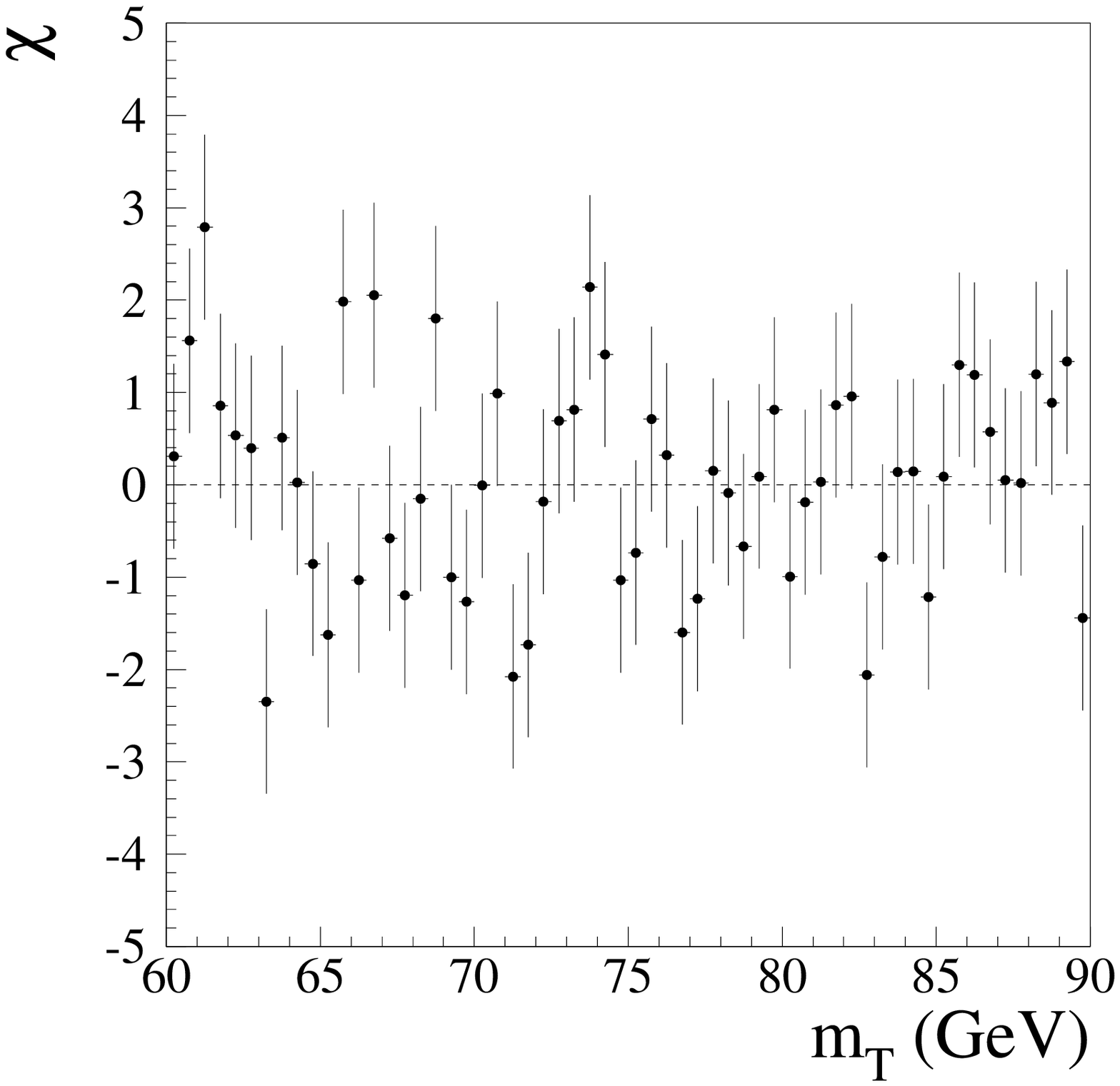,width=3.5in,height=3.5in}}
\vspace{-0.02in}
\caption{ The $\chi$ distribution for the fit to the \mt\ spectrum.}
\label{fig:chimt}
\end{figure}

\begin{figure}[htpb!]
\vspace{-0.3in}
\centerline{\psfig{figure=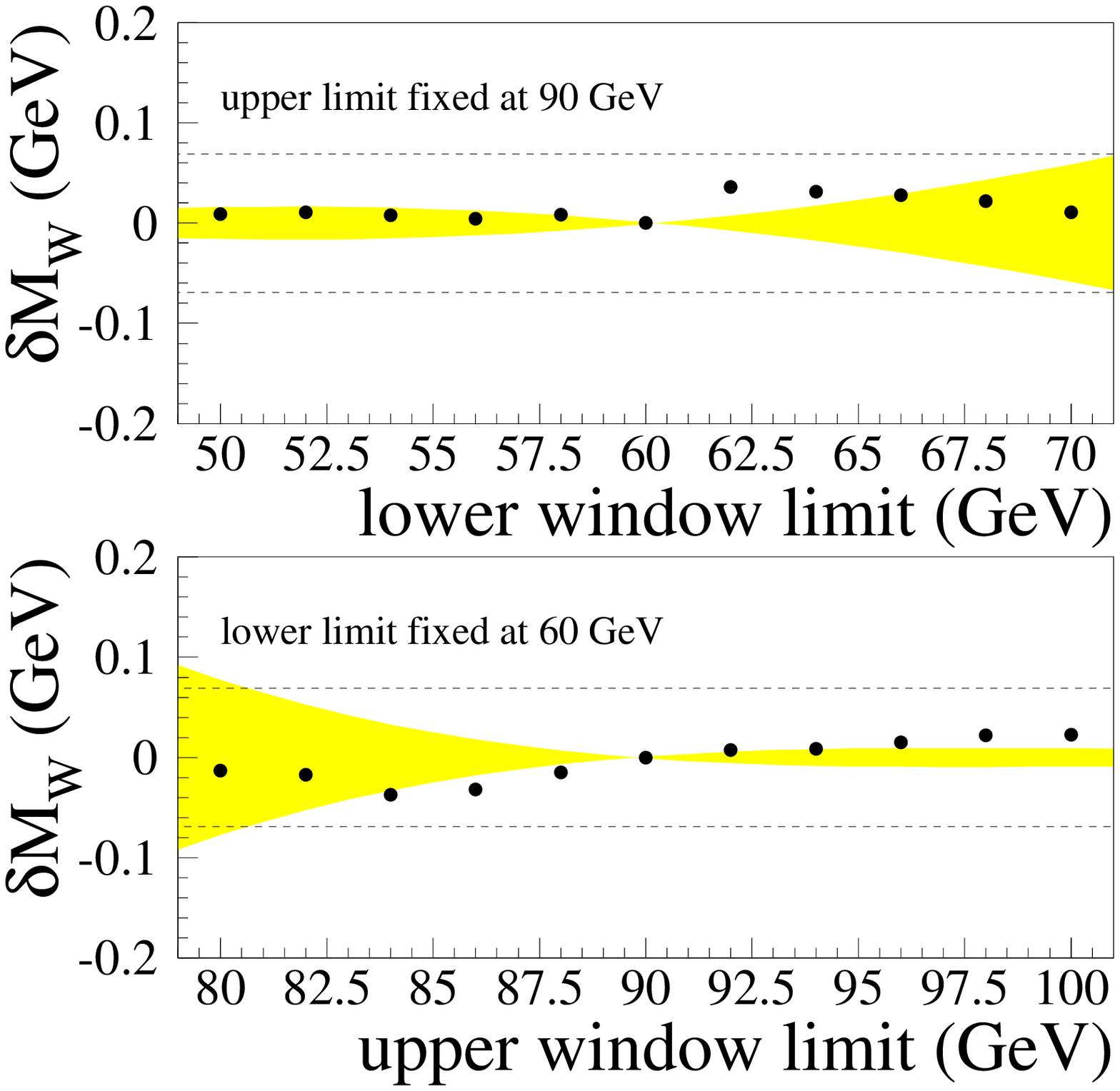,width=3.5in,height=3.5in}}
\vspace{-0.02in}
\caption{ Variation of the fitted mass with the \mt\ fit window limits. See
text for details. }
\label{fig:varmt}
\end{figure}

Figure~\ref{fig:chimt} shows the deviation of the data from the fit.
Summing over all bins in the fitting window, we get $\chi^2 =79.5$ for 60 bins.
For 60 bins there is a 3\% probability to obtain a larger value. The
Kolmogorov-Smirnov test gives $\kappa=0.25$ within the fit window and
$\kappa=0.84$ for the entire histogram.
Figure~\ref{fig:varmt} shows the sensitivity of the fitted mass to the choice of
fit interval.

In spite of the somewhat large value of $\chi^2$ there is no structure apparent
in Fig.~\ref{fig:chimt} that would indicate that there is a systematic
difference between the shapes of the observed spectrum and the probability
density function. The large $\chi^2$ can be attributed to a few
bins that are
scattered over the entire fit interval, indicating statistical fluctuations in
the data.  This is consistent with the good Kolmogorov-Smirnov probability which
is more sensitive to the shape of the distribution and insensitive to the
binning.

\section{ Consistency Checks}
\label{sec-checks}
\subsection {Neutrino \pt\ Spectrum}

\begin{figure}[htpb!]
\vspace{-0.3in}
\centerline{\psfig{figure=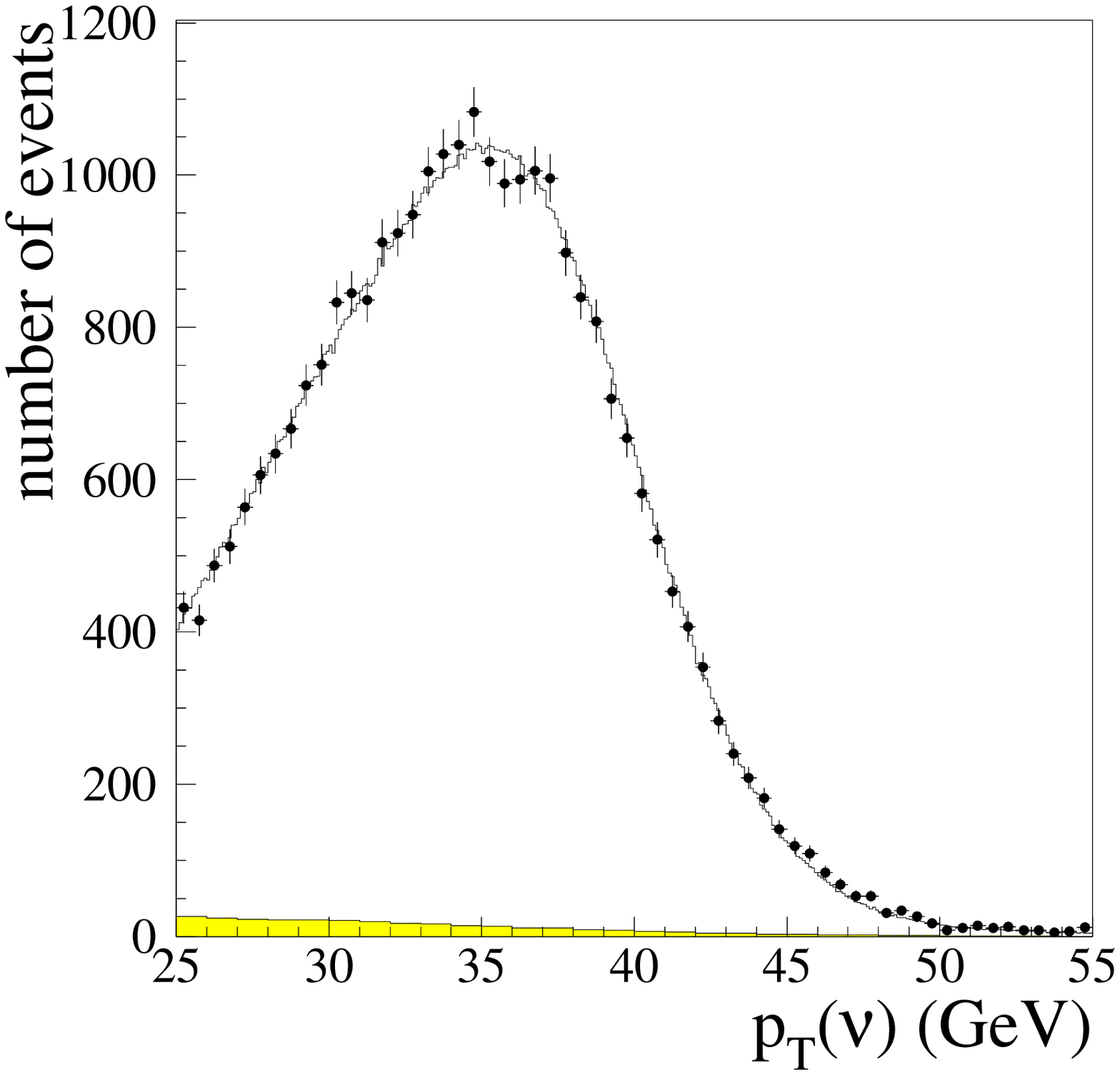,width=3.5in,height=3.5in}}
\vspace{-0.02in}
\caption{ Spectrum of \ptnu\ from the \wb\ data. The superimposed curve shows
the maximum likelihood fit and the shaded region shows the estimated
background.}
\label{fig:met}
\end{figure}

\begin{figure}[htpb!]
\vspace{-0.3in}
\centerline{\psfig{figure=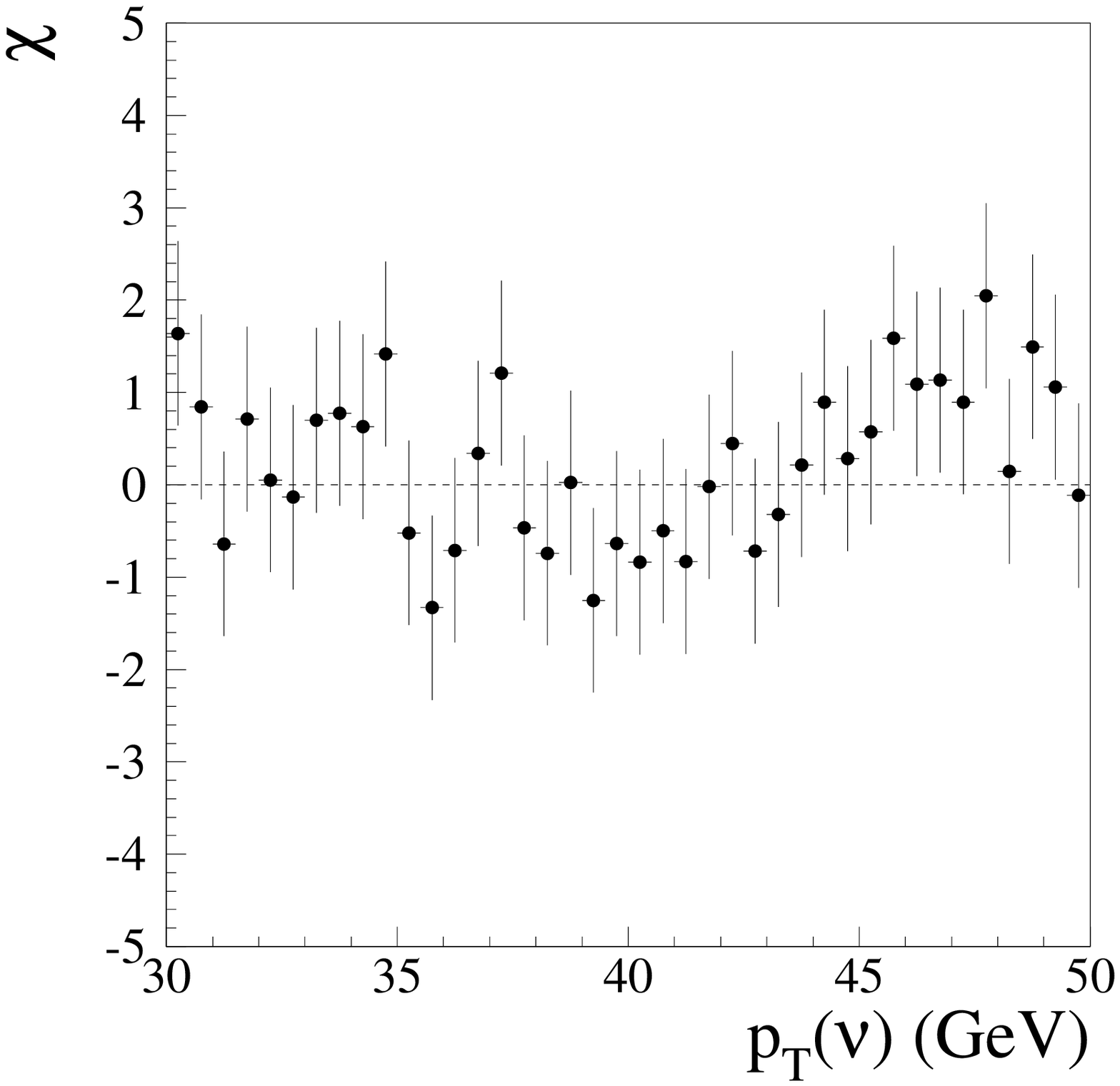,width=3.5in,height=3.5in}}
\vspace{-0.02in}
\caption{ The $\chi$ distribution for the fit to the \ptnu\ spectrum.}
\label{fig:chimet}
\end{figure}

As a consistency check, we also fit the \ptnu\ spectrum, although this
measurement is subject to much larger systematic uncertainties than the \mt\ and
\pte\ fits.  Figure~\ref{fig:met} shows the observed spectrum (points), signal
plus background for the best fit (solid line), and the estimated background
(shaded region). For the fit interval $30<\ptnu<50$ GeV the fitted mass is
$\mw = 80.37 \PM 0.11$ GeV, in good agreement with the \mt\ and \pte\ fits.
We compute $\chi^2=31.8$. The probability for a larger value is 75\%. The
Kolmogorov-Smirnov test gives $\kappa=0.20$ within the fit window and 
$\kappa=0.69$ for the entire histogram.
Figure~\ref{fig:chimet} shows the deviation $\chi$ between data
and fit. There is an indication of a systematic deviation between the observed
spectrum and the resolution function. This effect is not very significant.
For example, when we increase the hadronic resolution parameter \alphamb\ in
the simulation to 1.11, which corresponds to about 1.5 standard deviations,
this indication of a deviation between data and Monte Carlo vanishes.

\subsection { Luminosity Dependence}

We divide the \wb\ and \zb\ data samples into four luminosity bins

\begin{figure}[htpb!]
\vspace{-0.3in}
\centerline{\psfig{figure=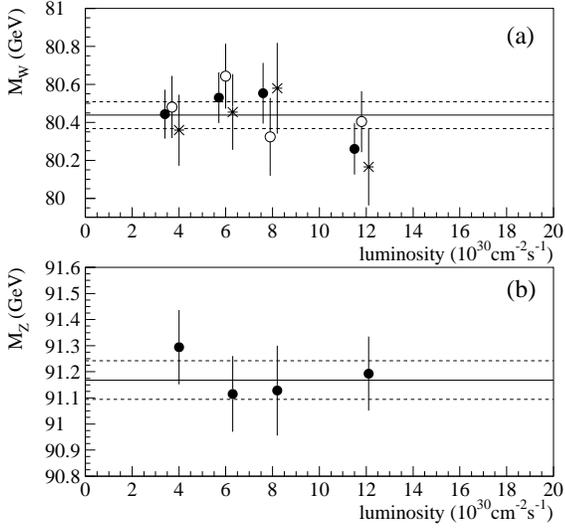,width=3.5in,height=3.5in}}
\vspace{-0.02in}
\caption{ The fitted \wb\ boson masses (a) in bins of luminosity from
the \mt\ ($\bullet$), \pte\ ($\circ$), and \ptnu\ ($\ast$) fits (the
points are offset for clarity) and the fitted \zb\ boson masses (b).  The solid
line is the central value for the \mt\ and \mee\ mass fits respectively over
the entire luminosity range and the dashed lines are the statistical errors.}
\label{fig:mwlum}
\end{figure}

\begin{center}
\begin{tabular} {rcl}
                 &${\cal L}$&$\le5\times10^{30}\hbox{cm}^{-2}\hbox{s}^{-1}$, \\
$5\times10^{30}<$&${\cal L}$&$\le7\times10^{30}\hbox{cm}^{-2}\hbox{s}^{-1}$, \\
$7\times10^{30}<$&${\cal L}$&$\le9\times10^{30}\hbox{cm}^{-2}\hbox{s}^{-1}$, \\
                 &${\cal L}$&$>9\times10^{30}\hbox{cm}^{-2}\hbox{s}^{-1}$ \\
\end{tabular}
\end{center}

\noindent
and generate resolution functions for
the luminosity distribution of these four subsamples.
We fit the transverse mass and lepton \pt\ spectra from the \wb\
samples and the dielectron invariant mass spectra from the \zb\ samples in each
bin. The fitted masses are plotted in Fig.~\ref{fig:mwlum}. The errors are
statistical only.
We compute the $\chi^2$ with respect to the \wb\ mass fit to the
\mt\ spectrum from the entire data sample.
The $\chi^2$ per degree of freedom (dof) for the \pte\ fit is 1.9/4 and for the
\ptnu\ fit is 2.4/4. The \mt\ fit has a $\chi^2$/dof of 2.7/3.
The solid and
dashed lines in the top plot indicate the \wb\ mass value and statistical
uncertainty from the fit to the $m_T$ spectrum of the entire data sample. All
measurements are in very good agreement with this value. In the bottom plot the
lines indicate the \zb\ mass fit to the \mee\ spectrum of the entire \zb\ data
sample. The measurements in the four luminosity bins have a $\chi^2$/dof of 
1.0/3.

\subsection { Dependence on $u_T$ Cut }

\begin{figure}[htpb!]
\vspace{-0.3in}
\centerline{\psfig{figure=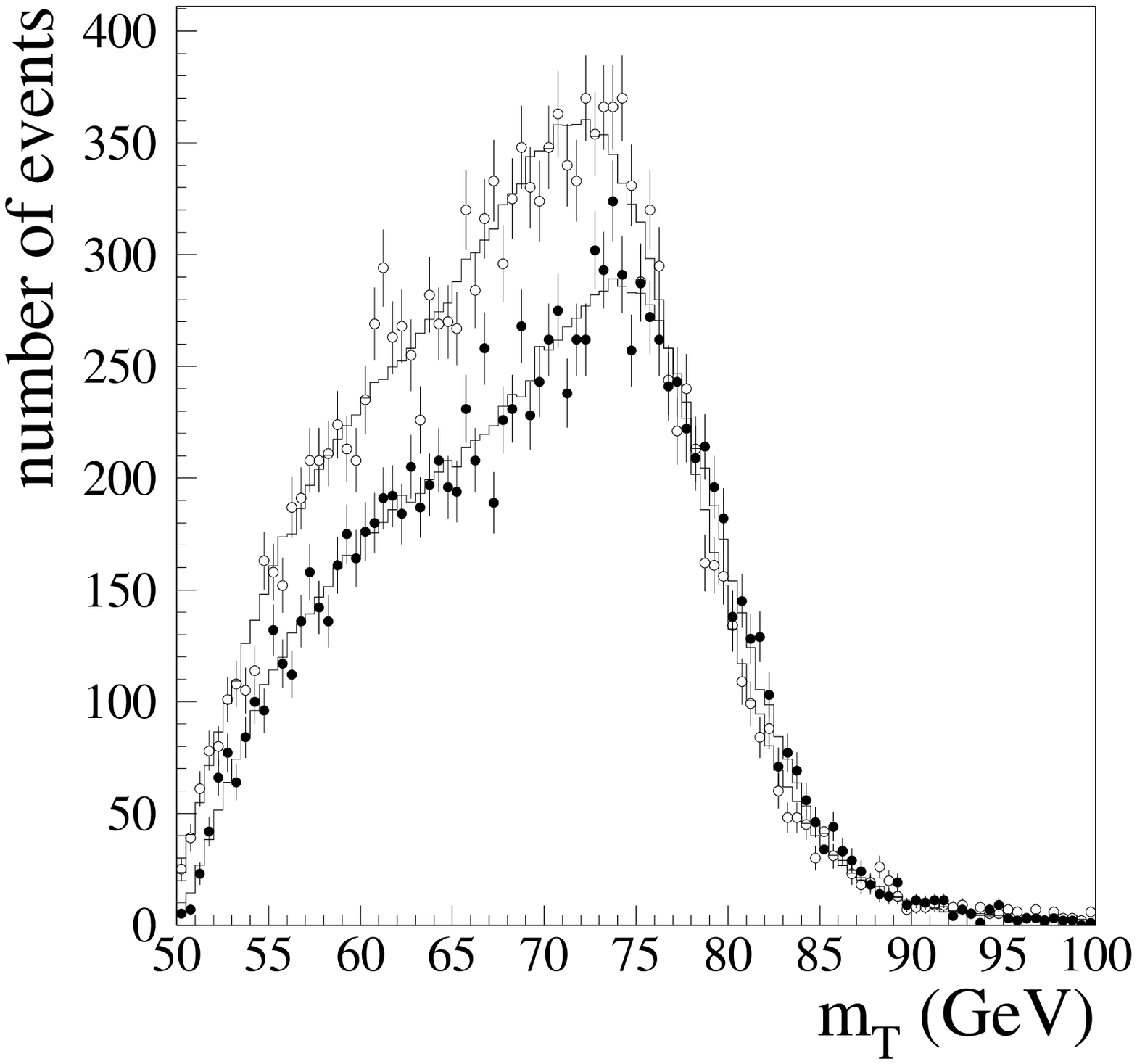,width=3.5in,height=3.5in}}
\vspace{-0.02in}
\caption{ Spectra of \mt\ from \wb\ data with $\upar<0$ ($\circ$) and $\upar>0$
($\bullet$) compared to Monte Carlo simulations (------).}
\label{fig:mt_upar_cut}
\end{figure}

\begin{figure}[htpb!]
\vspace{-0.3in}
\centerline{\psfig{figure=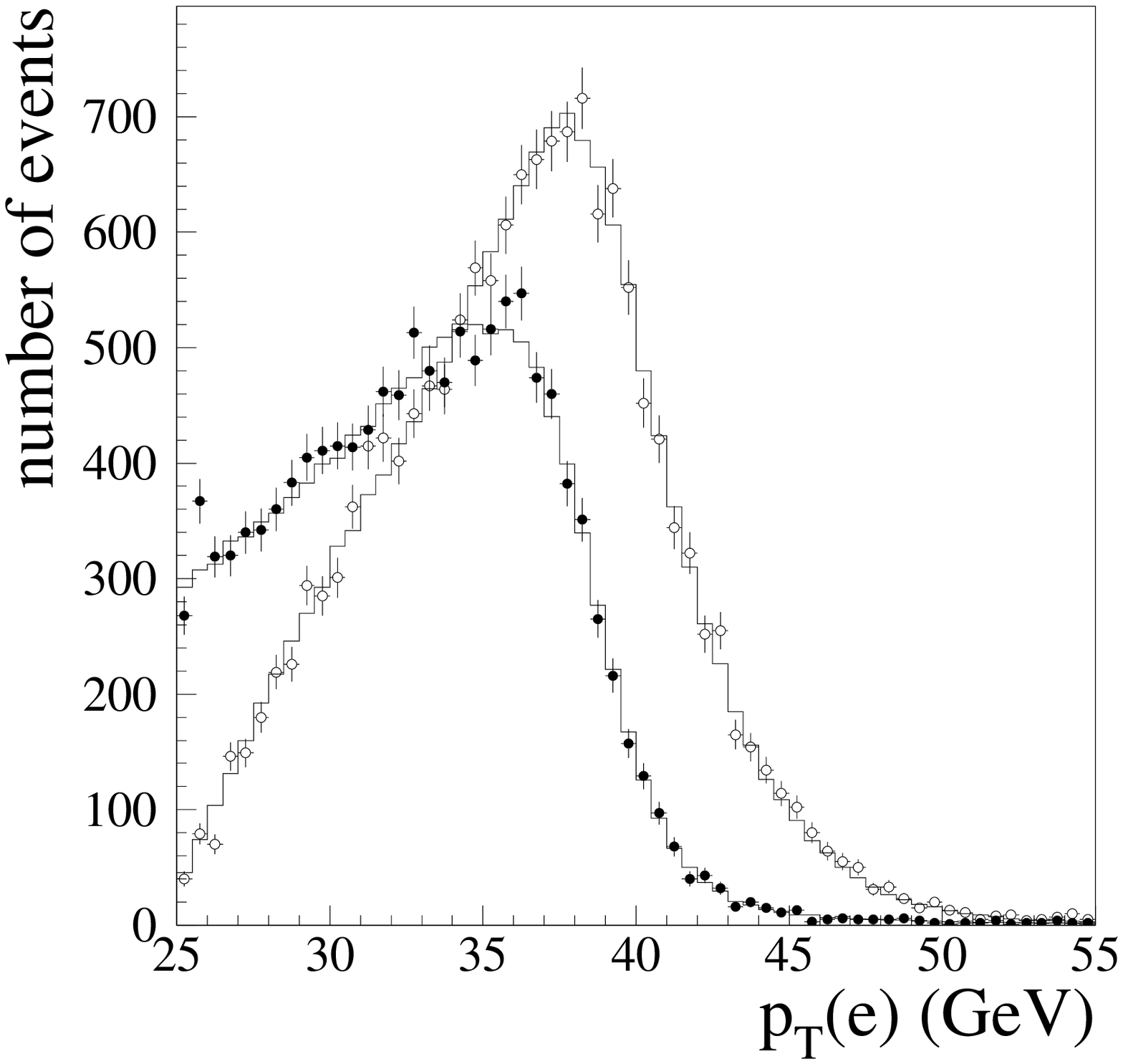,width=3.5in,height=3.5in}}
\vspace{-0.02in}
\caption{ Spectra of \pte\ from \wb\ data with $\upar<0$ ($\circ$)
and $\upar>0$ ($\bullet$) compared to Monte Carlo simulations (------).}
\label{fig:pte_upar_cut}
\end{figure}

\begin{figure}[htpb!]
\vspace{-0.3in}
\centerline{\psfig{figure=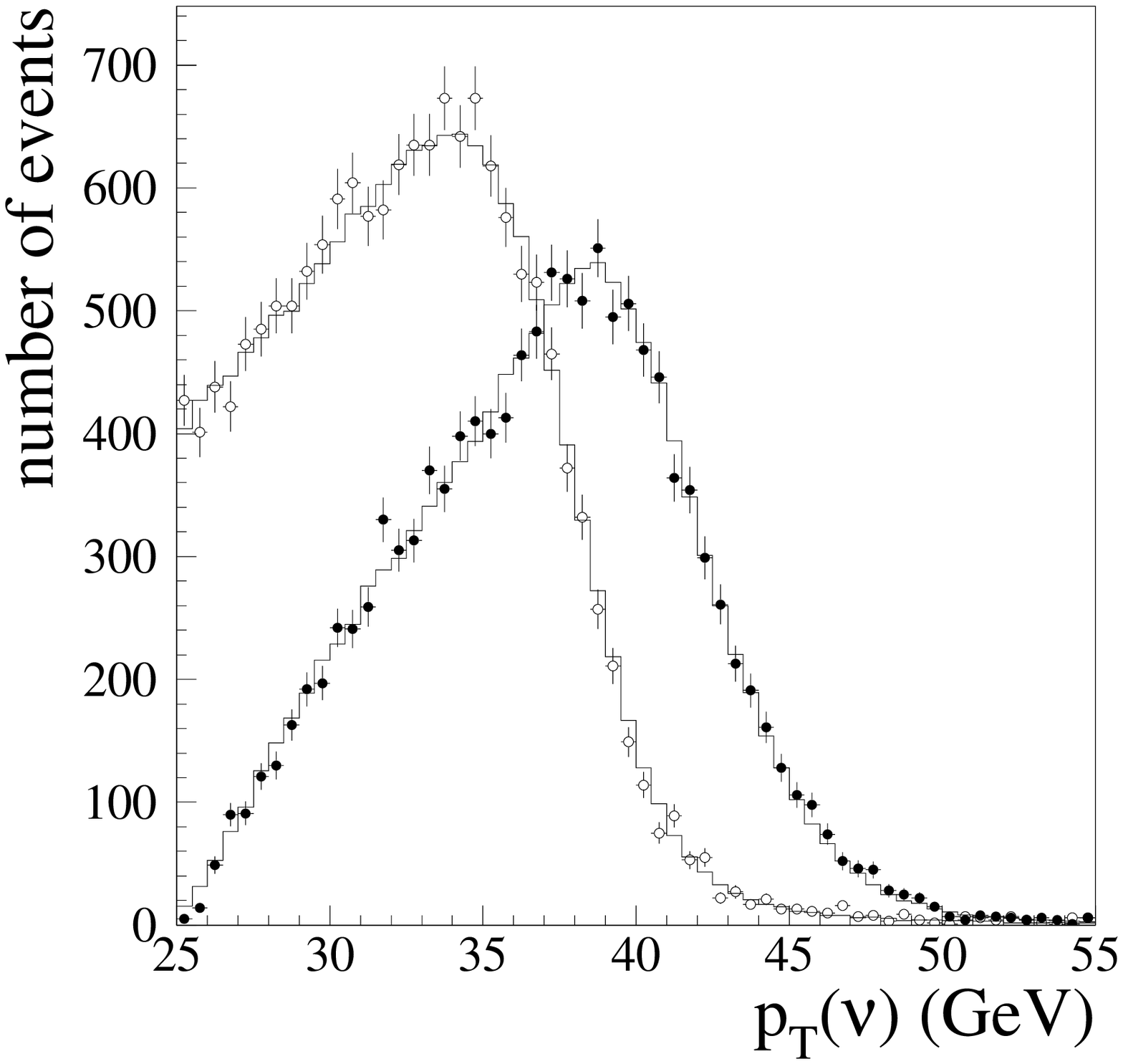,width=3.5in,height=3.5in}}
\vspace{-0.02in}
\caption{ Spectra of \ptnu\ from \wb\ data with $\upar<0$ ($\circ$)
and $\upar>0$ ($\bullet$) compared to Monte Carlo simulations (------).}
\label{fig:ptv_upar_cut}
\end{figure}

We change the cuts on the recoil momentum \ut\ and study how well the fast Monte
Carlo simulation reproduces the variations in the spectra. We split the \wb\
sample in two subsamples with $\upar\gt0$ and $\upar\lt0$.
In the simulation we
fix the \wb\ mass to the value from the \mt\ fit in Eq.~\ref{eq:mw}.
Figures~\ref{fig:mt_upar_cut}--\ref{fig:ptv_upar_cut} show the \mt, \pte, and
\ptnu\ spectra from the collider data for the subsamples with $\upar\gt0$ and
$\upar\lt0$ and the corresponding Monte Carlo predictions. 
Table~\ref{tab:kstest} lists the results of comparisons of collider data and
Monte Carlo spectra using the Kolmogorov-Smirnov test. Although there is
significant variation among the shapes of the spectra for the different cuts,
the fast Monte Carlo models them well.
Table~\ref{tab:kstest} also lists the results of comparisons of collider
data and Monte Carlo spectra for a \wb\ sample
selected with $\ut\lt30$ GeV which consists of 32{,}361 events.

\subsection {Dependence on Fiducial Cuts }

\begin{figure}[htpb!]
\centerline{\psfig{figure=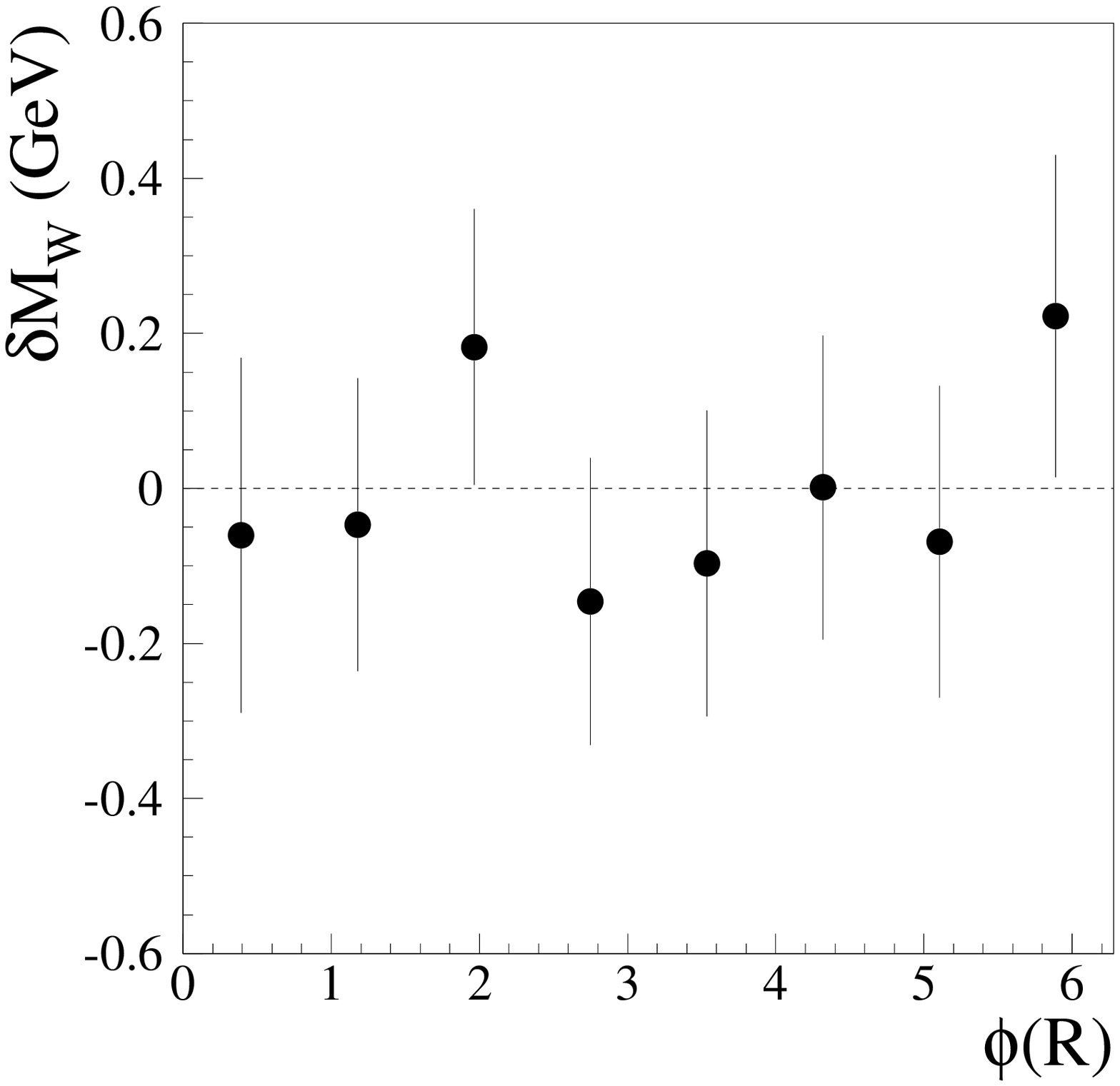,width=3.5in,height=3.5in}}
\vspace{-0.02in}
\caption{The variation in the \wb\ mass from the \mt\ fit as
a function of \phir.}
\label{fig:rec_phi}
\end{figure}

We divide the azimuth of the recoil momentum, $\phi(R)$, into eight bins.
This binning
is sensitive to azimuthal nonuniformities in the recoil momentum measurement,
\eg\ because of background from the Main Ring. Figure~\ref{fig:rec_phi}
shows the fitted \wb\ mass values versus \phir. The Main Ring is located at
$\phi\sim\pi/2$ and any biases caused by background from the Main Ring should
appear as structure in this direction or in the opposite direction. The rms of
the eight data points is 124 MeV, consistent with the statistical
uncertainty of 200 MeV for the data points. Thus the data are consistent with
azimuthal uniformity.

\begin{figure}[htpb!]
\centerline{\psfig{figure=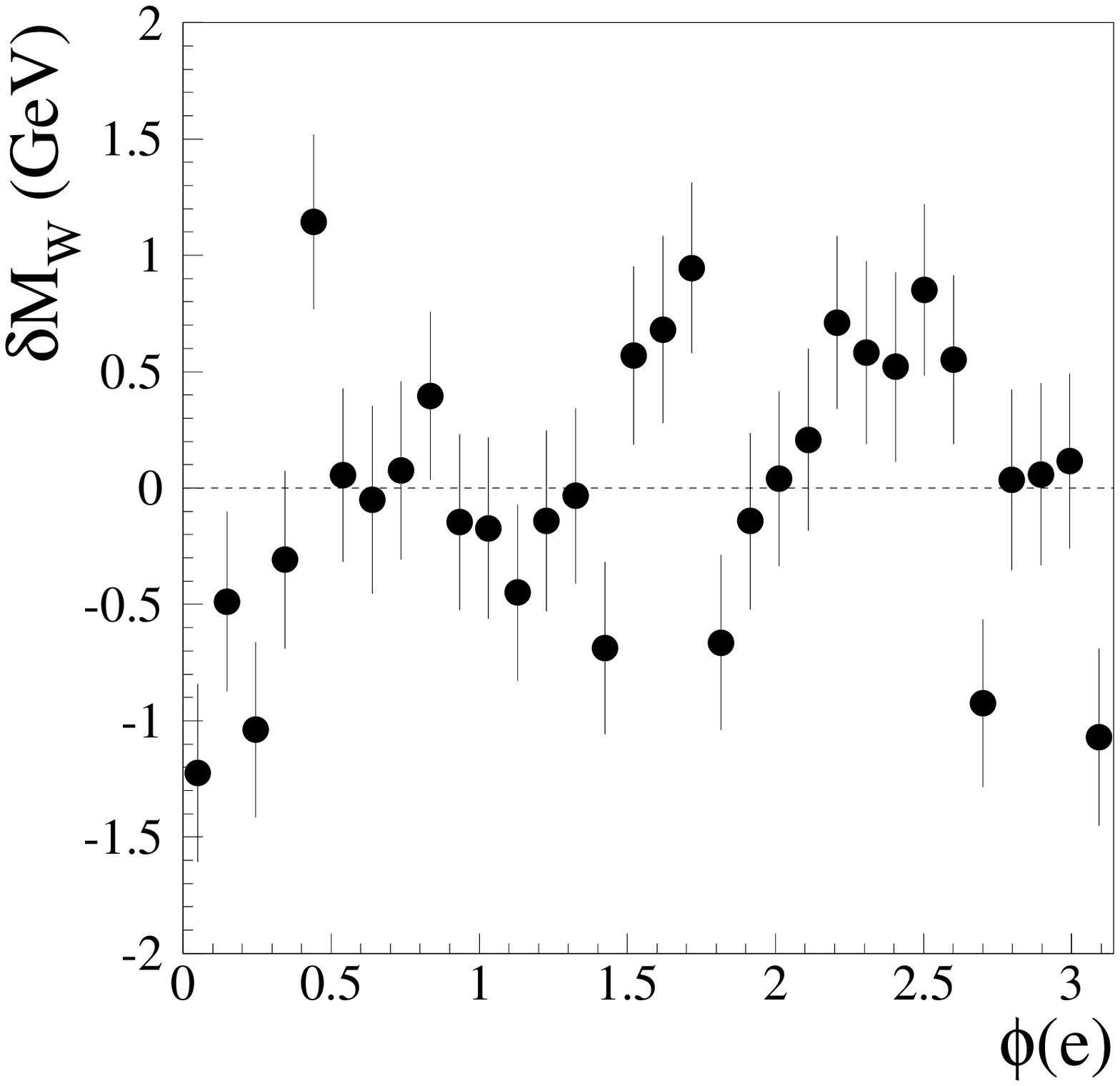,width=3.5in,height=3.5in}}
\vspace{-0.02in}
\caption{The variation in the \wb\ mass from the \mt\ fit as
a function of \phie.}
\label{fig:elc_phi}
\end{figure}

We divide the azimuthal direction of the electron, \phie,  into 32 bins
corresponding
to the 32 azimuthal modules of the CC-EM. Figure~\ref{fig:elc_phi} shows the
fitted \wb\ mass values versus \phie. The statistical uncertainty of the
data points is 400 \MeVm\ and the rms of the 32 points is 600 \MeVm. Thus
there is a 0.6\% nonuniformity in the response of the CC-EM, consistent with
the module-to-module calibration of 0.5\% \cite{Zhu_thesis}.

\begin{figure}[htpb!]
\centerline{\psfig{figure=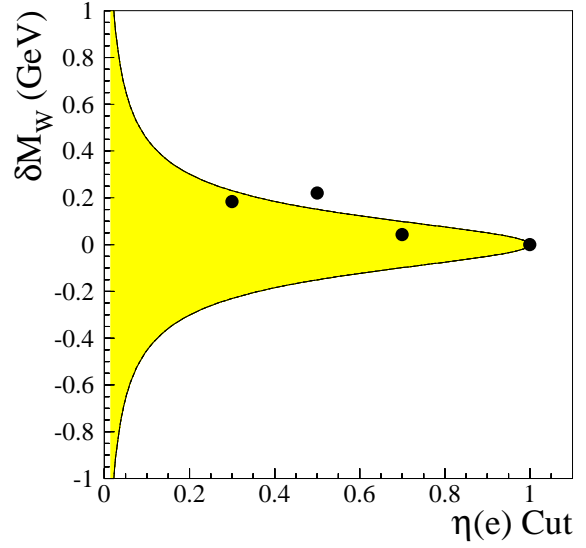,width=3.5in,height=3.5in}}
\vspace{-0.02in}
\caption{The variation in the \wb\ mass versus the $\etae$ cut.  The shaded
region is the expected statistical variation.  }
\label{fig:eta_cut}
\end{figure}

Finally, we fit the \mt\ spectrum from the \wb\ sample and the \mee\ spectrum
from the \zb\ sample for different pseudorapidity cuts on the electron
direction.
We use cuts of $|\etae|<$ 1.0, 0.7, 0.5, and 0.3.
Figure~\ref{fig:eta_cut}  shows the
change in the \wb\ mass versus the $\etae$ cut using the electron energy scale
calibration from the corresponding \zb\ sample.  The shaded region indicates the
statistical error. Within the uncertainties the mass is independent
of the $\etae$ cut.

\section{ Uncertainties in the Measurement }
\label{sec-syst}
\subsection { Statistical Uncertainties }

Table~\ref{tab:stat_errors} lists the uncertainties, rounded to the nearest 5
MeV, in the \wb\ measurement due to the finite sizes of the \wb\ and \zb\
samples used in the fits to the \mt, \pte, \ptnu, and \mee\ spectra. The
statistical uncertainty due to the finite \zb\ sample propagates into the \wb\
mass measurement through the electron energy scale \alphaem.

\subsection { \wb\ Production and Decay Model }
\label{sec-model_errors}

\subsubsection{Sources of Uncertainty}

Uncertainties in the \wb\ production and decay model arise from the following
sources: the phenomenological parameters in the calculation of the \ptw\
spectrum, the choice of parton distribution functions, radiative decays, and the
\wb\ boson width.
In the following we describe how we assess the size of the systematic
uncertainties introduced by each of these. We summarize the
size of the uncertainties in
Table~\ref{tab:theory_errors}, rounded to the nearest 5 MeV.

\subsubsection{\wb\ Boson \pt\ Spectrum}

In Sec.~\ref{sec-constraints} we determine $g_2$ so that the predicted \ptee\
spectrum agrees with the \zb\ data.
In order to quantify the uncertainty in the boson \pt\ spectra, we need to
consider variations in all four parameters, \lqcd, $g_1$, $g_2$, and
$g_3$. We use a series of modified CTEQ3M parton distribution functions fit with
\lqcd\ fixed at discrete values \cite{CTEQ3M_LQCD} to study the
variations in the \ptee\ spectrum and the fitted \wb\ boson mass with these
parameters.

We cannot constrain all these four parameters simultaneously by using only our
\zb\ data. We therefore introduce an external constraint on \lqcd. The CTEQ3M
fits prefer $\lqcd =158$ MeV but are also consistent with somewhat higher values
\cite{cteq3m}. Other measurements give a combined
value of $\lqcd =209^{+39}_{-33}$ MeV \cite{PDG}. All data are consistent
with \lqcd\ between 150 and 250 MeV, which we use as the range over which 
\lqcd\ is allowed to vary.  

\begin{figure}[htpb!]
\vspace{-0.3in}
\centerline{\psfig{figure=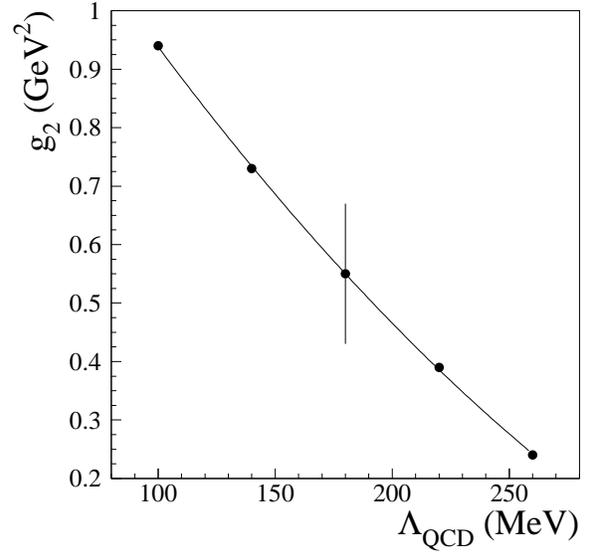,width=3.5in,height=3.5in}}
\vspace{-0.02in}
\caption{Value of $g_2$ as a function of $\lqcd$. The error bar
indicates the uncertainty in $g_2$ for fixed $\lqcd$.}
\label{fig:lqcd_g2_corr}
\end{figure}

The requirement that the fast Monte Carlo prediction for the
average \ptee\ over the range $\ptee<15$ GeV, corrected for background
contributions, must agree with the value observed in the \zb\ data, $\langle
\ptee\rangle=6.05\PM0.07$ GeV, couples the values of \lqcd\ and
$g_2$. Figure~\ref{fig:lqcd_g2_corr} shows a plot of $g_2$ versus \lqcd.
For any pair of values on the curve the fast Monte Carlo predicts a value of
$\langle\ptee\rangle$ that agrees with the \zb\ data. For any fixed value of
\lqcd, $g_2$ is determined to a precision of 0.12 GeV$^2$. This error includes
the statistical uncertainty (0.09 GeV$^2$) and the systematic uncertainty due to
normalization and shape of the background (0.07 GeV$^2$). All other
uncertainties, \eg\ due to electron momentum resolution and response or
selection biases, are negligible.

If we fix \lqcd\ and $g_2$, the requirement that the average \ptee\ predicted
by the fast Monte Carlo agree with the data allows an additional variation in
the parameters $g_1$ and $g_3$. The residual uncertainty in the measured \wb\
boson mass due to this variation, however, is small compared to the
uncertainty due to the variation allowed in $g_2$ and \lqcd\ and we neglect it.
Finally, we obtain the uncertainties in the fitted \wb\ boson
mass listed in Table~\ref{tab:theory_errors}.

\subsubsection{Parton Distribution Functions}

The choice of parton distribution function used to describe the momentum
distribution of the constituents of the proton and antiproton affects several
components of the model: the parton luminosity slope $\beta$, and the rapidity
and transverse momentum spectrum of the \wb.

Using several modern parton distribution function sets as input to the fast
Monte Carlo model, we generate \mt\ and lepton \pt\ spectra. In each case we
use the value of $g_2$ measured for that parton distribution function set
using our \zb\ data (Sec.~\ref{sec-constraints}). We then fit them in the
same way as the spectra from collider data, \ie\ using MRSA$'$ parton
distribution functions. Table~\ref{tab:pdf_error} lists the variation of the
fitted \wb\ mass values relative to MRSA$'$.

The MRSA$'$ and CTEQ3M parton distribution functions use the measured \wb\
charge asymmetry in \ppbar\ collisions \cite{CDFWasym} as input to the fit.
MRSD$-'$ and CTEQ2M do not explicitly use the asymmetry. The asymmetry
predicted by MRSD$-'$ agrees with the measurement; that of CTEQ2M disagrees at
the level of four standard deviations. We include CTEQ2M in our
estimate of the uncertainty to provide an estimate of the possible variations
with a rather large deviation from the measured asymmetry.

\subsubsection{Parton Luminosity}

The uncertainty of $10^{-3}\ \mbox{GeV}^{-1}$ in the parton
luminosity slope $\beta$ (Sec.~\ref{sec-mc}) translates into an uncertainty in
the fitted \wb\ mass. We
estimate the sensitivity in the fitted \wb\ mass by fitting Monte Carlo
spectra generated with different values of $\beta$.

\subsubsection{Radiative Decays}

We assign an error to the modeling of radiative decays based on varying
the detector parameters $E_0$ and $R_0$ (Sec.~\ref{sec-mc}).
$E_0$ defines the minimum photon energy generated and corresponds
to a cut-off below which the photon does not reach the calorimeter.
$R_0$ defines the maximum separation between the photon and
electron directions above which the photon energy is not included in the
electron shower. In general, radiation shifts the fitted mass down for the
transverse mass and electron fits, because for a fraction of the events the
photon energy is subtracted from the electron. Hence increasing
$R_0$ decreases the radiative shift. Similarly, decreasing
$E_0$ decreases the radiative shift. Both the fitted \wb\ and \zb\
masses depend on these parameters. Table~\ref{tab:radshifts} lists the change
in the fitted masses if radiative effects are turned off completely.
To estimate the systematic error, we fit Monte Carlo
spectra generated with different values for $E_0$ and
$R_0$. For the low value of $E_0=50$ MeV that we
use in the simulation, the dependence of the fits on this parameter is
negligible. The changes in the mass fits when varying $R_0$ by $\pm0.1$
are also listed in Table~\ref{tab:radshifts}. After propagating the change in
the \zb\ mass into the electron response the result of the \wb\ mass
measurement changes by about 15 MeV for all three spectra.

There are also theoretical uncertainties in the radiative decay calculation.
Initial state QED radiation is not included in the calculation of
Ref.~\cite{rad_decays_th}. However, initial state radiation does not affect the
kinematic distributions used to fit the mass in the final state.
Finally, the calculation includes only processes in which a single photon
is radiated. We use the code provided by the authors of
Ref.~\cite{baur_twophoton} to estimate the shift
introduced in the measured \wb\ mass
by neglecting two-photon emission. We find that two photons, with
$\pt>100$ MeV and separated by $\Delta R>0.3$ from the electron, are
radiated in about 0.24\% of all \wev\ decays. This reduces the mean value of
\mt\ within the fit window by 3 MeV. In 1.1\% of all \zee\ decays two photons,
with $\pt>100$ MeV and separated by $\Delta R>0.3$ from the electrons, are
radiated. We add the dielectron mass spectrum of these $\zee\gamma\gamma$ events
to our simulated \zb\ boson lineshape and fit the modified lineshape. The fitted
mass decreases by 10 MeV. This shift requires an adjustment
of the energy scale calibration factor \alphaem\ by $10^{-4}$. Neglecting
two-photon emission in both \wb\ and \zb\ boson decays then increases the
measured \wb\ mass by about 5 MeV.
Since this effect is an order of
magnitude smaller than the statistical uncertainty in our measurement we do not
correct for it, but add it in quadrature to the uncertainty due to radiative
corrections.

\subsubsection{\wb\ Boson Width}

To determine the sensitivity of the fitted \wb\ mass to the \wb\
width, we generate \mt\ and lepton \pt\ spectra using the fast Monte Carlo
model with a range of widths and fit them with the nominal templates.
The uncertainty on the fitted \wb\ mass correspond to the uncertainty in the
measured value of \wwidth\ = 2.062\PM0.059~GeV~\cite{Wwidth}.

\subsection { Detector Model Parameters }

The uncertainties on the parameters of the detector model determined in
Secs.~\ref{sec-elec}--\ref{sec-recoil} translate into uncertainties in the
\wb\ mass measurement.
We study the sensitivity of the \wb\ mass measurement to the values of the
parameters by fitting the data with spectra generated by the fast Monte Carlo
with modified input parameters.

Table~\ref{tab:detector_errors} lists the uncertainties in the measured \wb\
mass, caused by the individual parameters. We assign sets of correlated
parameters to the same item in the table. Correlations between items are
negligible. For each item the uncertainty is determined to typically 5~MeV for
the \mt\ fit and 10~MeV for the lepton \pt\ fits. We therefore round them
to the nearest 5 MeV in the table. To achieve this precision
10--20 million \wev\ decays are simulated for each item.

The residual calorimeter nonlinearity is parametrized by the offset
\deltaem. Calorimeter uniformity refers to a possible nonuniformity in
response as a function of $\eta$. It is limited by the test beam data \cite{D0}.
The electron momentum resolution is parametrized by \cem. The electron angle
calibration includes the effects of the parameters \alphacdc\ and
\alphacc, discussed in
Appendices~\ref{app:pos_cdc} and \ref{app:pos_cc}. The recoil resolution is
parametrized by \alphamb\ and \srec\ and the response by \alpharec\
and \betarec. Electron removal refers to the bias \dupar\ introduced in the
\upar\ measurement by the removal of the cells occupied by the electron.
Selection bias refers to the \upar\ efficiency.

\subsection { Backgrounds }

We determine the sensitivity of the fit results to the assumed background
normalizations and shapes by repeating the fits to the data with varied
background shapes and normalizations. Table~\ref{tab:bkg_errors} lists the
uncertainties rounded to the nearest 5 MeV.

\section{ Results }
\label{sec-results}

We present a precision measurement of the mass of the \wb\ boson.
From a fit to the transverse mass spectrum, we measure
\begin {equation}
\mw = 80.44 \pm 0.10(stat) \pm 0.07(syst)\ \hbox{GeV}.
\end {equation}
Adding all errors in quadrature gives 115 MeV. Since we calibrate the electron
energy scale against the known \zb\ mass, we effectively measure the \wb\ and
\zb\ mass ratio
\begin{equation}
{\mw\over\mz} = 0.8821\pm0.0011(stat)\pm0.0008(syst).
\end {equation}
A fit to the transverse momentum spectrum of the decay electrons gives
\begin {equation}
\mw = 80.48 \pm 0.11(stat) \pm 0.09(syst)\ \hbox{GeV}.
\end {equation}
Adding all errors in quadrature gives 140 MeV. As expected, the measurement from
the \mt\ spectrum has a larger uncertainty from detector effects (65 MeV) than
that from the \pte\ spectrum (50 MeV). On the other hand the \mt\ fit is less
sensitive to the \wb\ production model (30 MeV) than the \pte\ fit (75 MeV).
The good agreement between the two results indicates that we understand the
ingredients of our model and their uncertainties. In the end, the \mt\ fit gives
the more precise result and we quote this as our final result. However the fit
to the \pte\ spectrum may become more competitive in the future with larger data
samples and better constraints on the \wb\ production dynamics.

Table~\ref{tab:sum} lists the \Dzero\ \wb\ mass measurements from fits to the
\mt\ spectra from the 1992--1993~\cite{D0} and the 1994--1995 data sets
and their uncertainties. As indicated in Table~\ref{tab:sum},
some errors are common to the two measurements. Since both analyses use the same
\wb\ production and decay model we assign the uncertainties quoted in
Sec.~\ref{sec-model_errors} to both measurements. The precision of the electron
angle calibration has improved compared to Ref.~\cite{D0} and we use the 
reduced uncertainty for both measurements.
All uncertainties due to detector model parameters, which
were measured using statistically independent data sets, are uncorrelated
because their precision is dominated by statistical fluctuations. In order to
combine the two measurements we weight them by their uncorrelated errors
$\delta_{a}$ and $\delta_{b}$
\begin {equation}
\mw = {M_{a}/\delta_{a}^2+M_{b}/\delta_{b}^2 \over
1/\delta_{a}^2+1/\delta_{b}^2}.
\end {equation}
The uncertainty is then given by
\begin {equation}
\delta \mw = \sqrt{{1\over 1/\delta_{a}^2+1/\delta_{b}^2} +
\delta^2},
\end {equation}
where $\delta$ is the common uncertainty from the third column in
Table~\ref{tab:sum}.
The combination of the \Dzero\ measurements from the 1992--1993
and 1994--1995 data gives
\begin {equation}
\mw = 80.43\pm0.11\ \hbox{GeV}.
\label {eq:combined}
\end {equation}

\begin{figure}[htpb!]
\centerline{\psfig{figure=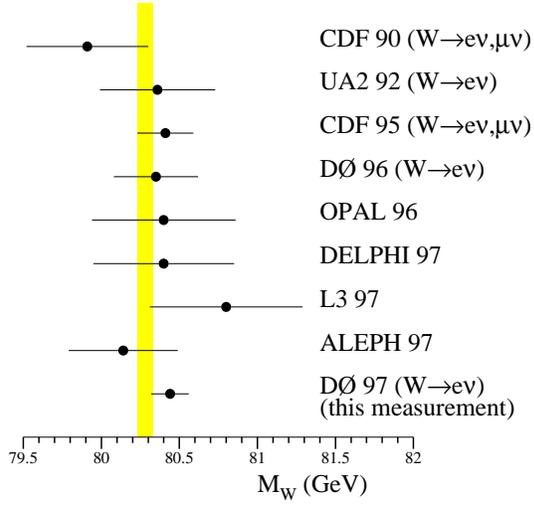,width=3.5in,height=3.5in}}
\vspace{-0.02in}
\caption{A comparison of this measurement with previously published \wb\ mass
measurements (Table~\ref{tab:mw}). The shaded region indicates the predicted
\wb\ mass value from global fits to the \zb\ lineshape data \protect\cite{mz}.}
\label{fig:mw_world}
\end{figure}

The \Dzero\ measurement is in good agreement with previous measurements and is
more precise than all the previously published measurements combined.
Table~\ref{tab:mw} lists previously published measurements with uncertainties
below 500 MeV. A global fit to all electroweak measurements from the LEP
experiments predicts $\mw = 80.278\pm0.049$ GeV \cite{mz}.
Figure~\ref{fig:mw_world} gives a graphical representation of these data.

\begin{figure}[htpb!]
\centerline{\psfig{figure=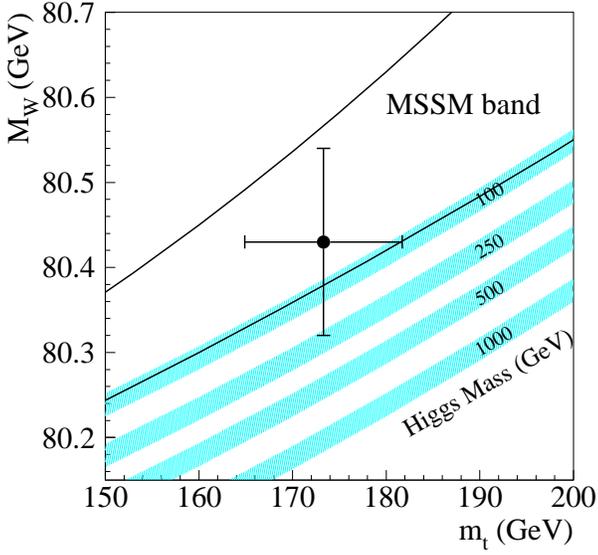,width=3.5in,height=3.5in}}
\vspace{-0.1in}
\caption{A comparison of the \wb\ and top quark mass measurements by the
\Dzero\ collaboration with the Standard Model predictions for different Higgs
boson masses~\protect\cite{mw_v_mt}. The width of the bands for each Higgs boson
mass value indicates the uncertainty due to the error in $\alpha(\mz^2)$. Also
shown is the range allowed by the MSSM~\protect\cite{susy}.}
\label{fig:mw_mt}
\end{figure}

We evaluate the radiative corrections $\Delta r$, defined in Eq.~\ref{eq:mw1}.
Our measurement of \mw\ from Eq.~\ref{eq:combined} leads to
\begin{equation}
\Delta r = -0.0288\pm0.0070,
\end{equation}
4.1 standard deviations from the tree level value. In Fig.~\ref{fig:mw_mt} we
compare the measured \wb\ and top quark masses~\cite{mtop_lj} to
the values predicted by the Standard Model for a range of Higgs mass values
\cite{mw_v_mt}. Also shown is the prediction from the calculation in
Ref.~\cite{susy} for a model involving supersymmetric particles
assuming the chargino, Higgs, and left-handed selectron masses are
greater than 90~GeV.
The measured values are in agreement with the prediction
of the Standard Model.

\section*{ Acknowledgements }
We wish to thank U. Baur for helpful discussions.
We thank the staffs at Fermilab and collaborating institutions for their
contributions to this work, and acknowledge support from the 
Department of Energy and National Science Foundation (U.S.A.),  
Commissariat  \` a L'Energie Atomique (France), 
State Committee for Science and Technology and Ministry for Atomic 
   Energy (Russia),
CNPq (Brazil),
Departments of Atomic Energy and Science and Education (India),
Colciencias (Colombia),
CONACyT (Mexico),
Ministry of Education and KOSEF (Korea),
CONICET and UBACyT (Argentina),
and CAPES (Brazil).

\appendix
\section{ Track Position Calibration }
\label{app:pos_cdc}

\begin{figure}[htpb!]
\vskip 0.5 cm
\begin{tabular}{c}
\epsfxsize = 6.0cm \epsffile[20 20 400 400]{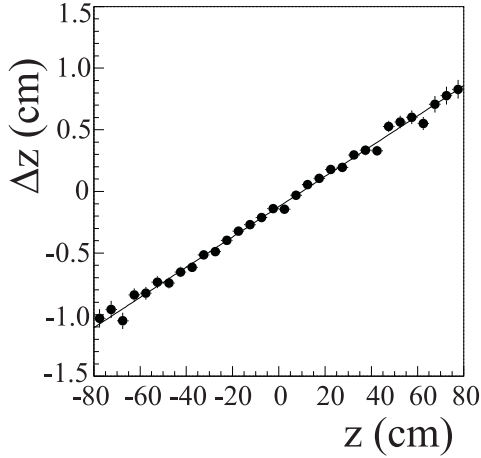}
\end{tabular}
\vspace{-0.02in}
\caption{The difference between the predicted
and the actual $z$-positions of the track center of gravity.}
\label{fig:cdc_sc_cos}
\end{figure}

We use cosmic ray muons which traverse the entire detector and pass close to
the beam position to calibrate the $z$-measurement of the track in the CDC.
We predict the trajectory
of the muon through the central detector by connecting the incoming and
outgoing hits in the innermost muon chambers by a straight line.
The center of gravity of the
incoming and outgoing CDC tracks are then calibrated relative to this
line. Figure~\ref{fig:cdc_sc_cos} shows the difference between the predicted
and the actual $z$-positions of the track centers of gravity.
These data are fit to a straight line.
We find the track position must be scaled
by the fitted slope, $\alphacdc= 0.9868\pm0.0004$.

\begin{figure}[htpb!]
\begin{tabular}{c}
\centerline{\psfig{figure=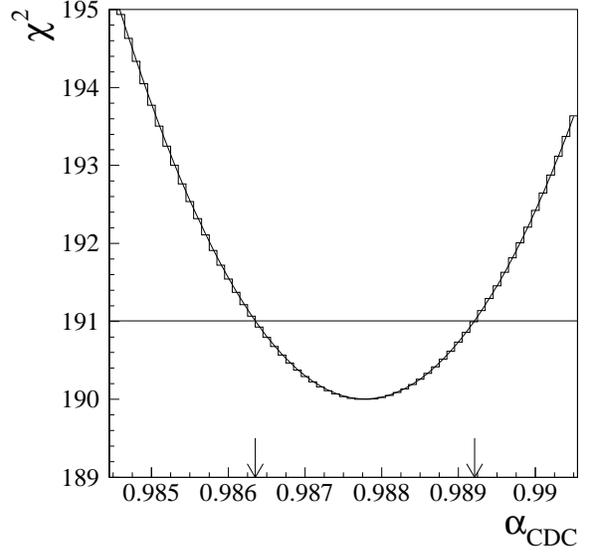,width=3.5in,height=3.5in}}
\end{tabular}
\vspace{-0.02in}
\caption{The $\chi^2$ versus \alphacdc\ value.  The arrows indicate
the statistical error on the fit.}
\label{fig:dimuon}
\end{figure}

We also use a sample of low-\pt\ dimuon events
 from \ppbar\ collisions where
both muons originate from the same interaction vertex. We
reconstruct
the muon trajectories from their hits in the innermost
muon chambers and the CDC. For
both muons we determine the point of closest approach of the trajectory to the
beam, $\zvtx (\mu)$.
We then scale the $z$-position of the CDC track to minimize
\begin{eqnarray}
\chi^2 = \sum_{\mbox{events}} \left({\zvtx (\mu_1) - \zvtx (\mu_2) \over
\sigma_\mu} \right)^2,
\label{eq:chisq}
\end{eqnarray}
where $\sigma_\mu$ is chosen so that the minimum value of $\chi^2$ equals
the number of events minus one.
The minimum occurs at $\alphacdc=0.9863\pm0.0011$.
The same analysis applied to a $Z\to\mu\mu$ sample gives
$\alphacdc=0.9878\pm0.0014$ and is shown in
Fig.~\ref{fig:dimuon}.

A scintillating fiber detector was inserted between the CDC and the CC to
calibrate the track
$z$-position.  The detector is built from 20 modules, each constructed on an
aluminum support plate 93.4 cm long and 16.5 cm wide.
Scintillating fibers, 12.7 cm long, were laid across the width of the module
every 11.43 cm along the support plate.  The eight scintillating fibers on each
module were connected to a clear waveguide and read out with a photomultiplier
tube.
The modules are mounted lengthwise along the cylinder of the CDC with half of
the modules covering $+z$ and the other half $-z$.  In the $r$-$\phi$ view
each module subtends $\pi/16$ radians with the fibers running azimuthally.
Because of spacial constraints not the entire CDC was covered.

When a fiber is hit by a charged particle the
$z$-position of the associated track, at the fiber radius, is compared with
the fiber $z$-position.
The $z$-position of the track at the radial position of the fiber
is determined from the direction and center of gravity of the track.
By comparing the $z$-position of the track and the hit fiber,
we determine that a scale of $\alphacdc=0.989\pm0.001$ is needed to correct the
track.

Combining all measurements of $\alphacdc$ gives
$\alphacdc=0.988\pm0.001$, which we use in the reconstruction of the electrons
in the \wb\ and \zb\ data samples.

\section{ Electron Shower Position Algorithm}
\label{app:pos_cc}

We determine the position of the electron shower centroid 
$\vec x_{\rm cal}=(\xcal,\ycal,\zcal)$ in the calorimeter
from the energy depositions in the third EM layer by computing the weighted mean
of the positions $\vec x_i$ of the cell centers,
\begin{equation}
\vec x_{\rm cal} = {\sum_i w_i \vec x_i\over \sum_i w_i}.
\end{equation}
The weights are given by
\begin{equation}
w_i = \max\left(0,w_0 + \log\left({E_i\over\Ee}\right)\right),
\end{equation}
where $E_i$ is the energy in cell $i$, $w_0$ is a parameter which 
depends upon \etae, and E(e) is the energy of the electron.
We calibrate the algorithm using Monte Carlo electrons simulated using \GEAN\
and electrons from the \zee\ data.
We apply a polynomial correction as a function of \zcal\ and \te\ based on
the Monte Carlo electrons.
We refine the calibration with the \zee\ data by exploiting the fact that 
both electrons originate from the same vertex. Using the algorithm 
given by Eq.~\ref{eq:chisq} we determine a
vertex for each electron from the shower centroid and the track center of
gravity. We minimize the difference between
the two vertex positions as a function of a scale factor \alphacc. 
More complex correction functions do not improve the $\chi^2$.  
The correction factor is $\alphacc = 0.9980 \pm 0.0005$, where the error
includes possible variations of the functional form of the correction.


\clearpage

\clearpage

\clearpage

\clearpage

\clearpage

\begin{table}[ht]
\begin{center}
\caption{\small Electron selection criteria.}
\begin{tabular}{lccc}
variable      & CC (loose)          & CC (tight)          & EC (tight) \\ \hline
fiducial cuts & $|\Delta\phi|>0.02$ & $|\Delta\phi|>0.02$ & --- \\
              & $|\zcal|<108$ cm  & $|\zcal|<108$ cm & $15\leq|i_\eta|\leq25$ \\
              & ---                 & $|\ztrk|<80$ cm     & --- \\
shower shape  & $\chi^2<100$        & $\chi^2<100$        & $\chi^2<100$ \\
isolation     & $\fiso <0.15$       & $\fiso <0.15$       & $\fiso <0.15$ \\
track match   & ---                 & $\sigm <5$          & $\sigm <10$ \\
\end{tabular}
\label{tab:e_selection}
\end{center}
\end{table}

\begin{table}[ht]
\begin{center}
\caption{\small Number of \wb\ and \zb\ candidate events.}
\begin{tabular}{lccccc}
channel & \multicolumn{3}{c}{\zee} & \wev \\
fiducial region of electrons   & CC/CC & CC/CC & CC/EC &    CC \\
$e$ quality (t=tight, l=loose) &   t/l &   t/t &   t/t &     t \\ \hline
pass Main Ring Veto            &   537 &  1225 &  1116 & 28323 \\
fail Main Ring Veto            &   107 &   310 &   268 &   --- \\
\end{tabular}
\label{tab:sample}
\end{center}
\end{table}

\begin{table}[ht]
\begin{center}
\caption{\small Parton luminosity slope $\beta$ and fraction of sea-sea
interactions \fss\ in the \wb\ and \zb\ production model. The $\beta$ value is
given for \wev\ decays with the electron in CC and for \zee\ decays with both
electrons in CC.}
\begin{tabular}{lccc}
                     & \zb\ production & \multicolumn{2}{c} {\wb\ production} \\
                     & $\beta$ (GeV$^{-1}$) & $\beta$ (GeV$^{-1}$) & \fss \\
\hline
MRSA$'$ \protect\cite{mrsa}  & $3.6\times10^{-3}$ & $8.6\times10^{-3}$ & 0.207
\\ 
CTEQ3M \protect\cite{cteq3m} & $3.3\times10^{-3}$ & $8.7\times10^{-3}$ & 0.203
\\ 
CTEQ2M \protect\cite{cteq2m} & ---                & $8.8\times10^{-3}$ & 0.203
\\ 
MRSD$-'$ \protect\cite{mrsd} & $3.8\times10^{-3}$ & $9.6\times10^{-3}$ & 0.201
\\ 
\end{tabular}
\label{tab:wprod}
\end{center}
\end{table}

\begin{table}[ht]
\begin{center}
\caption{\small \dupar\ for the \wb\ and \zb\ event samples.}
\begin{tabular}{ccc}
Event Sample & $\sum E_{1\times5}/\cosh\etae$ (MeV) & \dupar\ (MeV) \\ \hline
\wb & 95.8\PM 0.4   & 479\PM2\PM6 \\
\zb & 93.6\PM 1.3   & 468\PM7\PM6 \\
\end{tabular}
\label{tab:ue}
\end{center}
\end{table}

\begin{table}[ht]
\begin{center}
\caption{\small Fitted values of \gtwo\ for different parton distribution
functions. Uncertainties are statistical only.}
\begin{tabular}{lccc}
            & $\ptee<15$ GeV & \dphiee \\ \hline
MRSA$'$     & 0.59\PM 0.10 GeV$^2$ & 0.64\PM 0.14 GeV$^2$ \\
MRSD$-'$    & 0.61\PM 0.10 GeV$^2$ & 0.70\PM 0.15 GeV$^2$ \\
CTEQ3M      & 0.54\PM 0.10 GeV$^2$ & 0.57\PM 0.13 GeV$^2$ \\
CTEQ2M      & 0.61\PM 0.10 GeV$^2$ & 0.67\PM 0.14 GeV$^2$ \\
\end{tabular}
\label{tab:g2fit_values}
\end{center}
\end{table}

\begin{table}[ht]
\begin{center}
\caption{\small The results of the Monte Carlo ensemble tests fitting the 
\mw\ mass for 105 samples of 28{,}323 events.}
\begin{tabular}{crrccc}
             & mean       & rms   & \multicolumn{3}{c}{correlation matrix} \\
             &            &       & \mt   & \pte    & \ptnu    \\
              & (\GeVm{})  & (\GeVm{}) &   &         &       \\ \hline
\mt           & 80.404     & 0.067     & 1   & 0.669   & 0.630    \\
\pte          & 80.415     & 0.091     & 0.669 & 1     & 0.180    \\
\ptnu         & 80.389     & 0.105     & 0.630 & 0.180   & 1      \\
\end{tabular}
\label{tab:MCens}
\end{center}
\end{table}

\begin{table}[ht]
\begin{center}
\caption{ \small The confidence levels from Kolmogorov- Smirnov tests
comparing collider data to Monte Carlo predictions for \mw=80.44 GeV.}
\begin{tabular}{lccc}
             & \mt\        & \pte\       & \ptnu\      \\ \hline        
interval     & 60--90 GeV  & 30--50 GeV  & 30--50 GeV  \\ \hline
$\ut<15$ GeV & 0.25        & 0.81        & 0.20        \\
$\upar<0$    & 0.19        & 0.78        & 0.25        \\
$\upar>0$    & 0.61        & 0.80        & 0.48        \\
$\ut<30$ GeV & 0.55        & 0.99        & 0.58        \\ \hline
interval     & 50--100 GeV & 25--55 GeV  & 25--55 GeV  \\ \hline
$\ut<15$ GeV & 0.84        & 0.83        & 0.69        \\
$\upar<0$    & 0.77        & 0.67        & 0.62        \\
$\upar>0$    & 0.60        & 0.66        & 0.73        \\
$\ut<30$ GeV & 0.92        & 0.80        & 0.28        \\
\end{tabular}
\label{tab:kstest}
\end{center}
\end{table}

\begin{table}[ht]
\begin{center}
\caption{\small Uncertainties in 
the \wb\ mass measurement due to finite sample sizes.}
\begin{tabular}{lrrr}
            & \mt\ fit & \pte\ fit & \ptnu\ fit \\ \hline
\wb\ sample & 70 MeV   & 85 MeV    & 105 MeV    \\
\zb\ sample & 65 MeV   & 65 MeV    & 65 MeV     \\ \hline
total       & 95 MeV   & 105 MeV   & 125 MeV    \\
\end{tabular}
\label{tab:stat_errors}
\end{center}
\end{table}

\begin{table}[ht]
\begin{center}
\caption{\small Uncertainties in the \wb\ mass
measurement due to \wb\ production and decay model.}
\begin{tabular}{lrrr}
                          & \mt\ fit & \pte\ fit & \ptnu\ fit \\ \hline
$\ptw$ spectrum           & 10 MeV   & 50 MeV    & 25 MeV     \\
parton distribution functions & 20 MeV   & 50 MeV    & 30 MeV     \\
parton luminosity $\beta$ & 10 MeV   & 10 MeV    & 10 MeV     \\
radiative decays          & 15 MeV   & 15 MeV    & 15 MeV     \\ 
\wb\ width                & 10 MeV   & 10 MeV    & 10 MeV     \\ \hline
total                     & 30 MeV   & 75 MeV    & 45 MeV     \\
\end{tabular}
\label{tab:theory_errors}
\end{center}
\end{table}

\begin{table}
\begin{center}
\caption{\small\label{tab:pdf_error}
Variation of fitted \wb\ mass with choice of parton distribution function. }
\begin{tabular}{lccc}
         & \mt\ fit  & \pte\ fit & \ptnu\ fit \\ \hline
MRSA$'$  & 0         & 0         &  0         \\
MRSD$-'$ & ~$20$ MeV & ~$19$ MeV & ~$20$ MeV  \\
CTEQ3M   & ~~$5$ MeV & ~$48$ MeV & ~$22$ MeV  \\
CTEQ2M   & $-21$ MeV & $-17$ MeV & $-30$ MeV  \\
\end{tabular}
\end{center}
\end{table}

\begin{table}
\begin{center}
\caption{\small Changes in fitted \wb\ and \zb\ masses if radiative effects are
varied. }
\begin{tabular}{lcccc} 
variation              &  \mt\ fit  & \pte\ fit  & \ptnu\ fit & \mee\ fit \\ 
\hline
no radiative effects    & 50 MeV     & 43 MeV     & 30 MeV     & 143 MeV   \\ 
vary $R_0$ by$\pm 0.1$  & ~3 MeV     & ~4 MeV     & ~0 MeV     & ~19 MeV   \\
\end{tabular}
\label{tab:radshifts}
\end{center}
\end{table}

\begin{table}[ht]
\begin{center}
\caption{\small Uncertainties in the \wb\ mass 
measurement due to detector model parameters.}
\begin{tabular}{lrrr}
                            & \mt\ fit & \pte\ fit & \ptnu\ fit \\ \hline
calorimeter linearity       & 20 MeV   & 20 MeV    &  20 MeV    \\
calorimeter uniformity      & 10 MeV   & 10 MeV    &  10 MeV    \\
electron resolution         & 25 MeV   & 15 MeV    &  30 MeV    \\
electron angle calibration  & 30 MeV   & 30 MeV    &  30 MeV    \\
electron removal            & 15 MeV   & 15 MeV    &  20 MeV    \\
selection bias              &  5 MeV   & 10 MeV    &  20 MeV    \\
recoil resolution           & 25 MeV   & 10 MeV    &  90 MeV    \\
recoil response             & 20 MeV   & 15 MeV    &  45 MeV    \\ \hline
total                       & 60 MeV   & 50 MeV    & 115 MeV    \\
\end{tabular}
\label{tab:detector_errors}
\end{center}
\end{table}

\begin{table}[ht]
\begin{center}
\caption{\small Uncertainties in the \wb\ mass measurement due to backgrounds.}
\begin{tabular}{lrrr}
            & \mt\ fit & \pte\ fit & \ptnu\ fit \\ \hline
hadrons     & 10 MeV   & 15 MeV    & 20 MeV     \\
\zee        &  5 MeV   & 10 MeV    &  5 MeV     \\
\wtv        & \multicolumn{3}{c}{negligible}    \\
cosmic rays & \multicolumn{3}{c}{negligible}    \\
 \hline
total       & 10 MeV   & 20 MeV    & 20 MeV     \\
\end{tabular}
\label{tab:bkg_errors}
\end{center}
\end{table}

\begin{table}[ht]
\begin{center}
\caption{\small Summary of results from the 1992--1993 and 1994--1995 data sets
with the common and uncorrelated errors.}
\begin{tabular}{lrrr}
    &\multicolumn{1}{c}{1992--1993} &\multicolumn{1}{c}{1994--1995} & common \\
    \hline 
\mw\ from \mt\ fit         & 80.35 \GeVm\ & 80.44 \GeVm\ & \\ \hline
\wb\ statistics            & 140 MeV & ~70 MeV &      \\
\zb\ statistics            & 160 MeV & ~65 MeV &      \\
calorimeter linearity      &         &         & ~20 MeV \\
calorimeter uniformity     &         &         & ~10 MeV \\
electron resolution        & ~70 MeV & ~20 MeV &      \\
electron angle calibration &         &         & ~30 MeV \\
recoil resolution          & ~90 MeV & ~25 MeV &      \\
recoil response            & ~50 MeV & ~20 MeV &      \\
electron removal           & ~35 MeV & ~15 MeV &      \\
selection bias             & ~30 MeV & ~~5 MeV &      \\
backgrounds                & ~35 MeV & ~10 MeV &      \\
\wb\ production/decay      &         &         & ~30 MeV \\ \hline
total uncertainty          & 255 MeV & 105 MeV & ~50 MeV \\
\end{tabular}
\label{tab:sum}
\end{center}
\end{table}

\begin{table}[ht]
\begin{center}
\caption{\small Previously published measurements of the \wb\ boson mass.}
\begin{tabular}{llc}
measurement & \mw\ (GeV)    & reference \\ \hline
CDF 90      & 79.91\PM0.39  & \protect\cite{CDF90} \\
UA2 92      & 80.36\PM0.37  & \protect\cite{UA2} \\
CDF 95      & 80.41\PM0.18  & \protect\cite{CDF} \\
\Dzero\ 96  & 80.35\PM0.27  & \protect\cite{D0} \\
OPAL 96     & $80.40^{+0.45}_{-0.42}$  & \protect\cite{OPAL} \\
DELPHI 97   & 80.40\PM0.45  & \protect\cite{DELPHI} \\
L3 97       & $80.80^{+0.48}_{-0.42}$ & \protect\cite{L3} \\
ALEPH 97    & 80.14\PM0.35  & \protect\cite{ALEPH}
\end{tabular}
\label{tab:mw}
\end{center}
\end{table}

\end{document}